\documentclass[prd,onecolumn,preprintnumbers,superscriptaddress,floatfix,nofootinbib,11pt]{revtex4}
\usepackage[utf8]{inputenc}
\usepackage{slashed}
\usepackage{color}
\usepackage{float}
\usepackage{bm,amsmath,amssymb,amsfonts}
\usepackage{epsfig,graphicx}
\usepackage{epstopdf}
\usepackage{subfigure}
\usepackage{appendix}
\allowdisplaybreaks
\usepackage{enumitem}
\setenumerate[1]{itemsep=0pt,partopsep=0pt,parsep=\parskip,topsep=0pt}
\setitemize[1]{itemsep=0pt,partopsep=0pt,parsep=\parskip,topsep=0pt}
\setdescription{itemsep=0pt,partopsep=0pt,parsep=\parskip,topsep=0pt}
\usepackage{longtable,lscape}
\usepackage{hyperref}
\hypersetup{colorlinks,citecolor=blue,anchorcolor=red,menucolor=red, linkcolor=red,filecolor=red,runcolor=red,urlcolor=blue,frenchlinks=true}
\usepackage{balance}
\usepackage{booktabs}
\usepackage{dcolumn}
\usepackage{makecell}
\usepackage{multirow}
\usepackage{soul}

\newcommand{\ket}[1]{\left| #1 \right\rangle}
\newcommand{\bra}[1]{\left\langle #1 \right|}

\begin{document}
	\title{Coupled-channel effects of the $\bm{\Sigma_c^{(*)}\bar{D}^{(*)}}$--$\bm{\Lambda_c(2595)\bar{D}}$ system and molecular nature of the $\bm{P_c}$ pentaquark states from one-boson exchange model }
	
	\author{Nijiati Yalikun}
	\email{nijiati@itp.ac.cn}
	\affiliation{CAS Key Laboratory of Theoretical Physics, Institute
		of Theoretical Physics,  Chinese Academy of Sciences, Beijing 100190, China}
	\affiliation{School of Physical Sciences, University of Chinese Academy of Sciences, Beijing 100049, China}
	
	\author{Yong-Hui Lin}
	\email{yonghui@hiskp.uni-bonn.de}
	\affiliation{Helmholtz-Institut~f\"{u}r~Strahlen-~und~Kernphysik~and~Bethe~Center~for
Theoretical~Physics, Universit\"{a}t~Bonn,  D-53115~Bonn,~Germany}
	
	\author{Feng-Kun~Guo}
	\email{fkguo@itp.ac.cn}
	\affiliation{CAS Key Laboratory of Theoretical Physics, Institute
		of Theoretical Physics,  Chinese Academy of Sciences, Beijing 100190, China}
	\affiliation{School of Physical Sciences, University of Chinese Academy of Sciences, Beijing 100049, China}
	
	\author{Yuki Kamiya}
	\email{yukikamiya@mail.itp.ac.cn}
	\affiliation{CAS Key Laboratory of Theoretical Physics, Institute
		of Theoretical Physics,  Chinese Academy of Sciences, Beijing 100190, China}
	
	\author{Bing-Song Zou}\email{zoubs@mail.itp.ac.cn}
	\affiliation{CAS Key Laboratory of Theoretical Physics, Institute
		of Theoretical Physics,  Chinese Academy of Sciences, Beijing 100190, China}
	\affiliation{School of Physical Sciences, University of Chinese Academy of Sciences, Beijing 100049, China}
	\affiliation{School of Physics and Electronics, Central South University, Changsha 410083, China}

\begin{abstract}
The effects of the $\Sigma_c\bar{D}^*$-$\Lambda_{c}(2595)\bar{D}$ coupled-channel  dynamics and various one-boson-exchange (OBE) forces 
for the LHCb pentaquark states, $P_c(4440)$ and $P_c(4457)$, are reinvestigated. Both the pion and $\rho$-meson exchanges are considered for the $\Sigma_c\bar{D}^*$-$\Lambda_{c}(2595)\bar{D}$ coupled-channel  dynamics. It is found that the role of the $\Lambda_{c}(2595)\bar{D}$ channel in the descriptions of the $P_c(4440)$ and $P_c(4457)$ states is not significant with the OBE parameters constrained by other experimental sources. The naive OBE models with the short-distance $\delta(\vec{r})$ term of the one-pion exchange (OPE) keep failing to reproduce the $P_c(4440)$ and $P_c(4457)$ states simultaneously. The OPE potential with the full $\delta(\vec{r})$ term results in a too large mass splitting for the $J^P=1/2^-$ and $3/2^-$ $\Sigma_c\bar{D}^*$ bound states with total isospin $I=1/2$.  The OBE model with only the OPE  $\delta(\vec{r})$ term dropped may fit the splitting much better but somewhat underestimates the splitting. Since the $\delta(\vec r)$ potential is from short-distance physics, which also contains contributions from the exchange of mesons heavier than those considered explicitly, we vary the strength of the $\delta(\vec r)$ potential and find that the masses of the $P_c(4312)$, $P_c(4440)$, and $P_c(4457)$ can be reproduced simultaneously with the $\delta(\vec r)$ term in the OBE model reduced by about 80\%.
While two different spin assignments are possible to produce their masses, in the preferred description, the spin parities of the $P_c(4440)$ and $P_c(4457)$ are $3/2^-$ and $1/2^-$, respectively.
		
\end{abstract}
	
\maketitle

\newpage 

\section{Introduction}~\label{sec:1}

The study of multiquark states began even before the birth of QCD and was accelerated with the development of QCD. 
It is speculated that, apart from well-known $qqq$ baryons and $q\bar q$ mesons, there would be multiquark states, glueballs and quark-gluon hybrids in the quark model notation, which are collectively called exotic hadrons.
Multiquark states can be categorized into tetraquark states ($qq\bar q \bar q$), pentaquark states ($qqq q\bar q$)
and so on. The study of multiquark states and how the quarks are grouped inside (i.e., in a compact multiquark state or as a hadronic molecule) plays a crucial role for understanding the low-energy QCD, and it is very important to search for them in experiments.

In this century, many candidates of exotic tetraquark and pentaquark states in the charm sector have been observed. A great intriguing fact is that most of them are located near hadron-hadron thresholds. This property can be understood as there is an $S$ wave attraction between the relevant hadron pair~\cite{Dong:2020hxe}, and naturally leads to the hadronic molecular interpretation for them (as reviewed in Refs.~\cite{Chen:2016qju,Guo:2017jvc,Brambilla:2019esw,Yamaguchi:2019vea,Dong:2021juy}). The validity of the hadronic molecular picture is also reflected by the successful quantitative predictions of some exotic states in early theoretical works based on the hadron-hadron  interaction dynamics~\cite{Tornqvist:1993ng,Wu:2010jy,Wu:2010vk,Wang:2011rga,Yang:2011wz,Wu:2012md,Xiao:2013yca,Uchino:2015uha,Karliner:2015ina}.

The first observation of pentaquark candidates with hidden charm, $P_c(4380)$ and $P_c(4450)$,  was reported by the LHCb Collaboration in 2015~\cite{Aaij:2015tga,Aaij:2016phn}. These $P_c$ states are located very close to the hidden-charm $N^*$ states predicted above 4~GeV~\cite{Wu:2010jy,Wu:2010vk,Wang:2011rga,Yang:2011wz,Wu:2012md,Yuan:2012wz,Xiao:2013yca,Uchino:2015uha,Karliner:2015ina} and stimulated many further theoretical studies based on various pictures, such as the charmed baryon and anticharmed meson molecules~\cite{Chen:2015moa,He:2015cea,Roca:2015dva,Chen:2015loa,Xiao:2015fia,Burns:2015dwa,Roca:2015dva,Mironov:2015ica,Meissner:2015mza,Lu:2016nnt,Shen:2016tzq,Kang:2016ezb,Shimizu:2016rrd,Yamaguchi:2016ote,Lin:2017mtz,Shimizu:2017xrg,Voloshin:2019aut,Gutsche:2019mkg},  compact pentaquark states~\cite{Ali:2016dkf,Ali:2019npk,Maiani:2015vwa,Li:2015gta,Mironov:2015ica,Anisovich:2015cia,Ghosh:2017fwg,Wang:2015epa,Hiyama:2018ukv}, baryocharmonia~\cite{Dubynskiy:2008mq,Kubarovsky:2015aaa,Perevalova:2016dln,Eides:2017xnt,Eides:2018lqg,Eides:2019tgv}, and triangle singularities~\cite{Guo:2015umn,Liu:2015fea,Bayar:2016ftu}.  
With about one-order-of-magnitude larger data sample from Run~II of the Large Hadron Collider, the $P_c(4450)$ peak has been reanalyzed by the LHCb Collaboration and found to be composed of two narrow overlapping peaks, $P_c(4440)$ and $P_c(4457)$~\cite{LHCb:2019kea}, and a new state, $P_c(4312)$, was also observed in their new analysis. 
Those states were observed in the $J/\psi p$ invariant mass spectrum, indicating that all the states contain a combination of the $uudc\bar c$ quark flavors. 
The masses of $P_c(4312)$ and $P_c(4440, 4457)$ are just below the mass thresholds of the $\Sigma_c \bar{D}$ and $\Sigma_c \bar{D}^*$ channels, respectively, suggesting a $\Sigma_c \bar D^{(*)}$ molecular structure for them~\cite{Guo:2019kdc,Xiao:2019mvs,Xiao:2020frg,Xiao:2019aya,Guo:2019fdo,Liu:2019tjn,He:2019ify,Liu:2019zoy,Shimizu:2019jfy,Weng:2019ynv,Wang:2019dsi,Cheng:2019obk,Voloshin:2019aut,Du:2019pij,PavonValderrama:2019nbk,Liu:2019zvb,Yang:2019rgw,Xu:2019zme,Xu:2020gjl,Yamaguchi:2019seo,Ke:2019bkf,Giachino:2020rkj,Yang:2020twg,Azizi:2020ogm,Liu:2020hcv,Peng:2020gwk,Chen:2021htr,Phumphan:2021tta,Du:2021fmf,Dong:2021juy}.  

A lot of theoretical works have been done to identify the spin parity of these three states. Chen $et~al.$~\cite{Chen:2019asm} study them with the one-boson-exchange (OBE) model assisted with heavy quark spin symmetry (HQSS). In their work, by considering the coupled-channel effects and the $S$-$D$ wave mixing, the $P_c(4312)$, $P_c(4440)$ and $P_c(4457)$ are assigned to be $J^P=1/2^-$  $\Sigma_c\bar{D}$, $1/2^-$ $\Sigma_c\bar{D}^*$ and $3/2^-$ $\Sigma_c\bar{D}^*$ bound states, respectively. 
He~\cite{He:2019ify} uses the quasipotential Bethe-Salpeter approach to study the $\Sigma_c\bar{D}$-$\Sigma_c^*\bar{D}$-$\Sigma_c\bar{D}^*$-$\Sigma_c^*\bar{D}^*$ coupled-channel system and obtains the same quantum numbers for those states as given in Ref.~\cite{Chen:2019asm}.  
However, in the $\Sigma_c D^*$ molecular picture, there are two possibilities for the quantum numbers of the $P_c(4440)$ and $P_c(4457)$. In addition to the above assignment, the $P_c(4440)$ and $P_c(4457)$ may also be $3/2^-$ and $1/2^-$ states, respectively, as suggested in Refs.~\cite{PavonValderrama:2019nbk,Liu:2019tjn,Sakai:2019qph} considering contact term interactions. 
Analyses considering one-pion exchange in addition to the contact terms for the $\Sigma_c \bar D^{(*)}$ interactions prefer the assignment of  $3/2^-$ for $P_c(4440)$ and $1/2^-$ for $P_c(4457)$~\cite{Liu:2019zvb,Du:2019pij}, which is also the conclusion of the most sophisticated coupled-channel analysis in Ref.~\cite{Du:2021fmf}.
Furthermore, the analysis of Ref.~\cite{Du:2019pij} provides hints for the existence of a narrow $P_c(4380)$ in the new data~\cite{LHCb:2019kea}, which was also pointed out earlier in Ref.~\cite{Xiao:2019aya}, in contrast to the broad one reported by LHCb in 2015~\cite{Aaij:2015tga}.

In Ref.~\cite{Burns:2019iih}, Burns $et~al.$  investigated the role of the $\Lambda_{c1}\bar{D}$ channel, which has a threshold at 4457~MeV, in the $P_c(4440)$ and $P_c(4457)$ states by considering the $\Sigma_c\bar{D}^*$-$\Lambda_{c1}\bar{D}$ coupled-channel  dynamics with the one-pion-exchange (OPE) model. 
They suggest that the $P_c(4457)$ and $P_c(4440)$ have spin-parity quantum numbers $1/2^+$ and $3/2^-$. In this model, the $P_c(4457)$ state has a positive parity because the quantum numbers of the $\Lambda_{c1}$ are $1/2^-$, and it is bound by the interplay between the $S$ wave $\Lambda_{c1}\bar{D}$ and the $P$ wave $\Sigma_c\bar{D}^*$~\cite{Geng:2017hxc,Burns:2019iih}. 
The inclusion of the $\Lambda_{c1}\bar{D}$ channel is quite novel. It is argued~\cite{Burns:2019iih} to be naturally produced with the color-favored weak decay of the $\Lambda_b$, and does not contribute to the isospin breaking ratio proposed in Ref.~\cite{Sakai:2019qph} as a diagnosis of the internal structure of the $P_c(4457)$. 
However, to reproduce the binding energy of the $P_c(4457)$, this model requires the $\Lambda_{c1}\Sigma_c\pi$ coupling constant to be much larger than the physical value deduced from experimental measurements~\cite{Zyla:2020zbs}. 
If the physical value of that coupling constant is used, there would be only one bound state with spin-parity $J^P=3/2^-$ which is related to either $P_c(4440)$ or $P_c(4457)$. 
In this work, we will reinvestigate such a coupled-channel system by including more possible meson exchange interactions. 
This will enable us to estimate the effects of the $\Lambda_{c1}\bar{D}$ channel more comprehensively.

We will also investigate the role of the short-distance $\delta(\vec{r})$ term in the $\Sigma_c^{(*)}\bar D^{(*)}$ coupled-channel  systems. There are different treatments of the  $\delta(\vec{r})$ term in the literature of the phenomenological meson-exchange models.
In Refs.~\cite{He:2019rva,He:2019ify}, the coupled-channel  effects of $\Sigma_c^{(*)}\bar D^{(*)}$ and $\Lambda_c^{(*)}\bar D^{(*)}$ are studied by solving the Bethe–Salpeter equation within the OBE model that includes the $\delta(\vec{r})$ term. In their work, the $P_c(4312)$, $P_c(4380)$, $P_c(4440)$ and $P_c(4457)$ are reproduced with several cutoff parameters for the exchange of different light mesons ($\Lambda_\sigma, \Lambda_\pi, \Lambda_\eta, \Lambda_\rho$, and  $\Lambda_\omega$) as 
$$
\text{scenario I}
\left\{
\begin{aligned}
	P_c(4312)&:&J^P=1/2^-(\Sigma_c\bar D),\\
	P_c(4380)&:&J^P=3/2^-(\Sigma_c^*\bar D),\\
	P_c(4440)&:&J^P=1/2^-(\Sigma_c\bar D^*),\\
	P_c(4457)&:&J^P=3/2^-(\Sigma_c\bar D^*).
\end{aligned}
\right.
$$
These various cutoffs are related to one parameter $\alpha$ by means of the definition $\Lambda_{\rm{ex}}=m_{\rm{ex}}+ \alpha \times(0.22\,\rm{GeV})$. 
The same assignment for such four $P_c$ states is also obtained in Ref.~\cite{Chen:2019asm}, in which the authors solve the $\Sigma_c^{(*)}\bar D^{(*)}$ coupled-channel systems in the coordinate space with also the $\delta(\vec{r})$ contribution kept in their OBE potentials, and two different cutoffs are needed in that work. However, in Ref.~\cite{Liu:2019zvb}, the $P_c$ states are studied separately in the single-channel $\Sigma_c^{(*)}\bar D^{(*)}$ systems with the $\delta(\vec{r})$ term discarded in their OBE model. Four $P_c$ states are reproduced with the same cutoff and their spin-parity quantum numbers are given as
$$
 	\text{scenario II}
 	\left\{
 	\begin{aligned}
P_c(4312)&:&J^P=1/2^-(\Sigma_c\bar D),\\
P_c(4380)&:&J^P=3/2^-(\Sigma_c^*\bar D),\\
P_c(4440)&:&J^P=3/2^-(\Sigma_c\bar D^*),\\
P_c(4457)&:&J^P=1/2^-(\Sigma_c\bar D^*),
 	\end{aligned}
 	\right.
$$
where the spin parities of the $P_c(4440)$ and $P_c(4457)$ are interchanged compared to scenario I. It may imply that the $\delta(\vec{r})$ term in the OBE model plays an important role in the hadronic molecular descriptions of the $P_c(4312)$, $P_c(4380)$, $P_c(4440)$ and $P_c(4457)$ states. 

In principle, an effective field theory framework, which introduces counterterms to parametrize additional short-distance contributions, is needed for a consistent treatment of the  $\delta(\vec{r})$ term.
In this work, we rather follow a phenomenological path, and investigate the role of the  $\delta(\vec{r})$ term by adjusting its strength in the OBE model. 
We will take the OBE model with the potential derived from the Lagrangian with HQSS as given in Refs.~\cite{Cheng:1992xi,Yan:1992gz,Wise:1992hn,Liu:2011xc}. A percentage parameter $a$ is introduced to represent how much the $\delta(\vec{r})$ term is reduced in the potential of our OBE model, and thus mimics the variation of a constant contact term in a nonrelativistic effective field theory framework at leading order. The parameter $a$ is varied in the range of $[0,1]$, that is, $a=0(1)$ denotes fully including (excluding) the $\delta(\vec{r})$ term.          	
	
This paper is organized as follows. The details of the OBE model in the $\Sigma_c^{(*)}\bar{D}^{(*)}$-$\Lambda_{c1}\bar{D}$ coupled-channel  system are introduced in Sec.~\ref{sec:2}. Numerical results and discussions on the role of $\Lambda_{c1}\bar D$ channel and the $\delta(\vec{r})$ term in the OBE potential are given in Sec.~\ref{sec:3}. Finally, we draw our conclusion in Sec.~\ref{sec:4}.

\section{Formalism}\label{sec:2}

In this section, the phenomenological heavy hadron chiral Lagrangian is reviewed, and the potentials for the $\Sigma_c^{(*)}\bar{D}^{(*)}$-$\Lambda_{c1}\bar{D}$ system are constructed within the OBE framework.

\subsection{Lagrangian}\label{subsec_Lag}

To investigate the interactions between a charmed baryon or an anticharmed meson with light scalar, pseudoscalar and vector mesons, we employ the effective Lagrangian taking into account HQSS which has been developed in Refs.~\cite{Cheng:1992xi,Yan:1992gz,Wise:1992hn,Cho:1994vg,Casalbuoni:1996pg,Pirjol:1997nh,Liu:2011xc}. The light chiral and heavy quark spin symmetric effective Lagrangian, which describes the low-energy interactions between the $Qqq$ baryons/$\bar Q q$ mesons and the light bosons, can be written as
\begin{eqnarray}\label{lag}
	{\mathcal L}_{eff}&=&
	l_S\bar{S}_{a\mu} \sigma S^\mu_a
	-\frac{3}{2} g_1\varepsilon_{\mu\nu\lambda\kappa}v^\kappa \bar{S}_{ab}^\mu
	A_{bc}^\nu S_{ca}^\lambda
	+i\beta_S\bar{S}_{ab\mu} v_\alpha (\Gamma^\alpha_{bc}-\rho^\alpha_{bc}) S^\mu_{ca}
	+ \lambda_S\bar{S}_{ab\mu} F^{\mu\nu}(\rho_{bc})S_{ca\nu}\notag
	\\
	&-&ih_2[\bar S^\mu_{ab} v\cdot A_{bc} R_{ca\mu} +\bar R^\mu_{ab} v\cdot A_{cb} S_{ca\mu}]+h_3\varepsilon_{\mu\nu\lambda\kappa} iv^\kappa [ \bar S^\mu_{ab}(\Gamma^\nu_{bc}-\rho^\nu_{bc} )R^\lambda_{ca}+\bar R^\mu_{ab}(\Gamma^\nu_{bc}-\rho^\nu_{bc} )S^\lambda_{ca}]\nonumber\\
	&+&g_S{\rm Tr}[ \bar H_a^{\bar Q} \sigma H_a^{\bar Q}]+ig {\rm Tr}[ \bar H_a^{\bar Q} \gamma\cdot A_{ab}\gamma^5H_b^{\bar Q}]-i\beta{\rm Tr}[\bar H_a^{\bar Q} v_\mu(\Gamma^\mu_{ab}-\rho^\mu_{ab})H_b^{\bar Q}]\notag\\
	&+&i\lambda{\rm Tr}\left[ \bar H_a^{\bar Q} \frac{i}{2}[\gamma_\mu,\gamma_\nu]F^{\mu\nu}(\rho_{ab})H_b^{\bar Q}\right],
\end{eqnarray}
where flavor indices are denoted by $a,~b$ and $c$ and $\sigma$ is the lightest scalar meson, which is taken to be an SU(3) flavor singlet, 
\begin{align}
A^\mu &=\frac{1}{2}(\xi^\dagger\partial^\mu\xi-\xi\partial^\mu\xi^\dagger)=\frac{i}{f_\pi}\partial^\mu\mathbb{P}+\cdots, \notag\\
\Gamma^\mu &=\frac{i}{2}(\xi^\dagger\partial^\mu\xi+\xi\partial^\mu\xi^\dagger)=\frac{i}{2f_\pi^2}[\mathbb{P},\partial^\mu\mathbb{P}]+\cdots, 
\end{align}
with $\xi={\rm exp}(i\mathbb{P}/f_\pi)$ and $\rho^\alpha={ig_V}\mathbb{V}^\alpha/{\sqrt 2}$, where $f_\pi=132$~MeV is the pion decay constant. The symbols $\mathbb{P}$ and $\mathbb{V}^\alpha$ denote the light pseudoscalar octet and the vector nonet, respectively, 
\begin{eqnarray}
	\mathbb{P}=
	\begin{pmatrix}
		\frac{\pi^0}{\sqrt 2}+\frac{\eta}{\sqrt 6}&\pi^+&K^+\\
		\pi^-&-\frac{\pi^0}{\sqrt 2}+\frac{\eta}{\sqrt 6}&K^0\\
		K^-&\bar K^0&-\sqrt{\frac{2}{3}}\eta
	\end{pmatrix},\qquad
	\mathbb{V}^\alpha=
	\begin{pmatrix}
		\frac{\rho^0}{\sqrt 2}+\frac{\omega}{\sqrt 2}&\rho^+&K^{*+}\\
		\rho^-&-\frac{\rho^0}{\sqrt 2}+\frac{\omega}{\sqrt 2}&K^{*0}\\
		K^{*-}&\bar K^{*0}&\phi.
	\end{pmatrix}^\alpha .
\end{eqnarray}

Field operators for the $S$- (positive parity) and $P$ wave  (negative parity) heavy baryons $Qqq$ 
are denoted as interpolating fields $S^\mu_{ab}$ and $R^\mu_{ab}$, respectively. $H^{\bar Q}_a$ annihilates the antiheavy meson $\bar Qq$. They are defined as 
\begin{align}
	S^\mu_{ab}&=-\frac{1}{\sqrt3}(\gamma^\mu+v^\mu)\gamma^5 (B_{6Q})_{ab}+(B_{6Q}^{*\mu})_{ab},&
	\bar S^\mu_{ab}&=S^{\mu\dagger}_{ab}\gamma^0,\\
	R^\mu_{ab}&=-\frac{1}{\sqrt3}(\gamma^\mu+v^\mu)\gamma^5 (B'_{3Q})_{ab}+(B_{3Q}^{'*\mu})_{ab},&
	\bar R^\mu_{ab}&=R^{\mu\dagger}_{ab}\gamma^0,\\
	H_a^{\bar Q}&=(\tilde{\mathcal{P}}^*_{a\mu}\gamma^\mu-\tilde{\mathcal{P}}_a\gamma^5)\frac{1-\slashed v}{2},&
	\bar H_a^{\bar Q}&=\gamma^0H_a^{\bar Q\dagger}\gamma^0,
\end{align}
where the anticharmed pseudoscalar $\tilde{\mathcal{P}}_a$ and vector $\tilde{\mathcal{P}}^*_{a\mu}$ fields\footnote{Note that here $\tilde{\mathcal{P}}_a$ and $\tilde{\mathcal{P}}^*_{a\mu}$ are the heavy meson fields satisfying the normalization relations $\bra{0}\tilde{\mathcal{P}}\ket{\bar Q q(0^-)}=\sqrt{M_{\tilde{\mathcal{P}}}} $ and $\bra{0}\tilde{\mathcal{P}}^*_\mu\ket{\bar Q q(1^-)}=\epsilon_\mu\sqrt{M_{\tilde{\mathcal{P}}^*}} $~\cite{Wise:1992hn}. With this convention, all physical effective couplings for the heavy meson pair ($h_1$, $h_2$) interacting with light mesons are related to the couplings in the Lagrangian \eqref{lag} by multiplying an additional factor $\sqrt{M_{h_1}M_{h_2}}$ on the latter.  } are defined in flavor/isospin space as $(\bar D^0, D^-, D_s^{-} )$ and $(\bar D^{*0}, D^{*-}, D_s^{*-})$, respectively, with the subscript $a$ is the light flavor index, and $v^\mu=(1,0,0,0)$ is the 4-velocity of the heavy hadron. The charmed baryon fields $B_{6c}$ and $B'_{3c}$ in the SU(3) flavor space are written as 
\begin{eqnarray}
	B_{6c}=
	\begin{pmatrix}
		\Sigma_c^{++}&\Sigma_c^+/\sqrt 2&\Xi_c^{'+}/\sqrt 2\\
		\Sigma_c^+/\sqrt 2&\Sigma_c^{0}&\Xi_c^{'0}/\sqrt 2\\
		\Xi_c^{'+}/\sqrt 2&\Xi_c^{'0}/\sqrt 2&\Omega^0_c
	\end{pmatrix},\qquad
B'_{3c}=\left(\begin{array}{ccc}
	0&\Lambda_{c1}^+&\Xi^+_{c1}\\
	-\Lambda_{c1}^+&0&\Xi^0_{c1}\\
	-\Xi^+_{c1}&-\Xi^0_{c1}&0
\end{array}\right).
\end{eqnarray}
$B_{6c}^*$ and $B^{'*}_{3c}$ are the spin excited states of $B_{6c}$ and $B'_{3c}$, respectively, and have the same flavor matrix forms as above; that is, $B_{6c}^*$ and $B^{'*}_{3c}$ are spin-$3/2$ fields, and $B_{6c}$ and $B'_{3c}$ are spin-$1/2$ ones.

\subsection{Partial wave representation}

In our analysis, we consider three possible spin-parity states, $J^P=1/2^-,3/2^-$ and $1/2^+$ for the $\Sigma_c^{(*)}\bar{D}^{(*)}$-$\Lambda_{c1}\bar{D}$ coupled-channel system. The corresponding partial wave basis is listed in Table~\ref{part_waves} where we use the notation $^{2S+1}L_J$ to identify various partial waves. $S$, $L$ and $J$ stand for the total spin, orbital and total angular momenta,  respectively. 
\begin{table}[H]\centering
\caption{Partial waves for the $\Sigma_c^{(*)}\bar{D}^{(*)}$-$\Lambda_{c1}\bar{D}$ coupled-channel system.}\label{part_waves}
\begin{ruledtabular}
	\begin{tabular}{c|c|c|c|c|c}
		&$\Sigma_c\bar D$&$\Sigma_c^*\bar D$&$\Sigma_c\bar D^*$&$\Sigma_c^*\bar D^*$&$\Lambda_{c1}\bar D$\\\hline
		$J^P=1/2^-$&$^2S_{1/2}$&$^4D_{1/2}$&$^2S_{1/2}$, $^4D_{1/2}$&$^2S_{1/2},^4D_{1/2},^6D_{1/2}$&$^2P_{1/2}$\\
		$J^P=3/2^-$&$^2D_{3/2}$&$^4S_{3/2},^4D_{3/2}$&$^4S_{3/2},^2D_{3/2},^4D_{3/2}$&$^4S_{3/2},^2D_{3/2},^4D_{3/2},^6D_{3/2}$&$^2P_{3/2}$\\
		$J^P=1/2^+$&$^2P_{1/2}$&$^4P_{1/2}$&$^2P_{1/2},^4P_{1/2}$&$^2P_{1/2},^4P_{1/2}$&$^2S_{1/2}$ 
	\end{tabular}
\end{ruledtabular}
\end{table}
The partial wave function $|^{2S+1}L_J\rangle$ is explicitly written in the standard form as
\begin{eqnarray}\label{def_state}
	|^{2S+1}L_J\rangle=|LSJm\rangle=\sum\limits_{m_lm_s}\mathbb{C}_{Lm_l,Sm_s}^{Jm}|Lm_l\rangle|Sm_s\rangle,
\end{eqnarray}
where $\mathbb{C}_{Lm_l,Sm_s}^{Jm}$ is the Clebsch-Gordan (CG) coefficient, $|Sm_s\rangle$ is the spin wave function and $|Lm_l\rangle$ is the spherical harmonics.

One critical point in the partial wave implementation is the spin-orbital ordering convention which is hardly discussed in earlier works. 
Note that a change of the ordering of spin and orbital angular momenta that converts $\mathbb{C}_{Lm_l,Sm_s}^{Jm}$ into $\mathbb{C}_{Sm_s,Lm_l}^{Jm}$  will lead to an additional sign on the matrix elements of the spin-orbital operators, such as $S(\hat r,\sigma,i\epsilon_4^\dagger\times\epsilon_2)$ and $\epsilon_2\cdot\hat r$, which are obtained after being sandwiched between the $|^{2S+1}L_J\rangle$ states. 
A detailed illustration can be found in Appendix~\ref{app_spin}. However, as long as the same convention is used throughout the calculation, the derived potentials would not depend on the convention.
In this paper, all matrix elements are obtained with the $\mathbb{C}_{Lm_l,Sm_s}^{Jm}$ convention.

\subsection{Potentials}\label{subsec_potentials}

To get the OBE potentials for the $\Sigma_c^{(*)}\bar{D}^{(*)}$-$\Lambda_{c1}\bar{D}$ system, we need to derive the $t$-channel scattering amplitude $\mathcal{M}$ in the center-of-mass frame first. 
Note that the nonrelativistic approximation for the charmed hadrons is implemented in our calculation. 
Potentials in the momentum space can be obtained from the $t$-channel scattering amplitudes with the Breit approximation, that is, $V(\vec q)=-\mathcal{M}/\sqrt{(\Pi_i2M_i)(\Pi_f2M_f)}$, where $M_{f(i)}$ are the masses of particles in the final (initial) states.

 It is convenient to label the five channels considered in our work for the $\Sigma_c^{(*)}\bar{D}^{(*)}$-$\Lambda_{c1}\bar{D}$ system, i.e., $\Sigma_c\bar{D}$, $\Sigma_c^*\bar{D}$, $\Lambda_{c1}\bar{D}$, $\Sigma_c\bar{D}^*$, $\Sigma_c^*\bar{D}^*$, as the first, second, third, fourth, and
 fifth channels, respectively. They are sorted simply by their thresholds. The OBE potential in the momentum space for this five-channel coupled system can be written as
 \begin{eqnarray}
 	V=V_\sigma+V_{\pi}+V_{\eta}+V_{\rho}+V_{\omega},
 \end{eqnarray}
where all components are given as, 
 \begin{align}
 	V^{ij}_\sigma(\vec q)&= -A^{ij}_\sigma\mathcal{O}_1^{ij}\frac{1}{\vec q^{\,2}+\mu^2_{ij}},\label{p_sigma_pot}\\
 	V^{ij}_{\pi/\eta}(\vec q)&= B^{ij}_{\pi/\eta} (\mathcal{\vec O}_3^{ij}\cdot\vec q)(\mathcal{\vec O}_4^{ij}\cdot\vec q)\frac{1}{\vec q^{\,2}+\mu^2_{ij}},\label{pi_p_pot}\\
 	V^{ij}_{\rho/\omega}(\vec q)&= -C^{ij}_{\rho/\omega} \mathcal{O}_2^{ij}\frac{1}{\vec q^{\,2}+\mu^2_{ij}}+D^{ij}_{\rho/\omega}(\mathcal{\vec O}_3^{ij}\times\vec q)\cdot(\mathcal{\vec O}_4^{ij}\times\vec q)\frac{1}{\vec q^{\,2}+\mu^2_{ij}},\label{p_rho_pot}\\
 	V^{3k}_{\pi}(\vec q)&= E^{3k}_\pi (\mathcal{\vec O}_5^{3k}\cdot\vec q)\frac{1}{\vec q^{\,2}+\mu^2_{3k}},\label{p_pix3_pot}\\
    V^{3k}_{\rho}(\vec q)&= F^{3k}_\rho \mathcal{\vec O}_{6}^{3k}\cdot(\mathcal{\vec O}_{5}^{3k}\times\vec q)\frac{1}{\vec q^{\,2}+\mu^2_{3k}}.\label{p_vx3_pot}
 \end{align}
Here $i,j$ and $k$ are the channel indies with $i,j=1,2,4,5$ and $k=4,5$. $\mathcal{O}_{1},\cdots,\mathcal{O}_{6}$ are the spin operators in the $t$-channel transition. They are given explicitly in Appendix~\ref{app_spin}.  $\vec q$ and $\mu_{ij}=\sqrt{m^2_\text{ex}-\omega_{ij}^2}$ are the 3-momentum and effective mass for the exchanged meson, where $\omega_{ij}$ is the energy of exchanged meson in the $t$-channel transition $i\to j$. Note that there is no energy exchange in the elastic transition ($i=j$); that is, $\omega_{ii}=0$. $A,B,C,D,E$ and $F$ denote symmetric constant matrices consisting of the coupling constants and the flavor factors. All nonzero elements are listed below, 
\begin{eqnarray}
	&\quad&A^{11}_\sigma=A^{22}_\sigma=A^{44}_\sigma=A^{55}_\sigma=l_Sg_S,\nonumber\\
	&\quad&B^{14}_{\pi/\eta}=T_{\pi/\eta}\frac{gg_1}{f_\pi^2},\quad	B^{15}_{\pi/\eta}=B^{24}_{\pi/\eta}=B^{45}_{\pi/\eta}=T_{\pi/\eta}\frac{\sqrt{3}gg_1}{2f_\pi^2},\nonumber\\
	&\quad&B^{25}_{\pi/\eta}=B^{55}_{\pi/\eta}=-T_{\pi/\eta}\frac{3gg_1}{2f_\pi^2},\quad	B^{44}_{\pi/\eta}=T_{\pi/\eta}\frac{gg_1}{f_\pi^2},\nonumber\\
	&\quad&C_{\rho/\omega}^{11}=C_{\rho/\omega}^{22}=C_{\rho/\omega}^{44}=C_{\rho/\omega}^{55}=T_{\rho/\omega}\frac{1}{2}\beta\beta_Sg_V^2,\quad\nonumber\\
	&\quad&D_{\rho/\omega}^{14}=T_{\rho/\omega}\frac{2}{3}\lambda\lambda_Sg_V^2,\quad 
	D_{\rho/\omega}^{15}=D_{\rho/\omega}^{24}=D_{\rho/\omega}^{45}=T_{\rho/\omega}\frac{1}{\sqrt 3}\lambda\lambda_Sg_V^2,\nonumber\\
	&\quad&D_{\rho/\omega}^{25}=D_{\rho/\omega}^{55}=-T_{\rho/\omega}\lambda\lambda_Sg_V^2,\quad 
	D_{\rho/\omega}^{44}=T_{\rho/\omega}\frac{2}{3}\lambda\lambda_Sg_V^2,\quad\nonumber\\
	&\quad&E_{\pi}^{34}=-\tau_\pi\frac{h_2g\omega_{34}}{f_\pi^2},\quad F_{\rho}^{34}=-\tau_\rho\frac{1}{3}h_3\lambda g_V^2,\quad
	F_{\rho}^{35}=-\tau_\rho\frac{i}{\sqrt{3}}h_3\lambda g_V^2, 
\end{eqnarray} 
where $\{T_{\pi},T_{\eta},T_{\rho},T_{\omega}\}=\{-1,1/6,-1,1/2\}$ are the flavor factors and $\tau_{\pi/\rho}=\sqrt{3/2}$ for the isospin-$1/2$ system.

After implementing the Fourier transformation, we can obtain potentials in the coordinate space. With the dipole form factors included, they read as
 \begin{eqnarray}
 	V^{ij}_\text{ex}(r)=\frac{1}{(2\pi)^3}\int V^{ij}_\text{ex}(\vec q) \left(\frac{\Lambda^2_{ij}-\mu_{ij}^2}{\Lambda^2_{ij}+\vec q^{\,2}}\right)^2 \exp(i\vec q\cdot \vec r)d^3\vec q,
 \end{eqnarray}
with $\Lambda_{ij}=\sqrt{\Lambda^2-\omega_{ij}^2}$. For the $\sigma$ meson exchange, one gets
 \begin{eqnarray}
 		V^{ij}_\sigma(r)&=&-A^{ij}_\sigma\mathcal{O}_1^{ij}Y(r,\Lambda_{ij},\mu_{ij}),\\
 		Y(r,\Lambda_{ij},\mu_{ij})&=&	\frac{1}{(2\pi)^3}\int\frac{1}{\vec q^{\,2}+\mu_{ij}^2} \left(\frac{\Lambda^2_{ij}-\mu_{ij}^2}{\Lambda^2_{ij}+\vec q^{\,2}}\right)^2 \exp(i\vec q\cdot \vec r)d^3\vec q ,
 \end{eqnarray}
where $Y(r,\Lambda_{ij},\mu_{ij})$ is the attractive Yukawa potential. Before implementing the Fourier integral of the pseudoscalar-exchange potential, we decompose it into two terms as usually done in the literature, 
\begin{eqnarray}
	V^{ij}_{\pi/\eta}(\vec q)&=& B^{ij}_{\pi/\eta} \frac{1}{3}\left\{\mathcal{\vec O}_3^{ij}\cdot \mathcal{\vec O}_4^{ij}\left(1-\frac{\mu^2_{ij}}{\vec q^{\,2} +\mu_{ij}^2}\right)+\left(3\mathcal{\vec O}_3^{ij}\cdot\hat q \mathcal{\vec O}_4^{ij}\hat q-\mathcal{\vec O}_3^{ij}\cdot \mathcal{\vec O}_4^{ij}\right)\frac{\vec q^{\,2}}{\vec q^{\,2} +\mu_{ij}^2}  \right\}.
\end{eqnarray}
Note that the constant 1 inside the parentheses in the first term leads to a short-range 
$\delta$ potential [$\delta(\vec{r})$ term in the coordinate space]. 
As discussed in the Introduction, the short-distance contribution cannot be fully captured by the OBE model, which may be viewed as there can be contributions from exchanging heavier particles. Thus, we introduce a parameter $a$ to adjust the strength of the $\delta(\vec r)$ potential. It is introduced as               
 \begin{eqnarray}
 	V^{ij}_{\pi/\eta}(\vec q)&=& B^{ij}_{\pi/\eta} \frac{1}{3}\left\{\mathcal{\vec O}_3^{ij}\cdot \mathcal{\vec O}_4^{ij}(1-a-\frac{\mu^2_{ij}}{\vec q^{\,2} +\mu_{ij}^2})+S(\mathcal{\vec O}_3^{ij},\mathcal{\vec O}_4^{ij},\hat q)\frac{\vec q^{\,2}}{\vec q^{\,2} +\mu_{ij}^2}  \right\}\label{pi_p_pot2},
 \end{eqnarray}
with  $S(\mathcal{\vec O}_3^{ij},\mathcal{\vec O}_4^{ij},\hat q)\equiv3\mathcal{\vec O}_3^{ij}\cdot\hat q \mathcal{\vec O}_4^{ij}\hat q-\mathcal{\vec O}_3^{ij}\cdot \mathcal{\vec O}_4^{ij}$. 
Thus, $a=0$ corresponds to the case with the full $\delta(\vec r)$ potential of the exchanged meson, and $a=1$ corresponds to the case without it.
Then, the $V^{ij}_{\pi/\eta}(r)$ can be obtained as
  \begin{eqnarray}
 	V^{ij}_{\pi/\eta}(r)&=& -B^{ij}_{\pi/\eta} \frac{1}{3}\left\{\mathcal{\vec O}_3^{ij}\cdot \mathcal{\vec O}_4^{ij}C(r,\Lambda_{ij},\mu_{ij},a)+S(\mathcal{\vec O}_3^{ij},\mathcal{\vec O}_4^{ij},\hat r)T(r,\Lambda_{ij},\mu_{ij})  \right\}\label{pi_r_pot},
 \end{eqnarray}
 where 
  \begin{eqnarray}
    C(r,\Lambda_{ij},\mu_{ij},a) &=&\frac{1}{r^2}\frac{\partial}{\partial r}r^2\frac{\partial}{\partial r}Y(r,\Lambda_{ij},\mu_{ij})+a\frac{1}{(2\pi)^3}\int \left(\frac{\Lambda^2_{ij}-\mu_{ij}^2}{\Lambda^2_{ij}+\vec q^{\,2}}\right)^2e^{i\vec q \cdot \vec r}d^3\vec q, \\
 	T(r,\Lambda_{ij},\mu_{ij}) &=&r\frac{\partial}{\partial r}\frac{1}{r}\frac{\partial}{\partial r}Y(r,\Lambda_{ij},\mu_{ij}).
 \end{eqnarray}
With the same procedure, we can obtain the $r$-space potentials for Eqs.~(\ref{p_rho_pot}), ~(\ref{p_pix3_pot}) and ~(\ref{p_vx3_pot}), 
\begin{align}
 	V^{ij}_{\rho/\omega}(r)=& -C^{ij}_{\rho/\omega}\mathcal{O}_2^{ij}Y(r,\Lambda_{ij},\mu_{ij})-D^{ij}_{\rho/\omega} \frac{1}{3}\left\{2\mathcal{\vec O}_3^{ij}\cdot \mathcal{\vec O}_4^{ij}C(r,\Lambda_{ij},\mu_{ij},a)-S(\mathcal{\vec O}_3^{ij},\mathcal{\vec O}_4^{ij},\hat r)T(r,\Lambda_{ij},\mu_{ij})  \right\},\label{vec_r_pot}\\
 	 V^{3k}_{\pi}(r)=& -iE^{3k}_\pi \mathcal{\vec O}_5^{3k}\cdot \hat r\frac{\partial}{\partial r}Y(r,\Lambda_{3k},\mu_{3k}) ,\\
 	 V^{3k}_{\rho}(r)=& -iF^{ij}_\rho  \mathcal{\vec O}_{ 6}^{3k}\cdot(\mathcal{\vec O}_{5}^{3k}\times \hat r)\frac{\partial}{\partial r}Y(r,\Lambda_{3k},\mu_{3k}).
\end{align}

\begin{figure}[tb]\centering
	\includegraphics[width=0.6\textwidth]{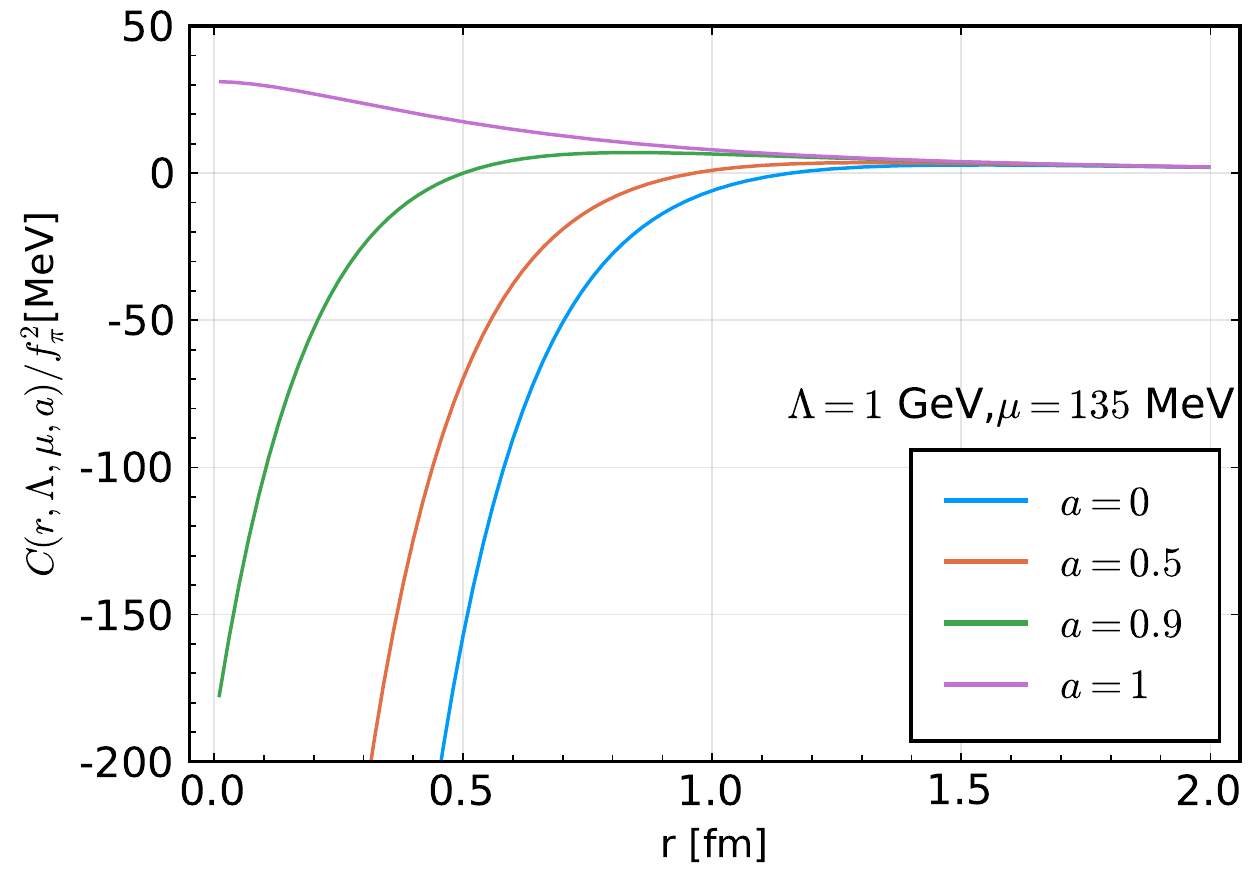}
	\caption{Central potential with different $a$ values.}\label{fig_deltapot}
\end{figure}

The $\delta(\vec{r})$ term in the central potential $C(r)$ appears not only in the pseudoscalar exchange potentials but also in the vector meson case. Figure~\ref{fig_deltapot} shows the contribution of the $\delta(\vec{r})$ term to the central potential, where $a=0(1)$ means a full inclusion (exclusion) of the $\delta(\vec{r})$ term. If the $\delta(\vec{r})$ term is fully removed, the central potential becomes very weak and repulsive.       

For the numerical analysis, we need some input parameters, such as masses of particles and coupling constants in Lagrangian Eq.~(\ref{lag}). All the masses of particles are referred to the isospin-averaged values of the experimental masses listed in the Review of Particle Physics~\cite{Zyla:2020zbs} and are collected in Table ~\ref{mass_parr}. The coupling constants are given in Table~\ref{coupl_constant}.   
\begin{table}[t]\centering
	\caption{Masses of the involved particles in units of MeV used in our calculation.}\label{mass_parr}
	\begin{ruledtabular}
	\begin{tabular}{c c c c c c c c c c}
$\sigma$&$\pi$&$\eta$&$\rho$&$\omega$&$\bar  D$&$\bar D^*$&$\Sigma_c$& $\Sigma_c^*$ & $\Lambda_{c1}$\\\hline
500.0&137.2&547.9&775.3&782.7&1867.2&2008.6&2453.9&2518.1&2592.3
	\end{tabular}
	\end{ruledtabular}
\end{table}
\begin{table}[t]\centering
	\caption{Coupling constants used in the calculation.}\label{coupl_constant}
	\begin{ruledtabular}
	\begin{tabular}{c c c c c c c c c c c}
		$g_S$~\cite{Ding:2008gr} & $l_S$~\cite{Liu:2011xc} & $g$~\cite{Meng:2019ilv} & $g_1$~\cite{Liu:2011xc} & $\beta$~\cite{Isola:2003fh} & $\beta_S$~\cite{Liu:2011xc} & $\lambda$~\cite{Isola:2003fh} & $\lambda_S$~\cite{Liu:2011xc} & $g_V$~\cite{Isola:2003fh}&$h_2$~\cite{Cheng:2015naa}&$h_3$\\\hline
		$0.76$ & $6.2$ & $-0.59$ & $0.94$ & $0.9$ & $-1.74$ & $0.56/\rm{GeV}$ & $-3.31/\rm{GeV}$ & $5.9$&$0.62$&$0.24$
	\end{tabular}
	\end{ruledtabular}
\end{table}
It should be mentioned that the scalar meson coupling constant $g_S=0.76$ for the $DD\sigma$ vertex is different from the values used in Refs.~\cite{Liu:2019zvb,GellMann:1960np,Liu:2011xc} by a factor of $2m_D$ due to the different conventions as introduced in Sec.~\ref{subsec_Lag}. 
The coupling $l_S$ 
in the $\Sigma_c^{(*)}\Sigma_c^{(*)}\sigma$ vertex is determined in Refs.~\cite{Ding:2008gr,Liu:2011xc} with the chiral multiplet assumption~\cite{Bardeen:2003kt}. 
The pseudoscalar meson couplings $g$ and $g_1$ are determined in Refs.~\cite{Meng:2019ilv,Liu:2011xc} from the experimental decay widths of $D^*\to D \pi $ and $\Sigma^*_c\to \Lambda_c \pi$~\cite{Zyla:2020zbs} (quark model relations are used to relate $g_1$ to the latter process). The vector meson couplings $\beta$, $\lambda$, $\beta_S$, $\lambda_S$ and $g_V$ are determined in Refs.~\cite{Liu:2011xc,Isola:2003fh} with the vector-meson dominance assumption. 
The coupling of the vector meson with the $P$- and $S$ wave baryons $h_3$ may be roughly estimated from the $\Lambda_{c1}\to\Sigma_c\gamma$ decay via the vector meson dominance assumption~\cite{Bando:1987br,Nagahiro:2008cv}. At the tree level, the radiative decay width of $\Lambda_{c1}\to\Sigma_c\gamma$ can be calculated as
\begin{eqnarray}
	\Gamma(\Lambda_{c1}\to\Sigma_c\gamma)&=&\frac{\alpha_{EM}h_3^2}{36m^5_{\Lambda_{c1}}m_\rho^2}(m_{\Lambda_{c1}}-m_{\Sigma_c})(m_{\Lambda_{c1}}+m_{\Sigma_c})^3(m^4_{\Lambda_{c1}}-2m^2_{\Lambda_{c1}}m^2_{\Sigma_c}+6m^2_{\Lambda_{c1}}m_\rho^2+m^4_{\Sigma_c})\nonumber\\
	&=&1.28h_3^2\text{ MeV},
\end{eqnarray}
where $\alpha_{EM}=1/137$ is the fine structure constant. The radiative decay width $\Gamma(\Lambda_{c1}\to\Sigma_c\gamma)$ has not been measured, but the prediction of it may help us to estimate the value of $h_3$. The decay width  $\Gamma(\Lambda_{c1}\to\Sigma_c\gamma)$ is  investigated in Ref.~\cite{Tawfiq:1999cf} with HQSS and predicted to be $71$~keV. One can infer the coupling constant $h_3$ from this to be $0.24$.

\section{Numerical results and discussion}\label{sec:3}

\subsection{Role of the $\Lambda_{c1}\bar{D}$ channel}

The $\Lambda_{c1}\bar{D}$ channel has its threshold very close to the mass of the $P_c(4457)$, and couples to $J^P=1/2^+$ in an $S$ wave due to the negative parity of the $\Lambda_{c1}$ baryon. 
It is interesting to study whether the $\Lambda_{c1}\bar{D}$ channel can trigger the formation of some $J^P=1/2^+$ molecular candidate near its threshold.  In Ref.~\cite{Burns:2019iih}, the $J^P=1/2^+$ system was investigated with the $\Sigma_c\bar{D}^*$-$\Lambda_{c1}\bar{D}$ coupled channels considering OPE with couplings from quark model. 
They demonstrated that if the nondiagonal potential of the $\Sigma_c\bar{D}^*$-$\Lambda_{c1}\bar{D}$ coupled channel is strong enough it is possible to reproduce  the $P_c(4457)$ and $P_c(4440)$ as $J^P=1/2^+$ and $3/2^-$ molecules simultaneously. 
Now, we revisit the role of $\Lambda_{c1}\bar{D}$ channel by including the vector meson exchange potential in the coupled channels of $\Lambda_{c1}\bar{D}$-$\Sigma_c\bar{D}^*$-$\Sigma_c^*\bar{D}^*$. The other channels below the $\Lambda_{c1}\bar{D}$ threshold are not considered because $\Sigma_c^{(*)}\bar{D}\to\Lambda_{c1}\bar{D}$ transitions, where only $\rho$ meson exchange is allowed, give zero amplitudes which can be deduced from the Lagrangian in Eq.~\eqref{lag}.

All potentials needed in our work are formulated in Sec.~\ref{subsec_potentials}. Here we use potentials of the $\Lambda_{c1}\bar{D}$-$\Sigma_c\bar{D}^*$-$\Sigma_c^*\bar{D}^*$ coupled channels to solve the Schr\"odinger equation to probe the molecular states with $J^P=1/2^+$ and $3/2^-$. 
To be consistent with Sect.~\ref{subsec_potentials}, we enumerate the three channels $\Lambda_{c1}\bar{D}$, $\Sigma_c\bar{D}^*$, and $\Sigma_c^*\bar{D}^*$, with channel labels $3,~4$ and $5$, which are ordered according to the channel thresholds $W_3,~W_4$ and $W_5$. The coupled-channel  Schr\"odinger equation for spherically symmetric potentials can be written as
\begin{eqnarray}
	\left[
	\begin{pmatrix}
		-\frac{\hbar^2}{2\mu_3}\nabla^2_3&0&0\\
		0&-\frac{\hbar^2}{2\mu_4}\nabla^2_4+\Delta_4&0\\
		0&0&-\frac{\hbar^2}{2\mu_5}\nabla^2_5+\Delta_5
	\end{pmatrix}+
	\begin{pmatrix}
	V_{33}&V_{34}&V_{35}\\
	V_{43}&V_{44}&V_{45}\\
	V_{53}&V_{54}&V_{55}
\end{pmatrix}
\right]
\begin{pmatrix}
	R_3(r)\\
	R_4(r)\\
	R_5(r)
\end{pmatrix}=
E\begin{pmatrix}
	R_3(r)\\
	R_4(r)\\
	R_5(r)
\end{pmatrix},\label{schro345}
\end{eqnarray}   
where $\nabla_i^2$, $\mu_i$ and $R_i(r)$ are the Laplacian, reduced mass and radial wave function of the $i$th channel ($i=3,4,5$); $\Delta_4=W_4-W_3$ and $\Delta_5=W_5-W_3$ are the mass differences; and $E$ is the binding energy. 
In the spherical coordinate, the Laplacian and radial wave function may be written as $\nabla_i^2=r^{-2}\frac{\partial}{\partial r}r^2\frac{\partial}{\partial r} -\frac{l_i(l_i+1)}{r^2}$ and $R_i( r)=\frac{u_i(r)}{r}$, respectively.
The ground-state binding energies of the $J^P=1/2^+$ and $3/2^-$ systems are obtained by solving the Schr\"odinger equation with the help of the Gaussian-expansion method~\cite{Hiyama:2003cu}. 
  
\begin{figure}[tbh]
	\centering
	\subfigure[$\Lambda_{c1}\bar D(^2S_{1/2})\to\Sigma_c\bar D^*(^2P_{1/2})$]{\label{V340}\includegraphics[width=0.3\textwidth]{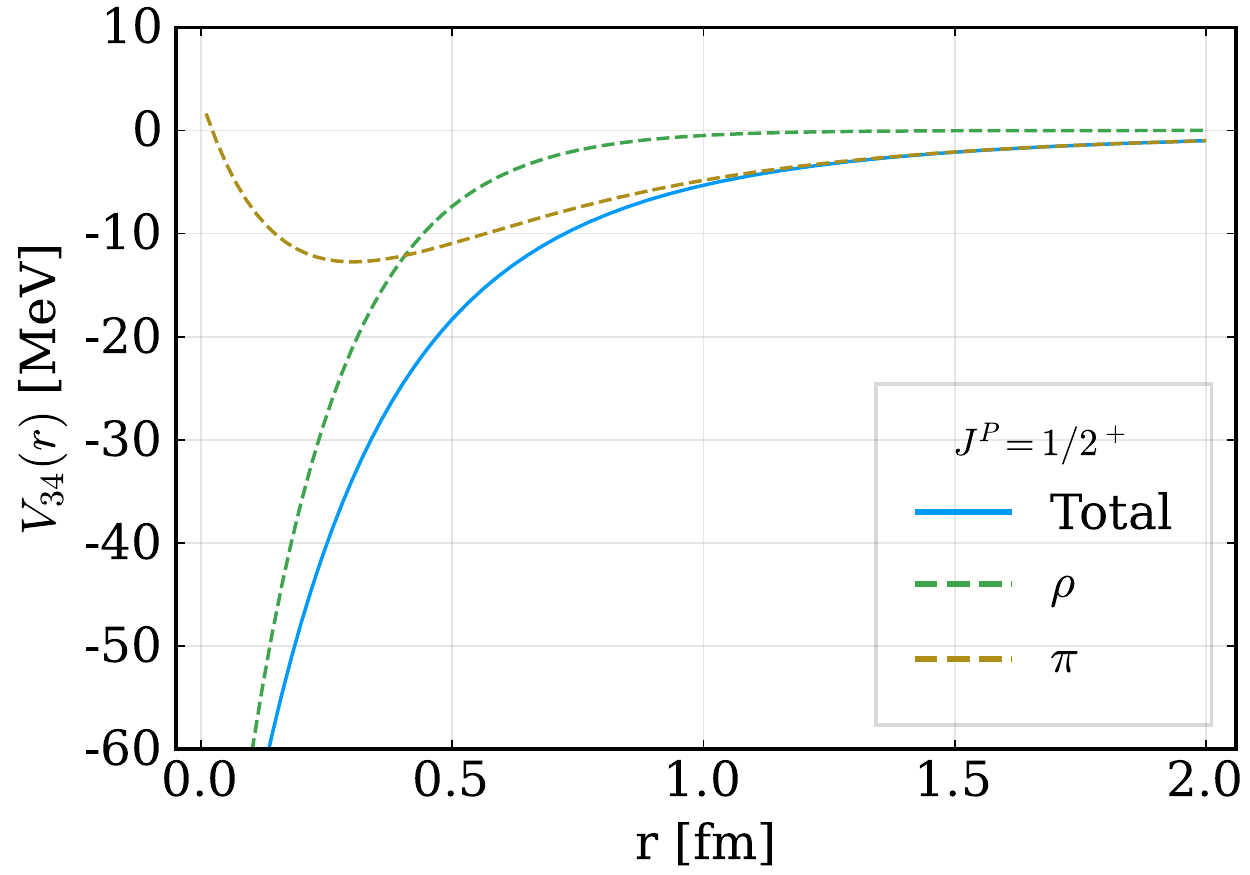}} 
	\subfigure[$\Lambda_{c1}\bar D(^2P_{3/2})\to\Sigma_c\bar D^*(^4S_{3/2})$]{\label{V3402}\includegraphics[width=0.3\textwidth]{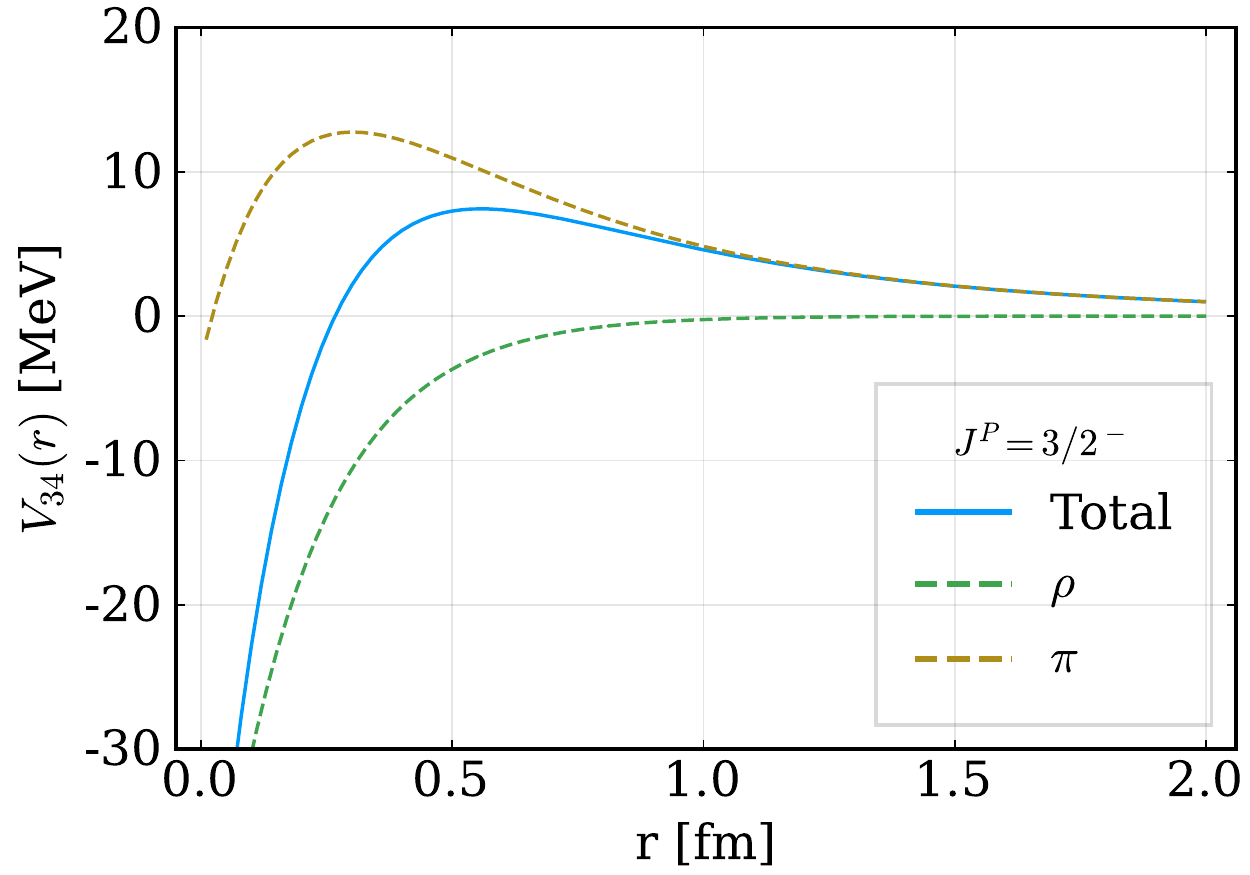}} 
	\subfigure[$\Lambda_{c1}\bar D(^2S_{1/2})\to\Sigma_c^*\bar D^*(^2P_{1/2})$]{\label{V350}\includegraphics[width=0.3\textwidth]{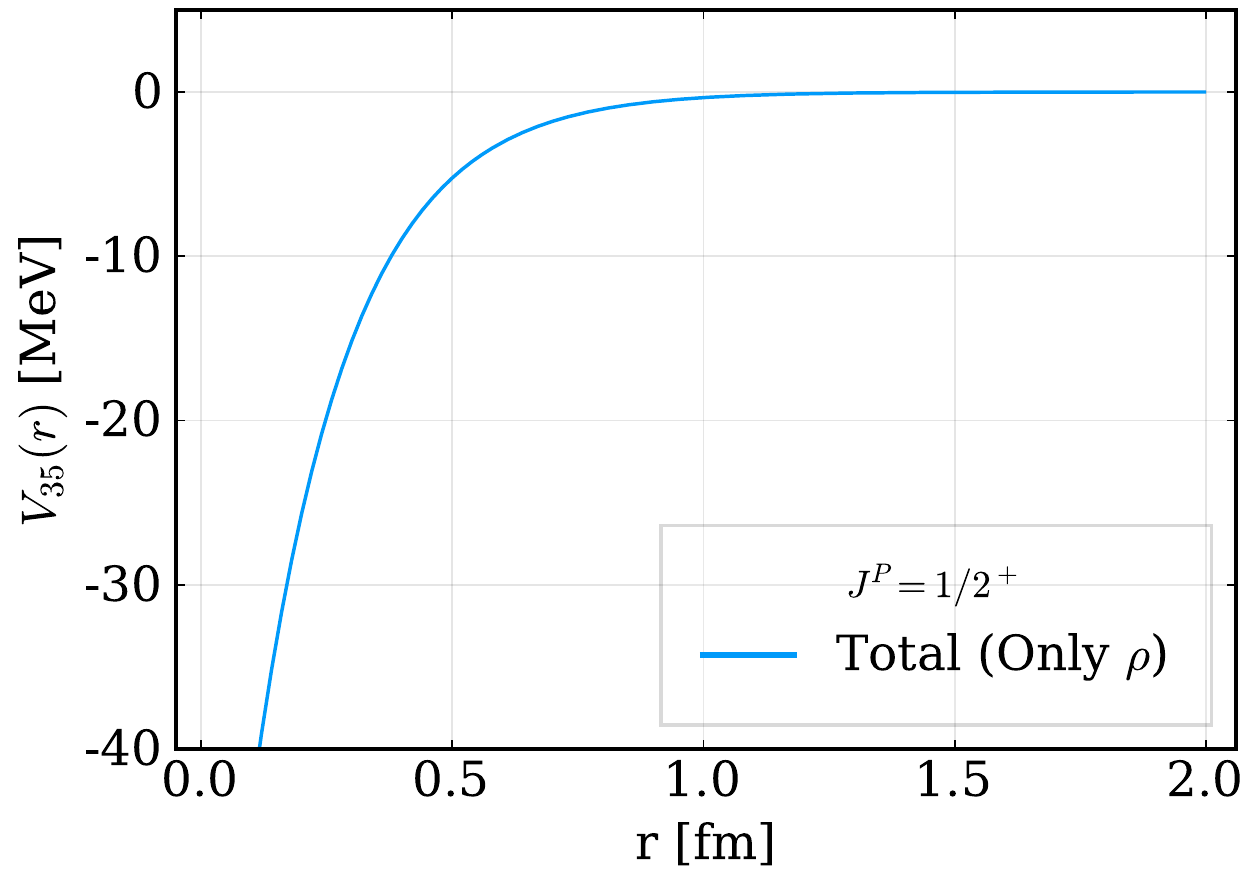}}
	\subfigure[$\Lambda_{c1}\bar D(^2P_{3/2})\to\Sigma_c^*\bar D^*(^4S_{3/2})$]{\label{V3502}\includegraphics[width=0.3\textwidth]{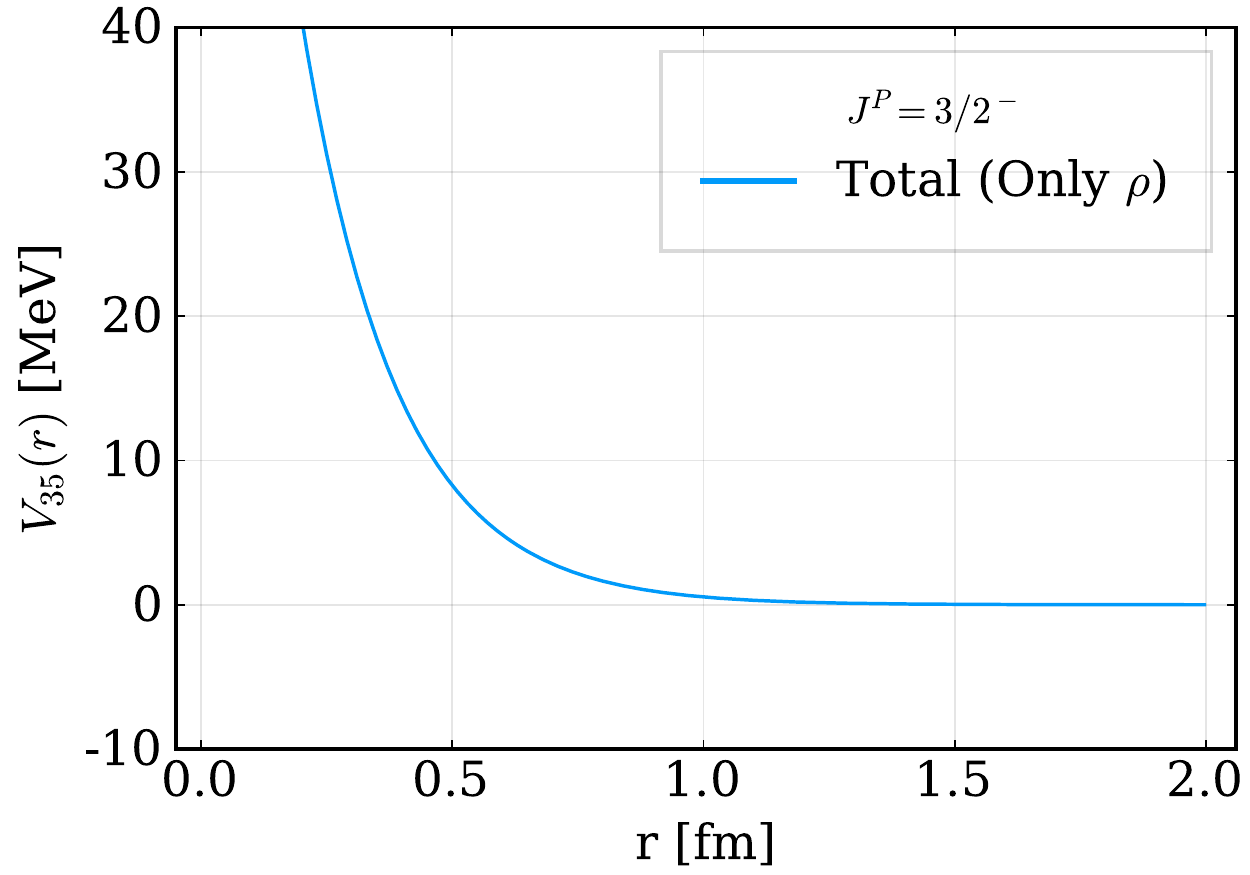}}
	\subfigure[$\Sigma_c\bar D^*(^2P_{1/2})\to\Sigma_c\bar D^*(^2P_{1/2})$]{\label{V440}\includegraphics[width=0.3\textwidth]{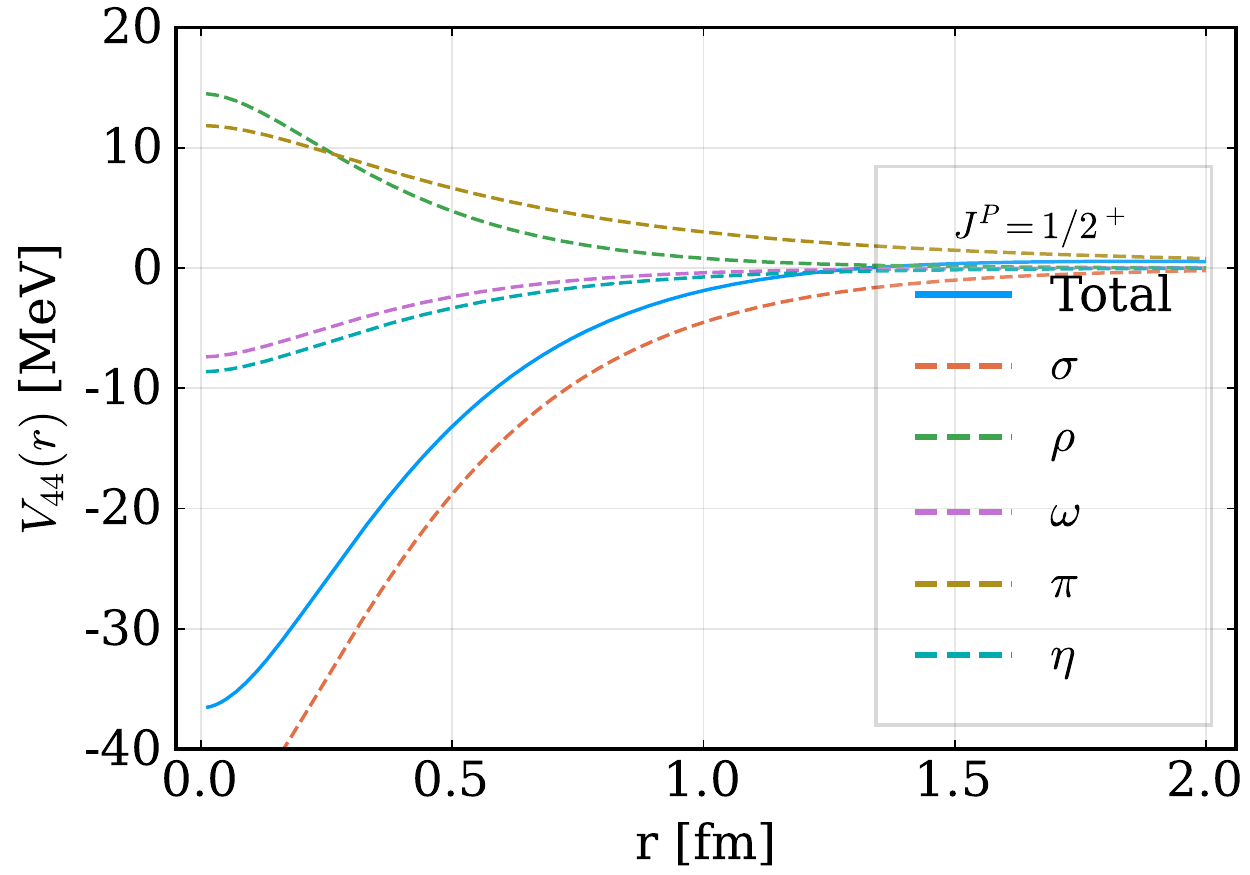}}
	\subfigure[$\Sigma_c\bar D^*(^4S_{3/2})\to\Sigma_c\bar D^*(^4S_{3/2})$]{\label{V4402}\includegraphics[width=0.3\textwidth]{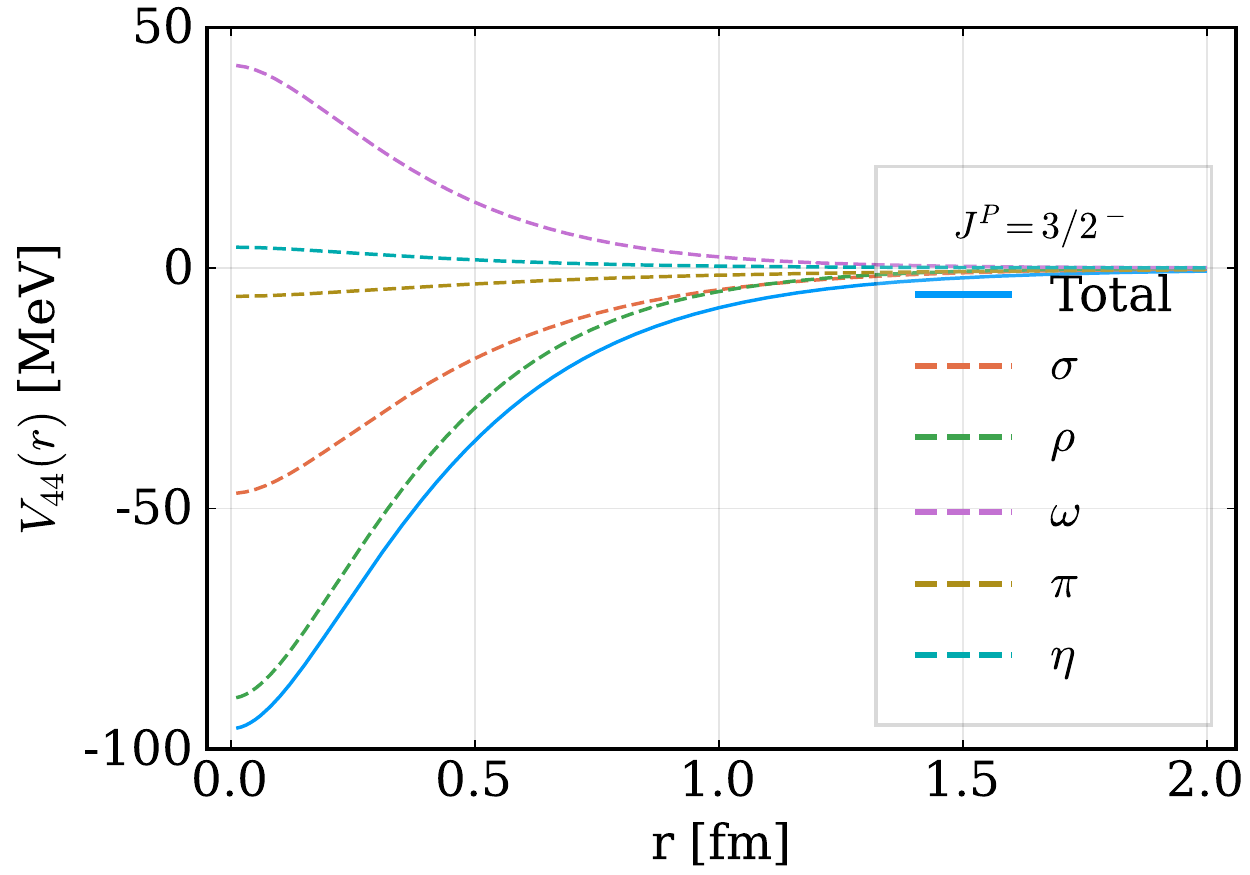}}
	\subfigure[$\Sigma_c\bar D^*(^2P_{1/2})\to\Sigma_c^*\bar D^*(^2P_{1/2})$]{\label{V450}\includegraphics[width=0.3\textwidth]{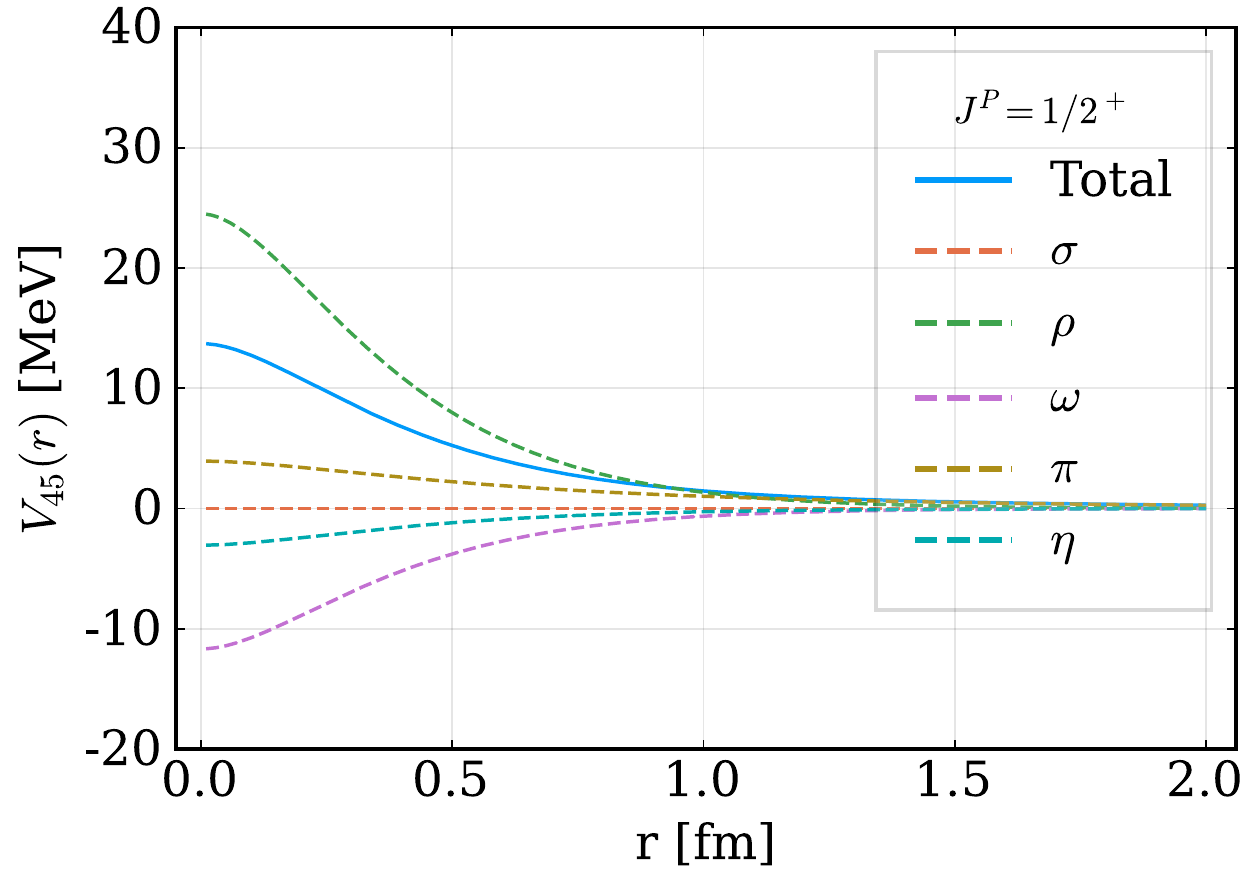}}
	\subfigure[$\Sigma_c\bar D^*(^4S_{3/2})\to\Sigma_c^*\bar D^*(^4S_{3/2})$]{\label{V4502}\includegraphics[width=0.3\textwidth]{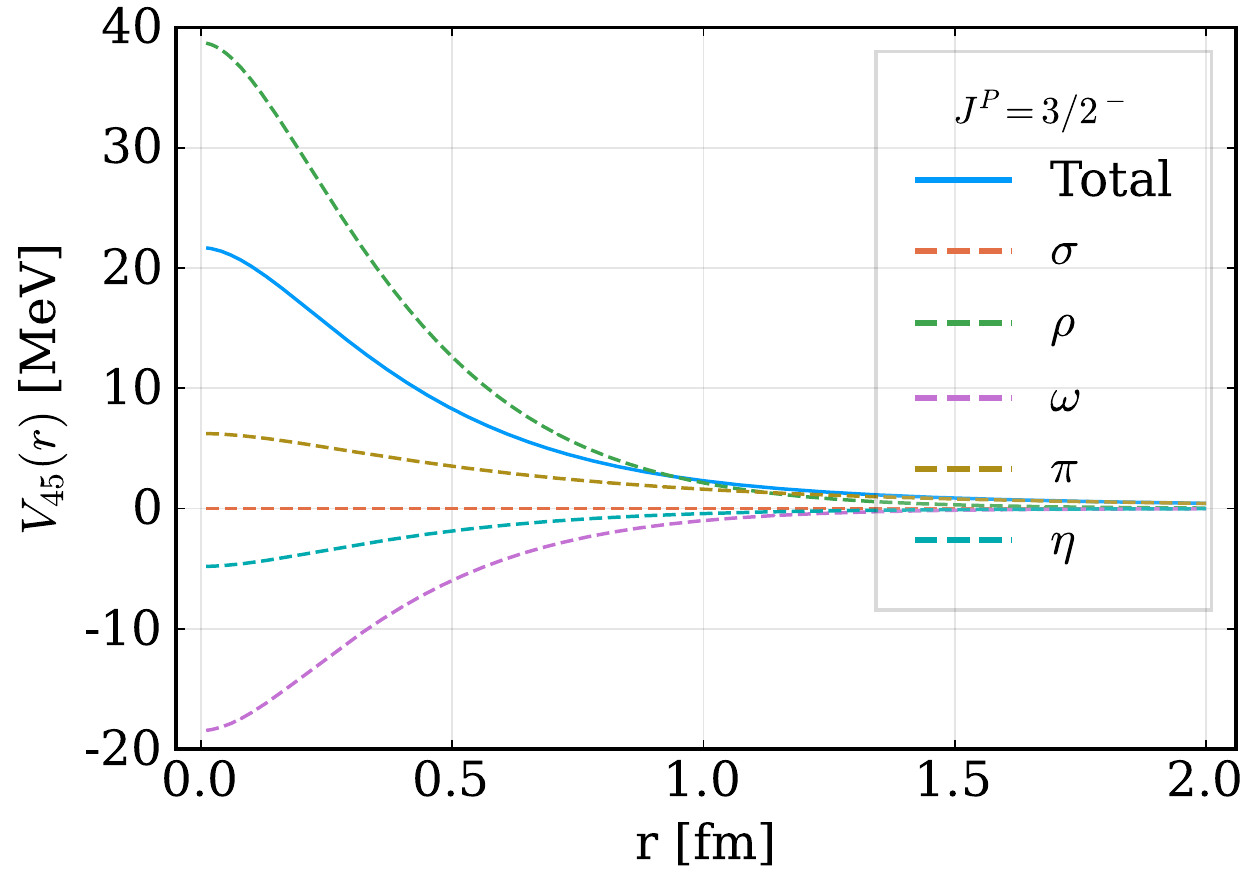}}
	\subfigure[$\Sigma_c^*\bar D^*(^2P_{1/2})\to\Sigma_c^*\bar D^*(^2P_{1/2})$]{\label{V550}\includegraphics[width=0.3\textwidth]{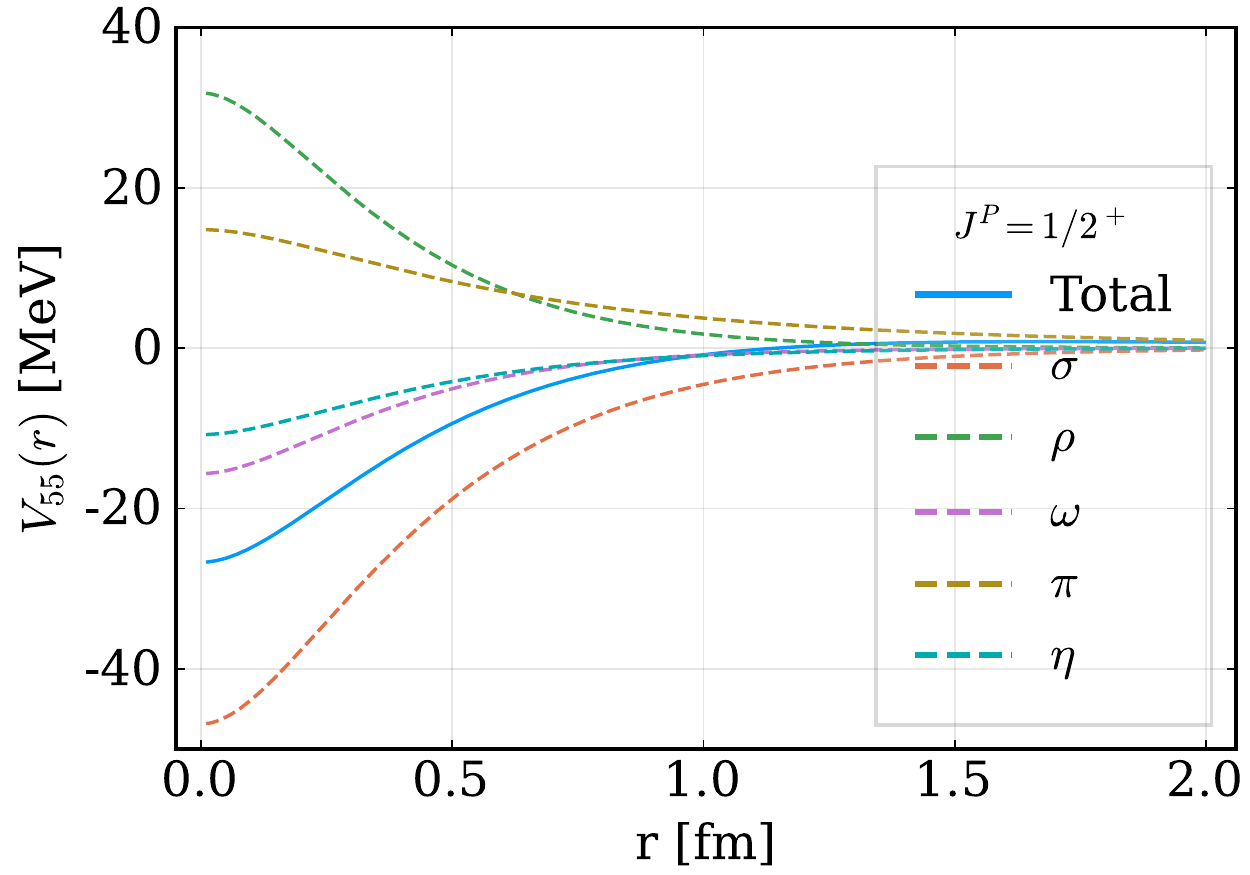}}
	\subfigure[$\Sigma_c^*\bar D^*(^4S_{3/2})\to\Sigma_c^*\bar D^*(^4S_{3/2})$]{\label{V5502}\includegraphics[width=0.3\textwidth]{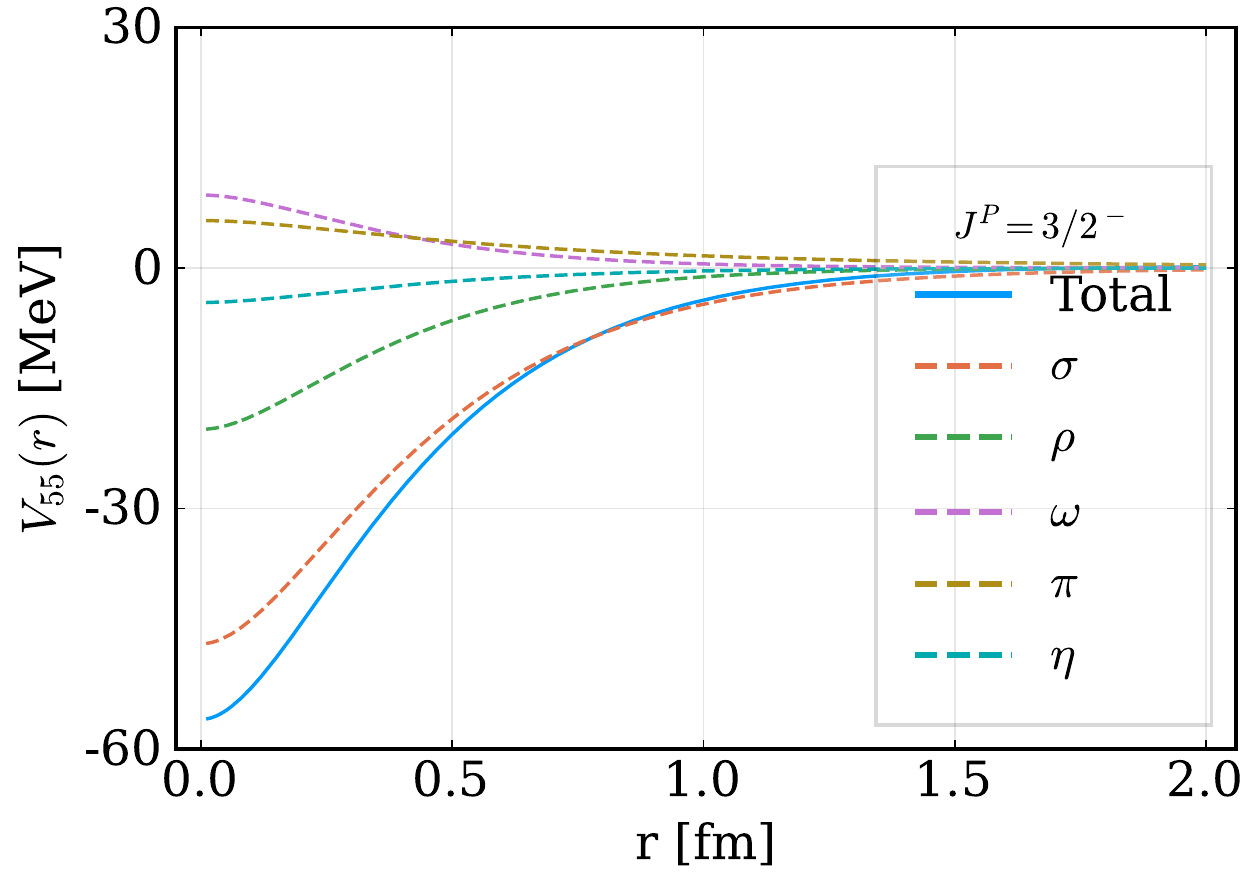}}
	\caption{Coupled-channel  potentials for the $J^P=1/2^+$ and $J^P=3/2^-$ isodoublet systems with $\Lambda=1$~GeV and without the $\delta(\vec{r})$ term ($a=1$).}
	\label{potfig_for_lambdac1}
\end{figure}

It is useful to take a look at the coupled-channel potentials. We plot the potentials with the coupling constants in Table~\ref{coupl_constant} for the $J^P=1/2^+$ and $3/2^-$ system with total isospin $I=1/2$ in Fig.~\ref{potfig_for_lambdac1}, in which we fully remove the $\delta(\vec{r})$ term by setting $a=1$ as is the case in Ref.~\cite{Burns:2019iih}.
The diagonal $\Lambda_{c1}\bar D$ potential is neglected in our formalism.
This is because our motivation is to check the mechanism proposed in Ref.~\cite{Burns:2019iih} where the $\Lambda_{c1} \bar{D}$-$\Sigma_c^{(\ast)}\bar{D}^{(\ast)}$ coupling forms a bound state lying below the $\Lambda_{c1} \bar{D}$ threshold.\footnote{Basically, the diagonal $\Lambda_{c1}\bar D$ component can be contributed by $\sigma$ and $\omega$ meson exchange. However, the $\omega$ exchange is repulsive which is similar to $\bar D\Lambda_c$ system as shown in Ref.~\cite{Dong:2021juy}. 
The $\sigma$ exchange gives the attractive force, which is considered to be small due to the weak $\bar{D}\bar{D}\sigma$ coupling.} 
The potentials for other exchanged mesons absent from the potential plots are forbidden due to the HQSS, isospin, and parity conservations. 
The exchange of the isoscalar mesons $\sigma,~\omega$ and $\eta$ for the $\Lambda_{c1}\bar D\to \Sigma_c \bar D^*$ is forbidden considering isospin symmetry, which is the case here, and since the $\Lambda_{c1}$ couples to $\Sigma_c^*\pi$ in te $D$ wave the pion exchange for $\Lambda_{c1}\bar D\to \Sigma_c^*\bar D^*$ is also not considered.
We can see that in the $1/2^+$ system the diagonal potentials $V_{44}$ and $V_{55}$ are attractive, but they couple in $P$ waves  and are largely canceled by the strong repulsive centrifugal potential. 
The off-diagonal elements involving the $S$ wave $\Lambda_{c1}\bar D$ channel contributes to the $J^P=1/2^+$ system and may trigger the $1/2^+$ system to form a bound state. 
The quark model calculation in Ref.~\cite{Burns:2019iih} indicates that the  $P_c(4457)$ and $P_c(4440)$  can be simultaneously reproduced as $J^P=1/2^+,3/2^-$ states within the $\Lambda_{c1}\bar D$-$\Sigma_c\bar D^*$ coupled channel  as long as the off-diagonal elements of the $\Lambda_{c1}\bar D$ potential are strong enough. Here, we further investigate such a scenario within the OBE model.  

The combined coupling of the OPE potential for the $\Lambda_{c1}\bar D\to\Sigma_c\bar D^*$ channel is defined in Ref.~\cite{Burns:2019iih} as 
 \begin{eqnarray}
 	\hat g=\frac{gh_2\omega}{2^{5/2}\pi f_\pi^2},
 \end{eqnarray}
where the values of $g$ and $h_2$ are given in Table~\ref{coupl_constant}, and the pion energy $\omega$ in the $t$-channel transition $\Lambda_{c1}\bar D\to\Sigma_c\bar D^*$ is given by $\omega=M_{\Lambda_{c1}}-M_{\Sigma_{c}}$. 
The physical value of $\hat g$ is $0.17(4)$ GeV$^{-1}$, as mentioned in Ref.~\cite{Burns:2019iih}, with coupling constants in Table ~\ref{coupl_constant}. In Ref.~\cite{Burns:2019iih}, it is required to be $\hat g=0.52$ GeV$^{-1}$ in order to set the lower $3/2^-$ state match the $P_c(4440)$. 
Note that quark model predictions of the OPE potential for the elastic channel $\Sigma_c\bar D^*$ which was studied in Ref.~\cite{Burns:2019iih} is roughly 1.7 times stronger than the one derived in our work.
Here, we take a simple ratio of the OPE potentials,   
\begin{eqnarray}
	\frac{V^{q}_{\pi}}{V_\pi^{44}}=-\frac{8}{3}\frac{g_q^2}{gg_1} \approx 1.7,
\end{eqnarray}                        
where the quark model potential $V^q_\pi$ is given in Ref.~\cite{Burns:2019iih} with $g^q=0.59$ and $V_\pi^{44}$ is obtained in our work. 
The ratio arises due to different determinations of the coupling constants at the quark and hadronic levels. As given in Table~\ref{coupl_constant}, the pseudoscalar couplings $g$ and $g_1$ are determined from the experimental data on the decays of $D^*\to D\pi$ and $\Sigma_c^*\to \Lambda_c \pi$, respectively, and used in our work.    

For the $\Lambda_{c1}\bar D$-$\Sigma_c\bar D^*$-$\Sigma_c^*\bar D^*$ coupled-channel  system, we remove the $\delta(\vec{r})$ term in the whole potentials by setting $a=1$. In this case, the $J^P=1/2^-$ state, which was a deep bound state when $a=0$, cannot be bound in this system with a value of $\Lambda$ up to almost 5~GeV. 
In the rest of this subsection, we will keep $a=1$ as the potential for the $\Lambda_{c1}\bar D\to \Sigma_c\bar D^{*}$, which triggers the $J^P=1/2^+$ state to be bound is independent of the $a$ parameter.
We mainly focus on the states below $\Lambda_{c1}\bar D$ and $\Sigma_c\bar D^*$ thresholds with spin parities $J^P=1/2^+$ and $J^P=3/2^-$, respectively, in order to discuss the effect of the $\Lambda_{c1}\bar D$ channel on these states, corresponding to the $P_c(4457)$ and the $P_c(4440)$. 
The other $P_c$ states at the resonance region above the $\Sigma_c\bar D^*$ threshold will be discussed in Sec.~\ref{sec_cc_delta}. 

In our manipulation of the coupled-channel  equation with threshold differences, the lowest threshold mass is always chosen to be the origin of energy and the $S$-$D$ mixing is also considered. Table~\ref{tab_phy_ghat} shows the binding energies of $J^P=1/2^+,3/2^-$ systems by varying the cutoff $\Lambda$, with the coupling constants given in Table~\ref{coupl_constant}, and $\hat g=0.17$ GeV$^{-1}$. 
The binding energies of $J^P=1/2^+,3/2^-$ states for the two-channel $\Lambda_{c1}\bar D$-$\Sigma_c\bar D^*$ case with the OPE potential are given in the column labeled case II, and compared with the results from quark model potentials as given in the column labeled case I.
With $g$ absorbed into the definition of $\hat g$, their difference lies in the value of $g_1$ ($g_1=  0.94$ for case II and $8g^q/3=1.57$ for case I).
The results including the scalar and vector meson exchange potentials are given in the column labeled case III. 
Considering the $\Lambda_{c1}\bar D$-$\Sigma_c\bar D^*$-$\Sigma_c^*\bar D^*$ coupled channels together with the OBE potentials, the results are given in the last column labeled case IV. 
One sees that the $J^P=1/2^+$ system, which couples to the $S$ wave $\Lambda_{c1}\bar D$ and the $P$ wave $\Sigma_c\bar D^*$, does not form a bound state because the $\Lambda_{c1}\bar D\to \Sigma_c\bar D^{(*)}$ transition potential is not strong enough, and there is only one bound state with $J^P=3/2^-$ in this system, which is mainly formed by the $S$ wave $\Sigma_c\bar D^*$. 
\begin{table}[tb]\centering
	\caption{Binding energies of the $J^P=1/2^+$  and $3/2^-$  $\Lambda_{c1}\bar D$-$\Sigma_c\bar D^*$-$\Sigma_c^*\bar D^*$ coupled-channel  systems with isospin $I=1/2$ as a function of cutoff $\Lambda$. We compare the binding energies obtained from several cases. 
	For all cases, we take $\hat g=0.17$~GeV$^{-1}$. In cases-I,-II,-III, we consider only the $\Lambda_{c1}\bar D$-$\Sigma_c\bar D^*$ coupled channels. 
	case I stands for the results obtained by considering only the OPE potentials with $g_1=8g^q/3=1.53$ from quark model. The results obtained using OPE potentials $V_\pi^{44}$ in our work are given as case II. 
	The results obtained by considering OBE potentials in our work are given as case III. 
	For case IV, we consider the  OBE potentials for the $\Lambda_{c1}\bar D$-$\Sigma_c\bar D^*$-$\Sigma_c^*\bar D^*$ coupled channels. Each entry with a ``$-$'' means that the potentials are not enough to form a bound state. The values of  $\Lambda$ and binding energy are in units of MeV.
	}\label{tab_phy_ghat}
	\begin{ruledtabular}
\begin{tabular}{c|cc|cc|cc|cc}
\multirow{2}{*}{$\Lambda$}&  \multicolumn{2}{c|}{case I}& \multicolumn{2}{c|}{case II} & \multicolumn{2}{c|}{case III}&\multicolumn{2}{c}{case IV}\\
&	$1/2^+$&	 $3/2^-$&$1/2^+$&	 $3/2^-$&$1/2^+$&	 $3/2^-$&$1/2^+$&	 $3/2^-$\\
\hline
1000	&$-$	&$-0.36$	&   $-$	&$-$	    &$-$	&$-$	    &   $-$	&$-0.23$ \\
1200	&$-$	&$-4.52$	&   $-$	&$-$	    &$-$	&$-5.32$	&   $-$	&$-8.47$ \\
1420	&$-$	&$-17.73$ &	$-$	&$-$	    &$-$	&$-21.91$ &	$-$	&$-31.64$\\
1600	&$-$	&$-38.68$ &	$-$	&$-0.67$	&$-$	&$-41.91$ &	$-$	&$-59.36$
\end{tabular}
	\end{ruledtabular}
\end{table}

As shown in Ref.~\cite{Burns:2019iih}, if $\hat g=0.52$~GeV$^{-1}$, then a $J^P=1/2^+$ bound state emerges, and masses of both  $P_c(4457)$ and $P_c(4440)$ would be reproduced with the same cutoff $\Lambda=1.42$~GeV. 
The results with $\hat g=0.52$~GeV$^{-1}$ are given in Table~\ref{tab_lar_ghat}, where the results in the column labeled case I are consistent with Ref.~\cite{Burns:2019iih}, as they should be. 
It is shown within the same table that it is more difficult to bind the $3/2^-$ state using the $g_1$ value from Table~\ref{coupl_constant} comparing with case I using $g_1$ from the quark model. Because of the larger value of $\hat g$ in contrast to Table~\ref{tab_phy_ghat}, the $1/2^+$ bound state appears with relatively large cutoffs.  
The inclusion of the scalar and vector meson exchange (case III) as well as the $\Sigma_c^*\bar D^*$ coupled channel (case IV) only makes minuscule contributions to the binding energy of the $1/2^+$ state. 

Another $S$ wave channel for the $1/2^+$ state is the $\Lambda_{c1}\bar{D}^*$. One may expect a sizeable role it would play, although its threshold is around 140~MeV above the
$\Lambda_{c1}\bar{D}$ one. However, its $S$ wave elastic potential is repulsive since the repulsive $\omega$-exchange force contributes dominantly. And it is found that the nondiagonal dynamics of the $\Lambda_{c1}\bar{D}^*$ channel is also negligible due to the experimental constraints on the $\Lambda_{c1}$ couplings similarly to the $\Lambda_{c1}\bar{D}$ case. 
An explicit inclusion of this channel into case III would only slightly increase the numerical values of the absolute values of the binding energies by less than 0.5~MeV for the cutoff in the range listed in Table~\ref{tab_phy_ghat}.
All other $S$ wave channels for the $1/2^+$ state with higher thresholds are expect to be irrelevant.

\begin{table}[tb]\centering
	\caption{Binding energies of the $J^P=1/2^+$ and $3/2^-$ $\Lambda_{c1}\bar D$-$\Sigma_c\bar D^*$-$\Sigma_c^*\bar D^*$ coupled-channel  systems with $I=1/2$ as a function of cutoff $\Lambda$. The difference from Table~\ref{tab_phy_ghat} is that here  $\hat g=0.52$~GeV$^{-1}$ is taken. Each entry with a ``$-$'' means that the potentials are not enough to form a bound state. The values of  $\Lambda$ and binding energy are in units of MeV.}\label{tab_lar_ghat}
	\begin{ruledtabular}
	\begin{tabular}{c|cc|cc|cc|cc}
		\multirow{2}{*}{$\Lambda$}&  \multicolumn{2}{c|}{case I}& \multicolumn{2}{c|}{case II} & \multicolumn{2}{c|}{case III}&\multicolumn{2}{c}{case IV}\\
		&	$1/2^+$&	 $3/2^-$&$1/2^+$&	 $3/2^-$&$1/2^+$&	 $3/2^-$&$1/2^+$&	 $3/2^-$\\
		\hline
1000	&$-$	    &$-0.99$	& $-$	     &$-$	     &	$-$    &$-0.55$	 &$-$	    &$-1.06$ \\
1200	&$-$	    &$-6.08$	& $-$	     &$-$	     &$-$	    &$-7.95$	 &$-$	    &$-11.17$\\
1420	&$-0.33$	&$-20.19$	&$-0.01$&	$-0.54$	&$-0.53$	&$-26.50$	&$-0.56$	&$-35.97$\\
1600	&$-1.33$	&$-41.75$	&$-0.42$&	$-2.66$	&$-1.64$	&$-48.09$	&$-1.70$	&$-64.98$
	\end{tabular}
\end{ruledtabular}
\end{table}

\subsection{Role of the $\delta(\vec{r})$ term in the OBE model}\label{sec_cc_delta}

In the hadronic molecular picture, the masses of the latest observed pentaquarks  $P_c(4312)$, $P_c(4440)$ and $P_c(4457)$ are claimed to be well reproduced as the $\Sigma_c^{(*)}\bar D^{(*)}$ bound states with various OBE models~\cite{He:2019rva,He:2019ify,Chen:2019asm,Wang:2019spc,Liu:2019zvb}. However, for the two higher states $P_c(4440,4457)$ close to the $\Sigma_c\bar D^*$ threshold, their spins are very model dependent to be either $J^P(4440,4457)=(1/2^-,3/2^-)$~\cite{Wang:2019spc} or $J^P(4440,4457)=(3/2^-,1/2^-)$~\cite{Liu:2019zvb}.
A recent analysis using an effective  field theory framework shows that the LHCb data can be well described with both quantum number assignments, while the latter is preferred because of its insensitivity on the cutoff values used in regularizing the coupled-channel Lippmann-Schwinger equation~\cite{Du:2021fmf}.
In this section, first, we use the OBE potentials derived in Sec.~\ref{sec:2} to simultaneously reproduce all the observed $P_c$ states by varying the cutoff $\Lambda$ and the magnitude of the $\delta(\vec{r})$ term without coupled channels. 
Then, we include the $\Sigma_c^{(*)}\bar D^{(*)}$ coupled-channel  effects  and try to distinguish the two spin-parity assignments for $P_c(4440)$ and $P_c(4457)$ with help of the $\delta(\vec{r})$ term. 
Here, we consider four channels $\Sigma_c\bar D$, $\Sigma_c^*\bar D$, $\Sigma_c\bar D^*$ and $\Sigma_c^*\bar D^*$ and exclude the $\Lambda_{c1}\bar D$ channel due to the fact that its  contributions are negligible for negative parity states with the physical value for $\hat g$.

\begin{figure}[tb]
	\centering
	\subfigure[$V_{11/22}$ potentials which are independent of spin and $a$. ]{\label{V1111}\includegraphics[width=0.3\textwidth]{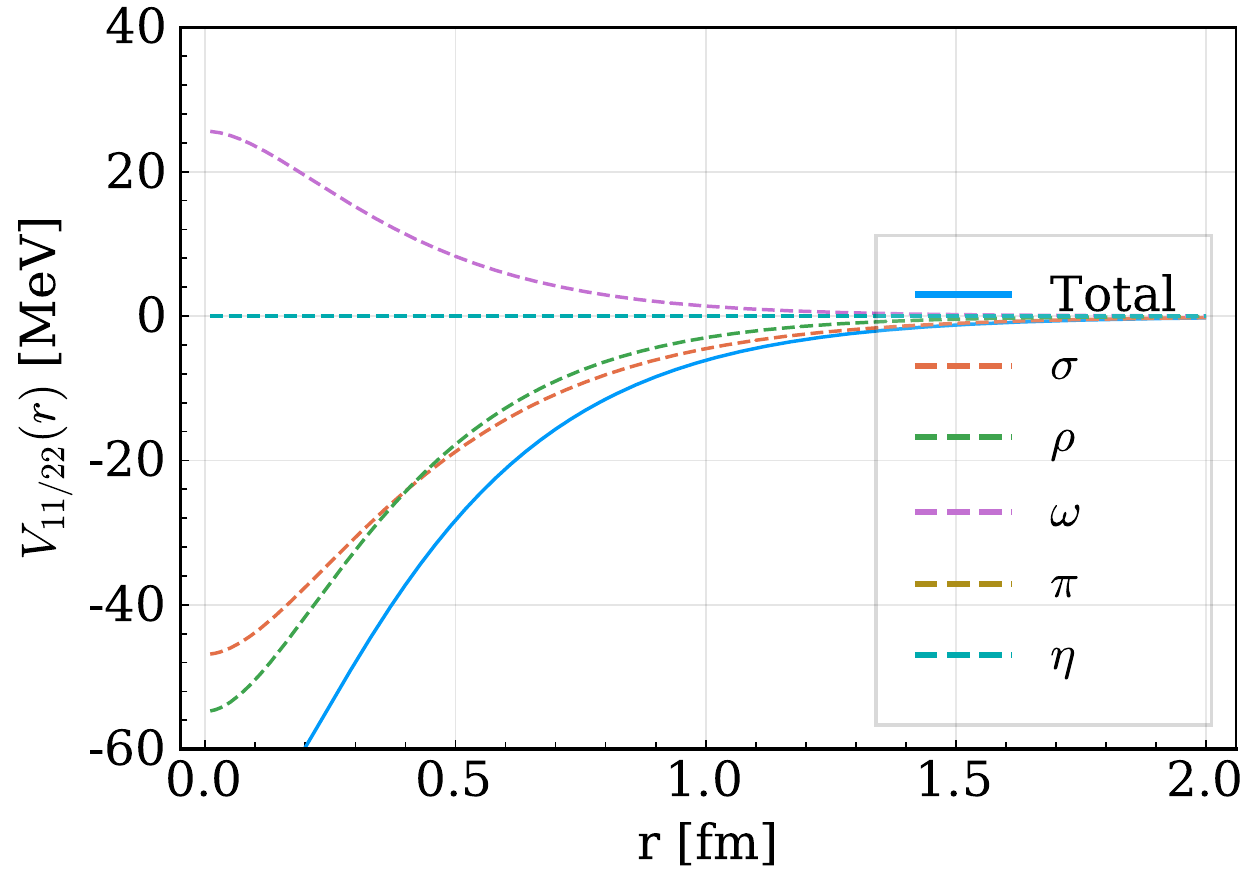}} 
	\subfigure[$V_{44}, J^P=1/2^-, a= 0$]{\label{V4410}\includegraphics[width=0.3\textwidth]{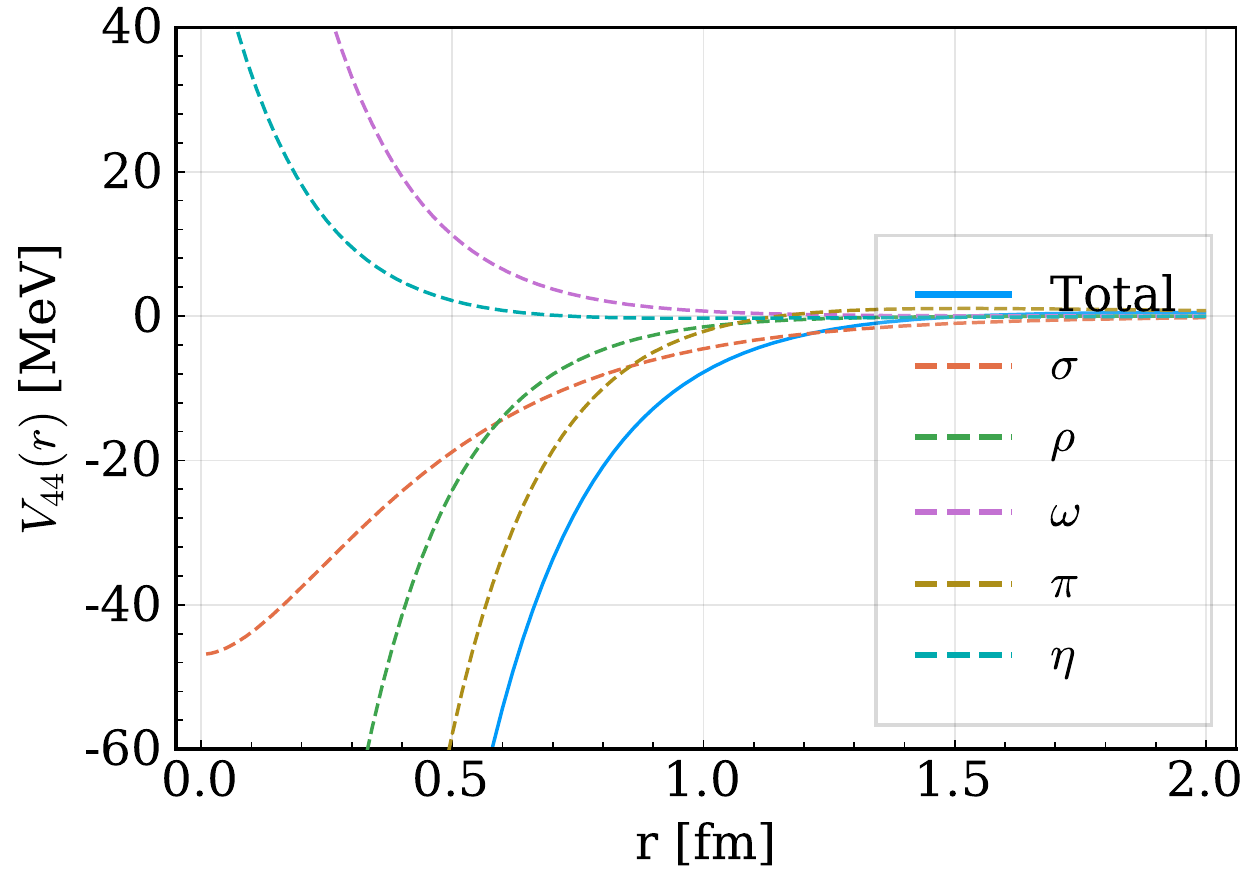}} 
	\subfigure[$V_{44}, J^P=1/2^-, a= 1$]{\label{V4411}\includegraphics[width=0.3\textwidth]{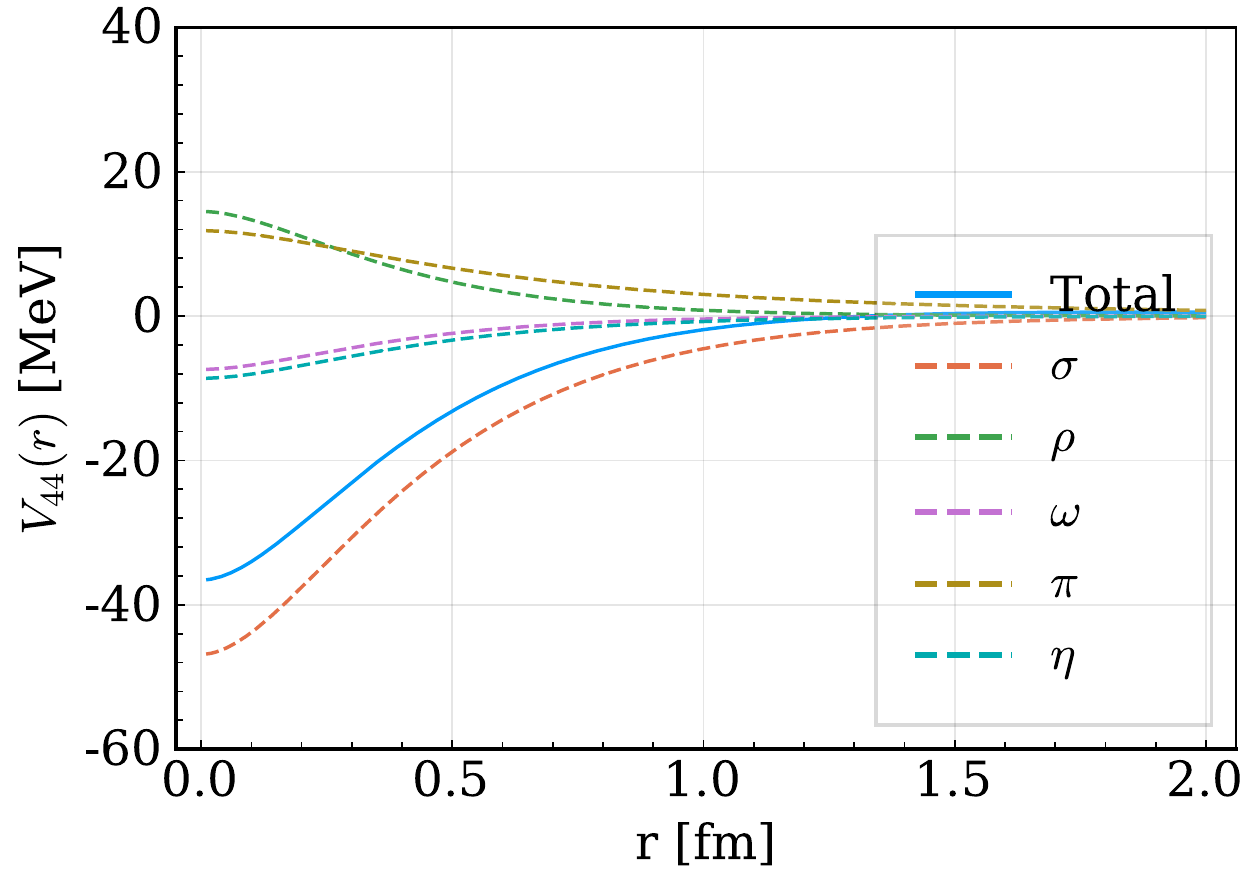}}
	\subfigure[$V_{44}, J^P=3/2^-, a= 0$]{\label{V4420}\includegraphics[width=0.3\textwidth]{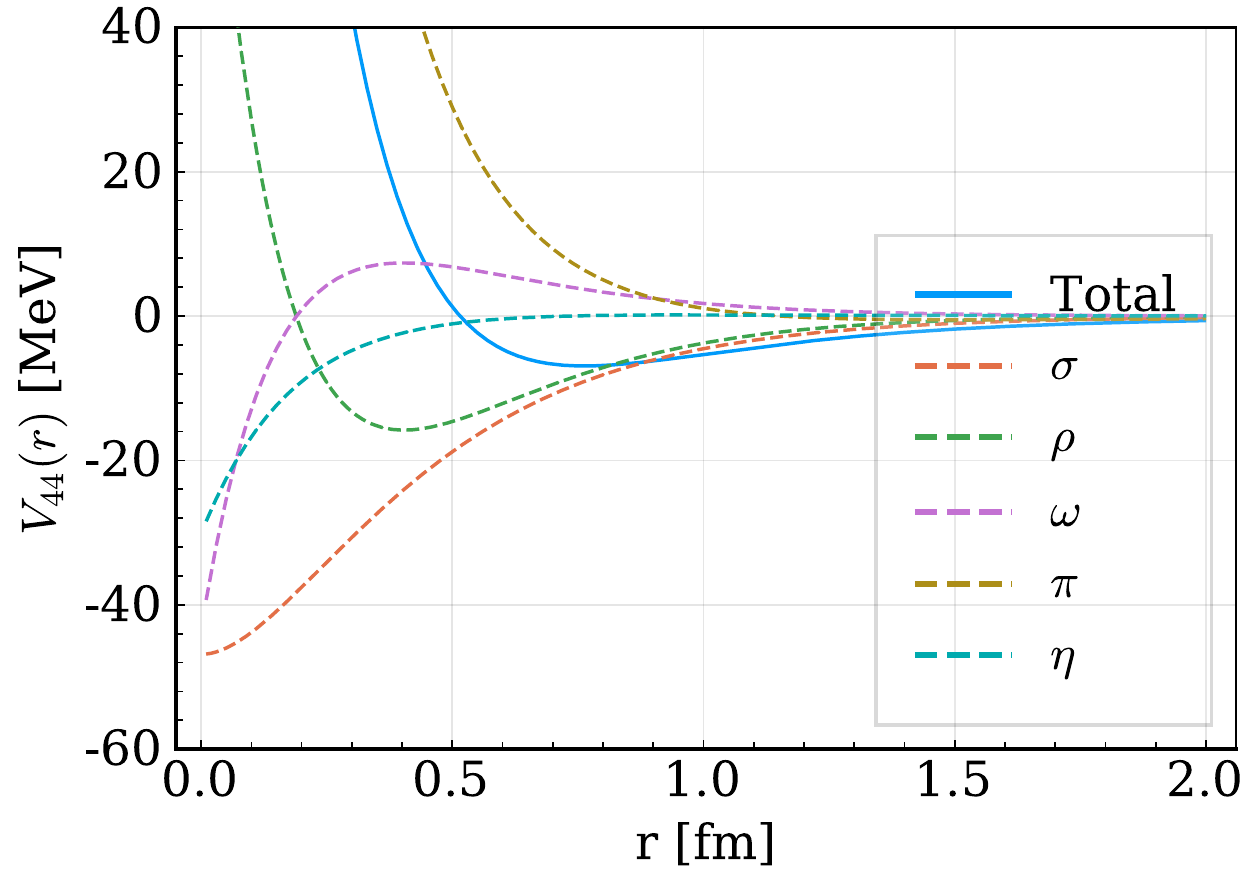}}
	\subfigure[$V_{44}, J^P=3/2^-, a= 1$]{\label{V4421}\includegraphics[width=0.3\textwidth]{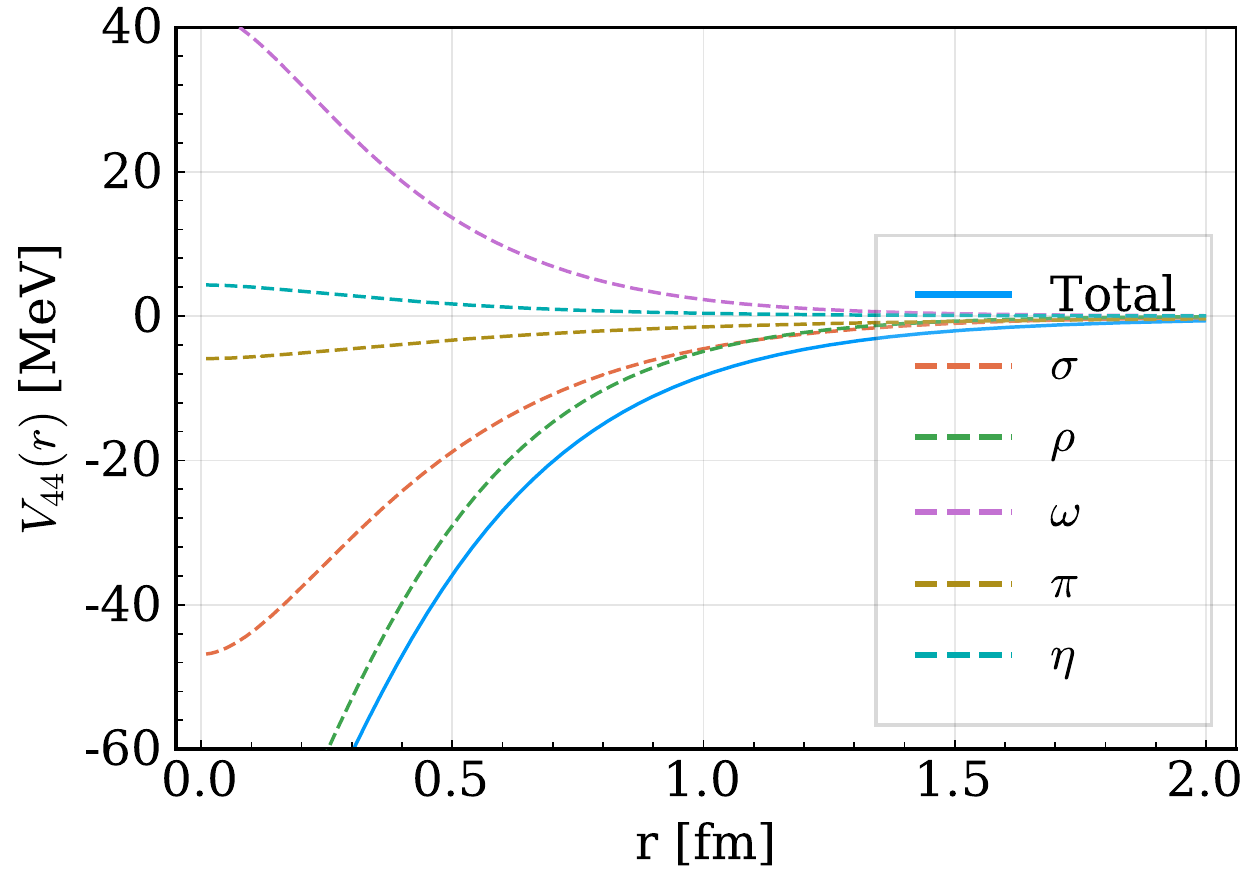}}
	\subfigure[$V_{55}, J^P=1/2^-, a= 0$]{\label{V5510}\includegraphics[width=0.3\textwidth]{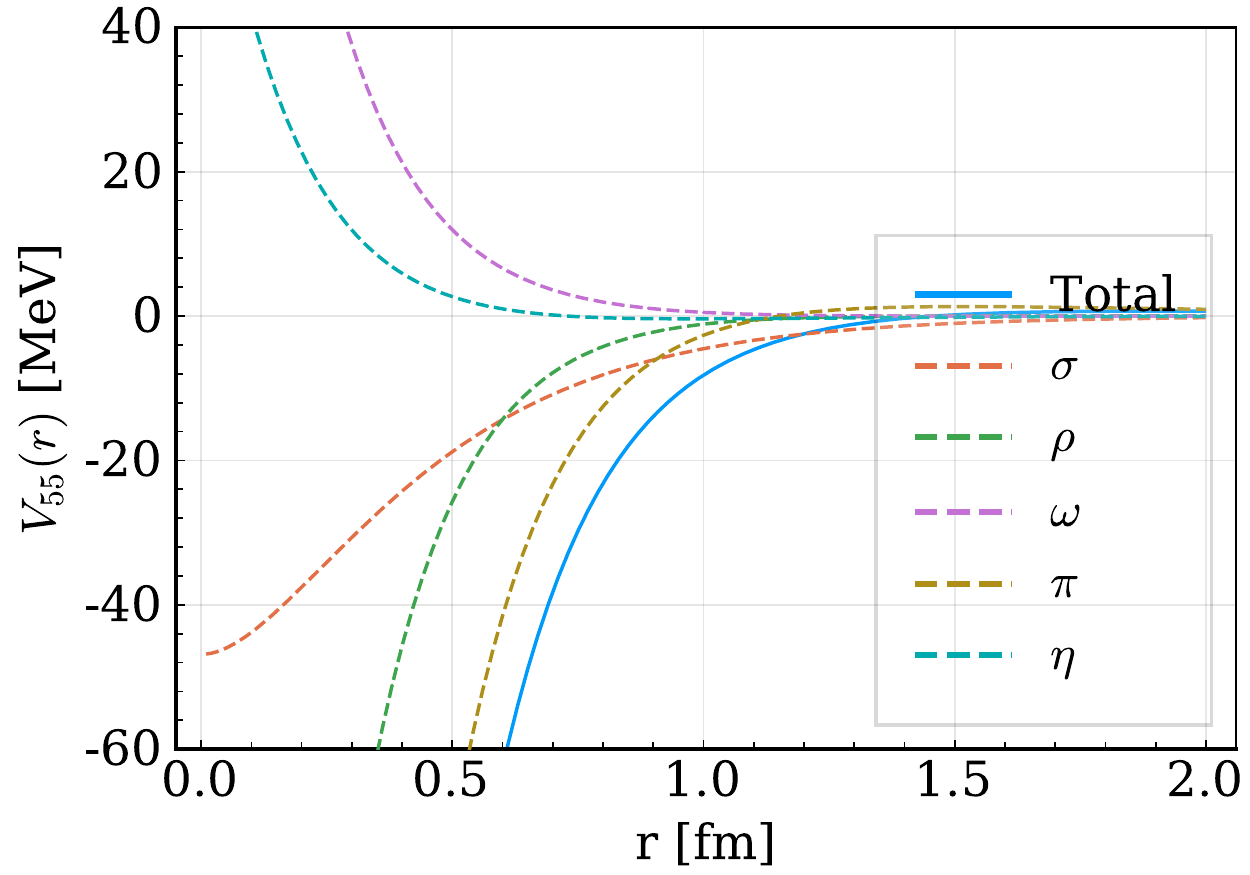}}
	\subfigure[$V_{55}, J^P=1/2^-, a= 1$]{\label{V5511}\includegraphics[width=0.3\textwidth]{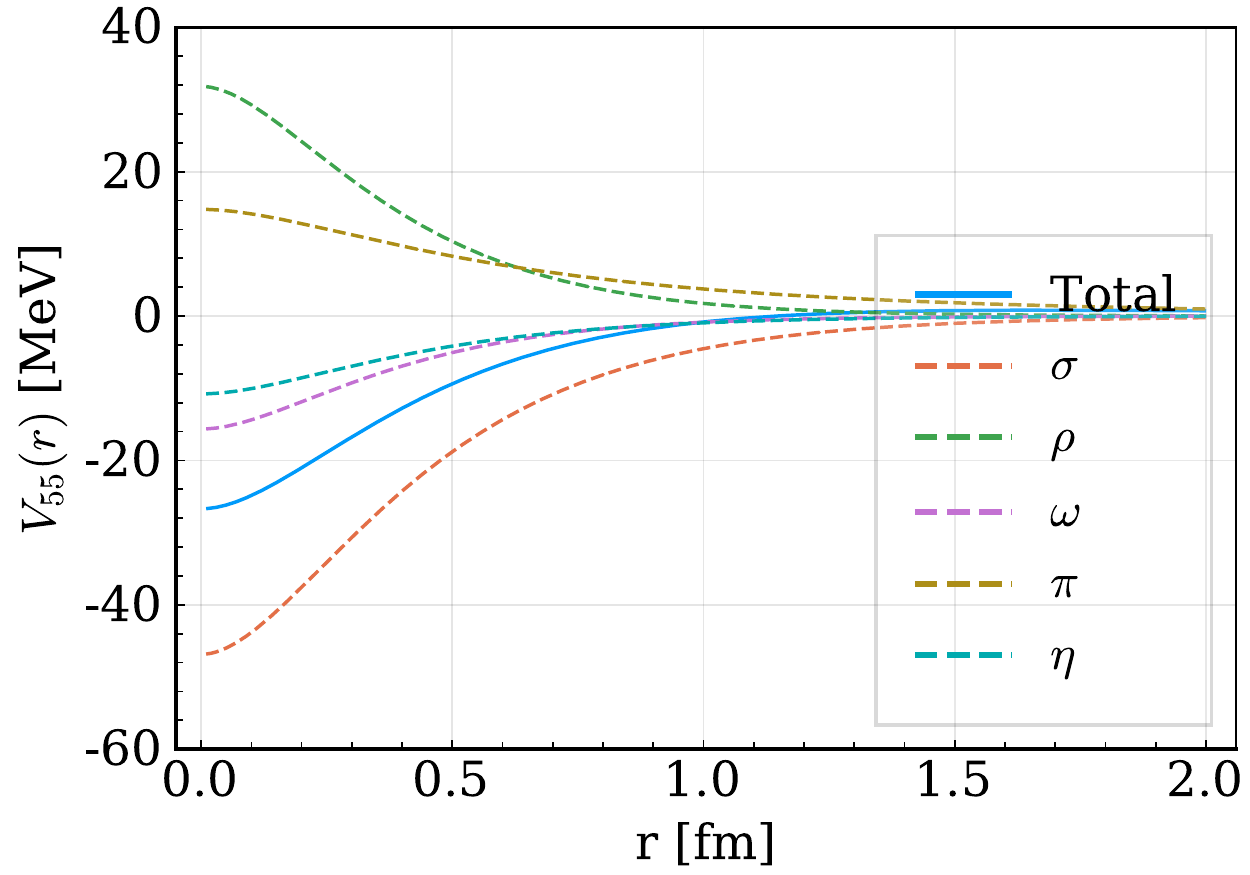}}
	\subfigure[$V_{55}, J^P=3/2^-, a= 0$]{\label{V5520}\includegraphics[width=0.3\textwidth]{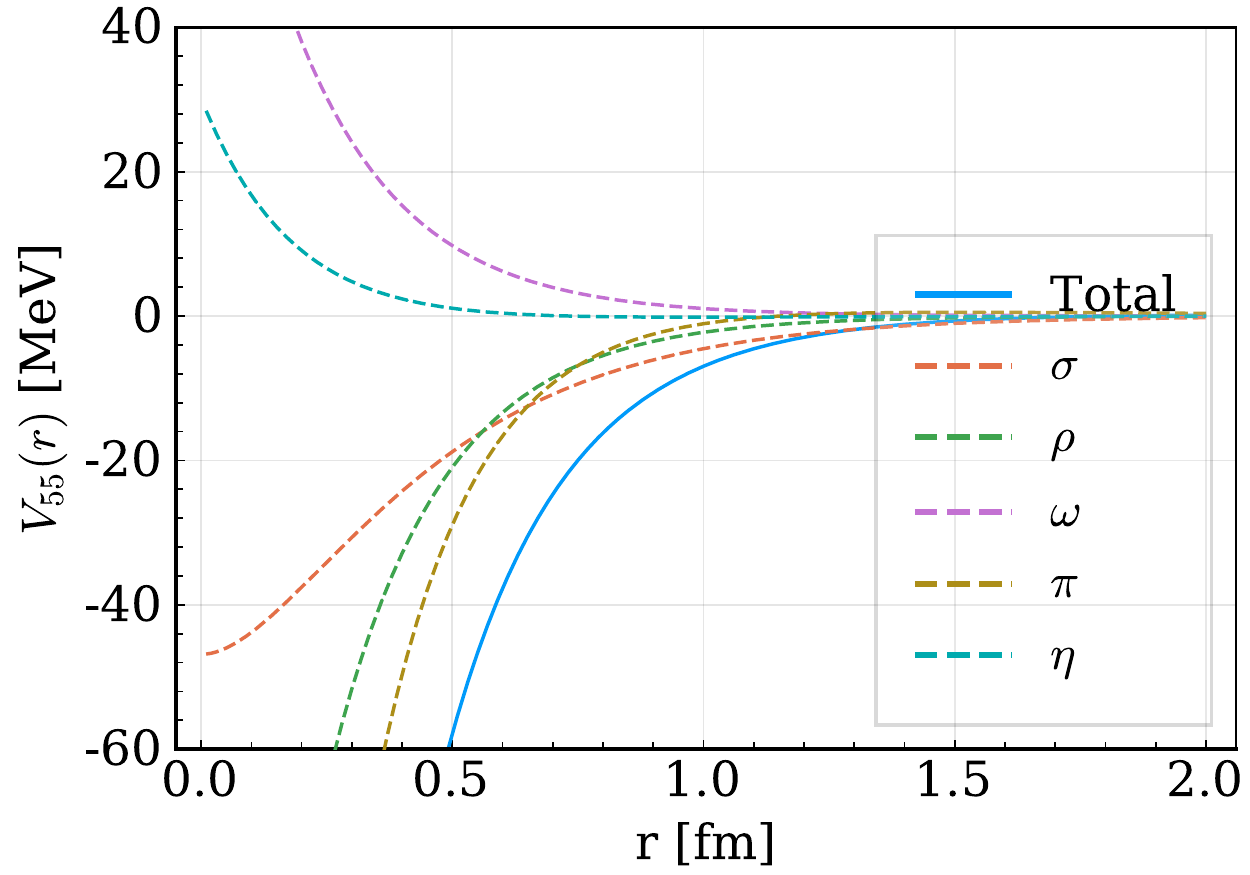}}
	\subfigure[$V_{55}, J^P=3/2^-, a= 1$]{\label{V5521}\includegraphics[width=0.3\textwidth]{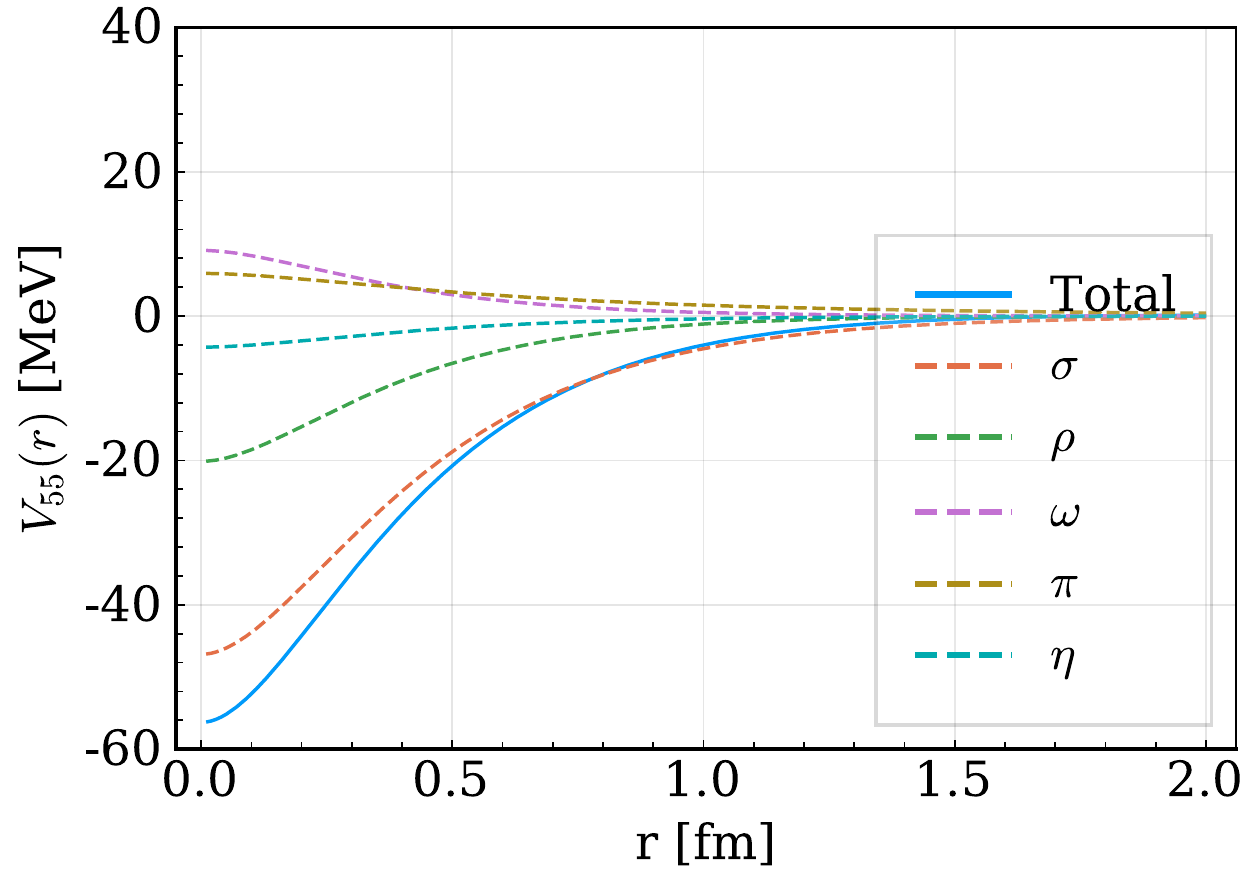}}
	\caption{Diagonal $S$ wave potentials for the $J^P=1/2^-$ and $J^P=3/2^-$ isodoublet systems with $\Lambda=1$ GeV. $a=1$ means without the $\delta(\vec{r})$ term.}\label{potfig_for_single}
\end{figure}

To be consistent with Sec.~\ref{subsec_potentials}, we enumerate the channels $\Sigma_c\bar D$, $\Sigma_c^*\bar D$, $\Sigma_c\bar D^*$, and $\Sigma_c^*\bar D^*$ with $1,~2,~4$ and $5$, respectively, which are ordered according to the channel thresholds $W_1,~W_2,~W_4 $ and $W_5$. 
First, we take a look at the potentials for the $J^P=1/2^-$ and $3/2^-$ systems. 
Figure~\ref{potfig_for_single} shows the diagonal $S$ wave potentials, where we compare the potentials with the $\delta(\vec r)$ term ($a=0$) and those without it ($a=1$) using a cutoff $\Lambda=1$~GeV. 
The $V^{11}$ and $V^{22}$ potentials are independent of spin and $a$. Both vertices in the $t$-channel transitions are in $S$ waves and there is no central term as discussed in Sec.~\ref{subsec_potentials} which leads to the $\delta(\vec{r})$ term. 
Both the pseudoscalar and vector meson exchange potentials in $V^{44}$ have the $\delta(\vec{r})$ term originated from central potentials. 
As shown in Figs.~\ref{V4410}, \ref{V4411}, \ref{V4420}, and~\ref{V4421}, when we fully remove the $\delta(\vec{r})$ term, the total potential for $J^P=1/2^-$ becomes very weakly attractive, while that of $J^P=(3/2^-)$ becomes strongly attractive. 
The potentials for $V^{55}$ also have the same behavior. We do not show the off-diagonal elements of the potentials. The analytic expressions can be found in  Sec.~\ref{subsec_potentials}, and the $\delta(\vec{r})$ term also has similar effects. 
In the following, we use these potentials to solve the  Schr\"odinger equation to reproduce the masses of the $P_c(4312)$, $P_c(4380)$, $P_c(4440)$, and $P_c(4457)$ states.                                

It is worth it to mention the method of solving the coupled-channel  Schr\"odinger equation with threshold differences in our approach. In the $\Sigma_c\bar D$-$\Sigma_c^*\bar D$-$\Sigma_c\bar D^*$-$\Sigma_c^*\bar D^*$ coupled channels, the  $P_c(4312)$, $P_c(4380)$, $P_c(4440)$, and $P_c(4457)$ masses determined in the experimental analyses are located as
\begin{eqnarray}
	M(4312)<W_1<M(4380)<W_2<M(4440)<M(4457)<W_4<W_5,
\end{eqnarray}      
where $W_i$ is the threshold energy of the $i$th channel. 
Here, we directly solve the coupled-channel  Schr\"odinger equation, 
\begin{eqnarray}
		\left[-\frac{\hbar^2}{2\mu_i}\frac{d^2}{d r^2}+\frac{\hbar^2 l_i(l_i+1)}{2\mu_ir^2} + \Delta_i\right]u_i+
\sum_j
	V_{ij}u_j=
E u_i,
\label{eq_schro_coupl}
\end{eqnarray}
where $i$ is the channel index, $u_i$ is defined by $u_i(r)=rR_i(r)$ with the radial wave function $R_i(r)$ for the $i$th channel, and $\mu_i$ is the corresponding reduced mass. 
The eigenmomentum for channel $i$ is given as $q_i=\sqrt{2\mu_i(E-\Delta_i)}$, where $\Delta_i$ is the threshold difference with respect to the threshold of the lowest channel.
By solving Eq.~\eqref{eq_schro_coupl}, we obtain the coupled-channel wave function, which is normalized to satisfy the boundary condition for the $j$th channel given as~\cite{osti_4661960}
\begin{eqnarray}
	u_{i}^{(j)}(r)\overset{r\rightarrow \infty}{\longrightarrow} \delta_{ij}e^{-iq_i r}-S_{ij}(E)e^{iq_i r}\label{asymp_wave},
\end{eqnarray}
where $S_{ij}(E)$ is the scattering matrix component. 
Bound states and resonances are represented as poles at $E_{\mathrm{pole}} $ of $S_{ij}(E)$ in the complex energy plane.
Among them, bound states emerge as poles on real energy axis ($E_{\mathrm{pole}}<0$) in the Riemann sheet where the momentum $q_i$ is purely positive imaginary for each channel.
While resonances are related to those poles of $S_{ij}(E)$ in the Riemann sheet closest to the real axis of the physical sheet corresponding to the scattering energy region ($\mathrm{Re} \ E_{\mathrm{pole}}>0$ and $\mathrm{Im} \ E_{\mathrm{pole}} < 0$).

First, we show the results of the single channels $\Sigma_c\bar D$, $\Sigma_c^*\bar D$, $\Sigma_c\bar D^*$, and $\Sigma_c^*\bar D^*$ with the $S$-$D$ wave mixing effects considered.\footnote{Note that the $S$-$D$ wave mixing effects on binding energy are calculated by solving the coupled-channel  Schr\"odinger equation with all partial wave channels for each single hadron pair.} The quantum numbers can be $J^P=1/2^-,3/2^-$, and $5/2^-$. 
Table~\ref{tab_signle_be} shows the binding energy of each single channel with various cutoffs. We list the results for two different $\delta(\vec{r})$-term contributions, that is, $a=0$ and $a=1$. 
In the single-channel   case, the binding energies of the $1/2^-(\Sigma_c\bar D)$ and $3/2^-(\Sigma_c^*\bar D)$ states are independent of the $\delta(\vec{r})$ term, and the two systems are loosely bound.
The corresponding potentials for these two systems are identical as given in Fig.~\ref{V1111}. The small difference between the binding energies is completely caused by the different reduced masses. 
However, for all the other states, the $\delta(\vec{r})$ term has an impressive effect on the binding energies. And the binding energies are heavily dependent on the cutoff $\Lambda$ when the $\delta(\vec{r})$ term is included because of the short-distance nature of the $\delta(\vec r)$ term. 
The single-channel results show that we cannot reproduce the $P_c(4312)$, $P_c(4380)$, $P_c(4440)$ and $P_c(4457)$ simultaneously by including or excluding fully the $\delta(\vec{r})$ term. 

\begin{table}[tb]\centering
	\caption{Binding energies of the $J^P=1/2^-,3/2^-,5/2^-$ isodoublet states in the single-channel   $\Sigma_c\bar D$, $\Sigma_c^*\bar D$, $\Sigma_c\bar D^*$, and $\Sigma_c^*\bar D^*$ systems by varying the cutoff $\Lambda$. The hadron pair inside each set of parentheses denotes the corresponding single channel. Each entry with a ``$-$'' means that the potentials are not enough to form a bound state. The values of  $\Lambda$ and binding energy are in units of MeV.}\label{tab_signle_be}
	\begin{ruledtabular}
	\begin{tabular}{c|cc|cc|cc|cc|cc|cc|cc}
		\multirow{2}{*}{$\Lambda$}&  \multicolumn{2}{c|}{$1/2^-(\Sigma_{c}\bar D)$}& \multicolumn{2}{c|}{$3/2^-(\Sigma_{c}^*\bar D)$} & \multicolumn{2}{c|}{$1/2^-(\Sigma_{c}\bar D^*)$}&\multicolumn{2}{c|}{$3/2^-(\Sigma_{c}\bar D^*)$}&\multicolumn{2}{c|}{$1/2^-(\Sigma_{c}^*\bar D^*)$}&\multicolumn{2}{c|}{$3/2^-(\Sigma_{c}^*\bar D^*)$}&\multicolumn{2}{c}{$5/2^-(\Sigma_{c}^*\bar D^*)$}\\
		&	$a=0$&	 $a=1$&$a=0$&	 $a=1$&$a=0$&	 $a=1$&$a=0$&	 $a=1$&$a=0$&	 $a=1$&$a=0$&	 $a=1$&$a=0$&	 $a=1$\\
		\hline
		1000    &\multicolumn{2}{c|}{$-$}          &\multicolumn{2}{c|}{$-$}            &$-23.12$     &$-$    & $-$       & $-$       & $-50.13$     & $-$    & $-2.44$   & $-$       & $-$       &$-0.48$\\
		1200    &\multicolumn{2}{c|}{$-$}          &\multicolumn{2}{c|}{$-$}            &$-117.27$    &$-$    & $-$       & $-4.99$   & $-351.24$    & $-$    & $-27.15$ & $-$       & $-$   &$-10.03$\\
		1400    &\multicolumn{2}{c|}{$-0.28$}      &\multicolumn{2}{c|}{$-0.36$}        &$-325.26$    &$-$    &$-0.04$    &$-19.42$   &$<-500$       & $-$    &$-88.16$ &$-0.21$    &$-0.24$   &$-31.65$\\
		1600    &\multicolumn{2}{c|}{$-3.73$}      &\multicolumn{2}{c|}{$-4.03$}        &$<-500$    &$-$    &$-0.98$    &$-41.04$  &$<-500$       & $-$    &$<-500$ &$-2.07$    &$-1.78$  &$-215.79$
	\end{tabular}
	\end{ruledtabular}
\end{table}

\begin{figure}[tb]
	\centering
	\subfigure[$J^P=1/2^-(\Sigma_c\bar D)$]{\label{wf4312}\includegraphics[width=0.3\textwidth]{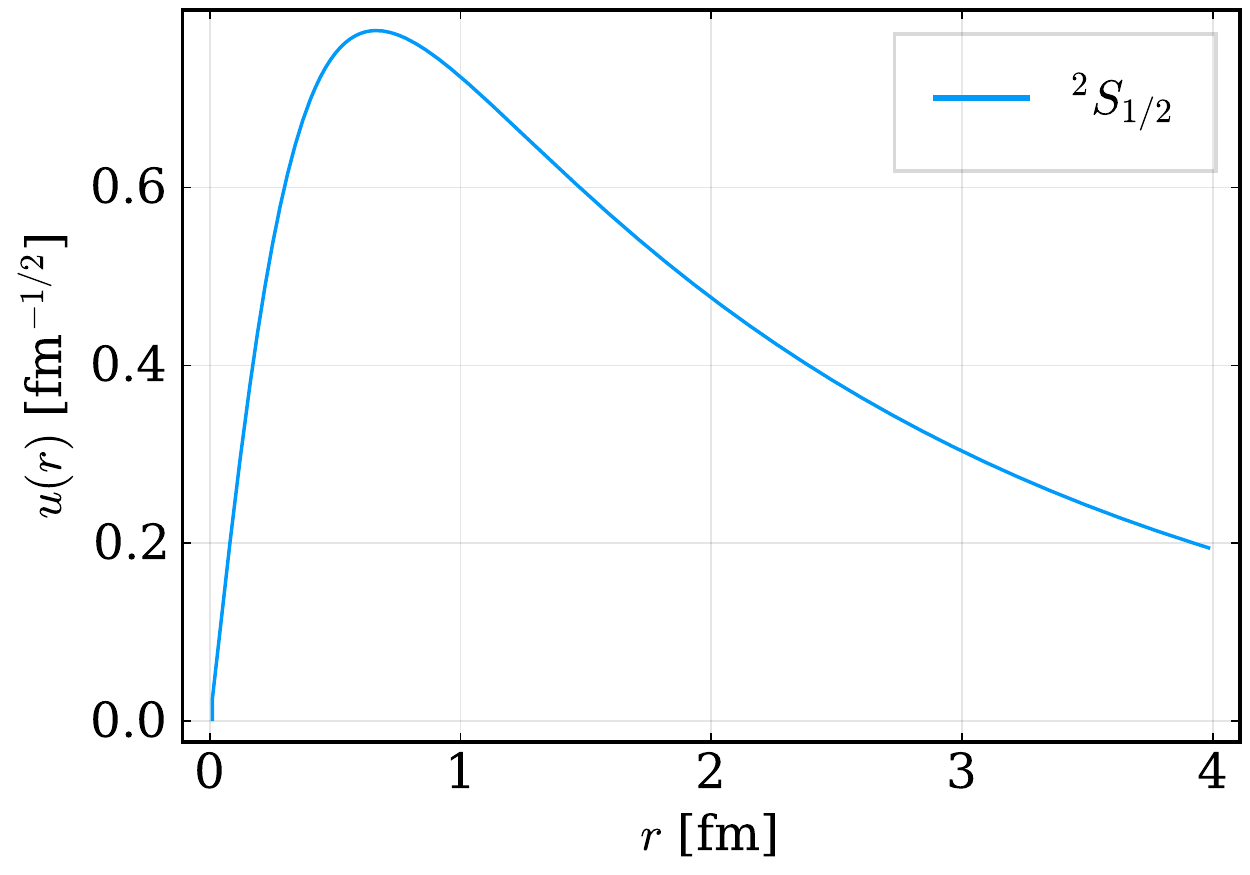}}
	\subfigure[$J^P=3/2^-(\Sigma_c^*\bar D)$]{\label{wf4380}\includegraphics[width=0.3\textwidth]{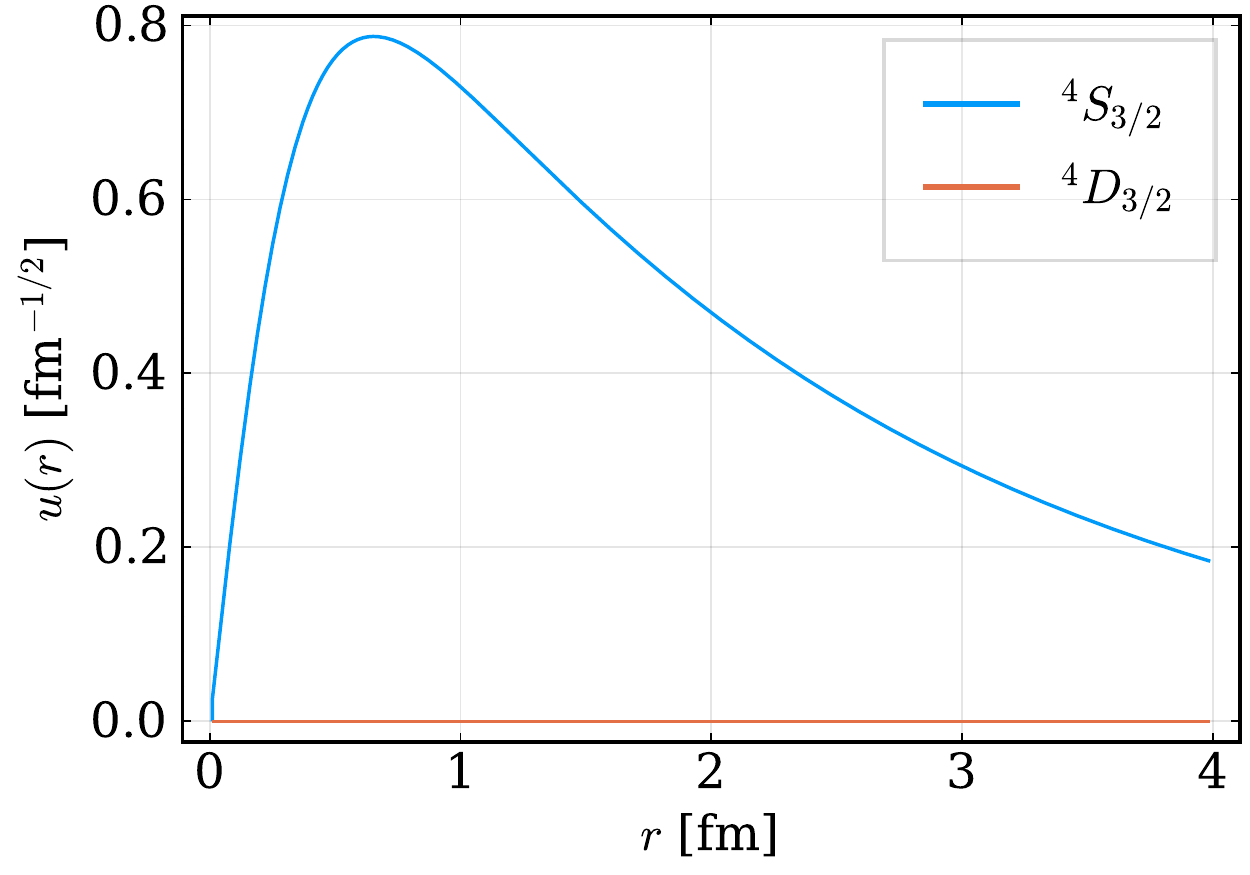}}
	\subfigure[$J^P=3/2^-(\Sigma_c\bar D^*)$]{\label{wf4440}\includegraphics[width=0.3\textwidth]{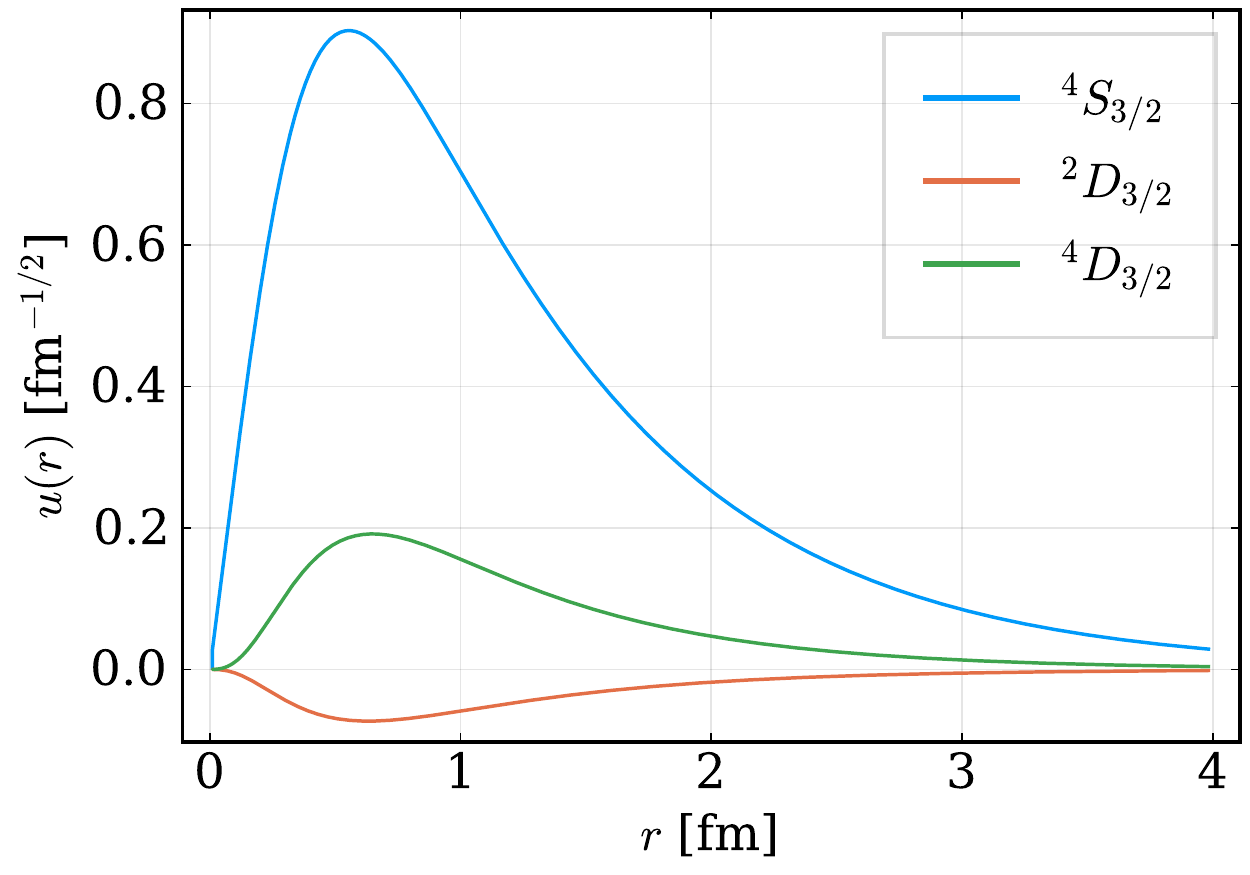}}
	\subfigure[$J^P=1/2^-(\Sigma_c\bar D^*)$]{\label{wf4457}\includegraphics[width=0.3\textwidth]{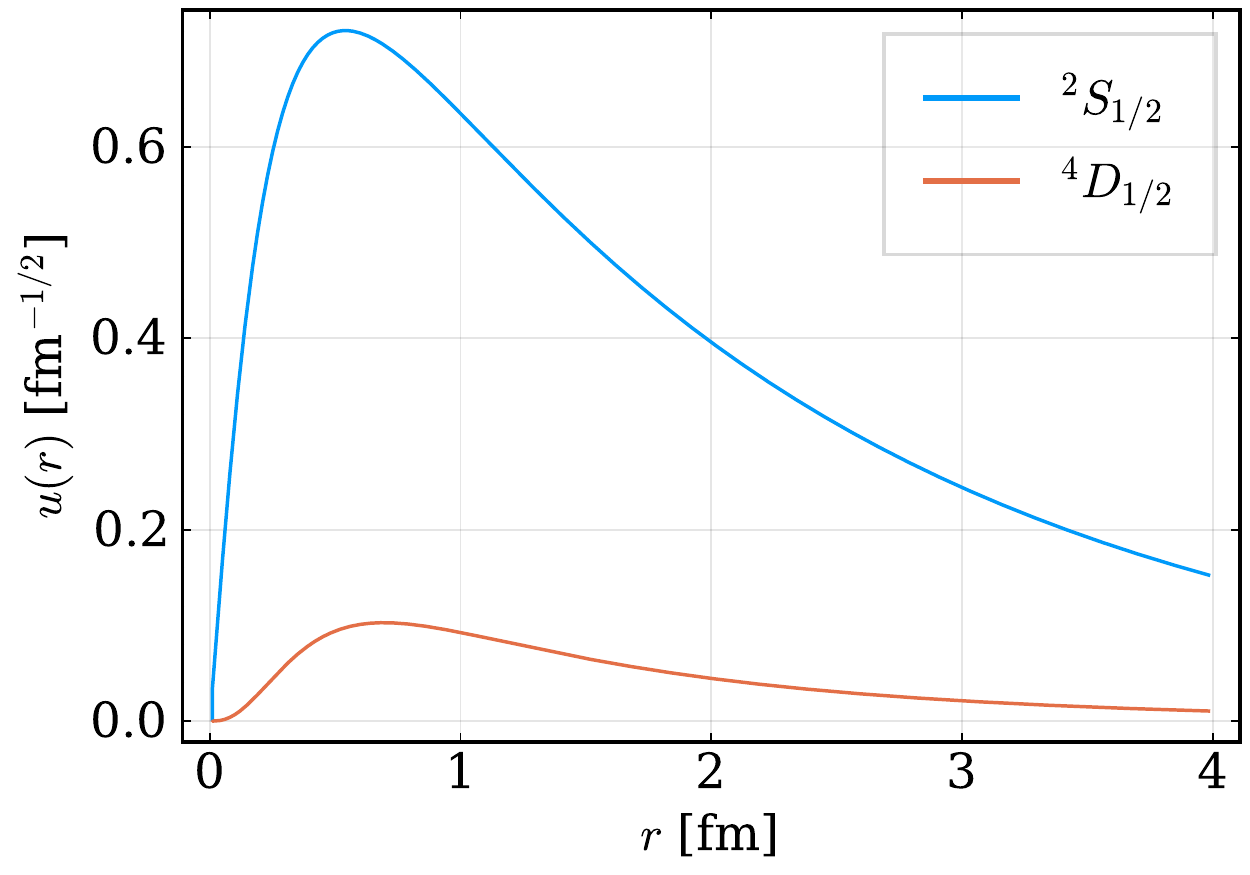}}
	\subfigure[$J^P=1/2^-(\Sigma_c^*\bar D^*)$]{\label{wf4518}\includegraphics[width=0.3\textwidth]{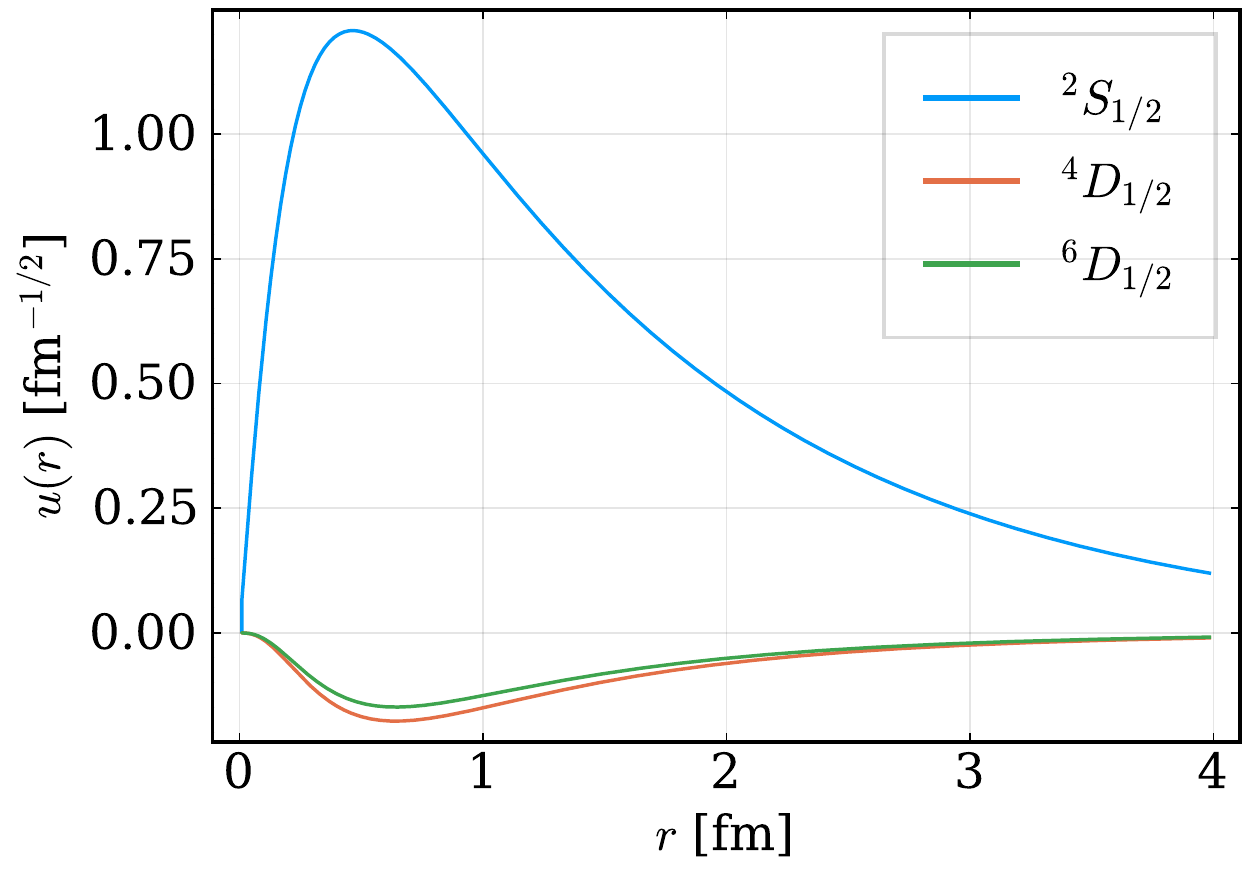}}
	\subfigure[$J^P=3/2^-(\Sigma_c^*\bar D^*)$]{\label{wf4514}\includegraphics[width=0.3\textwidth]{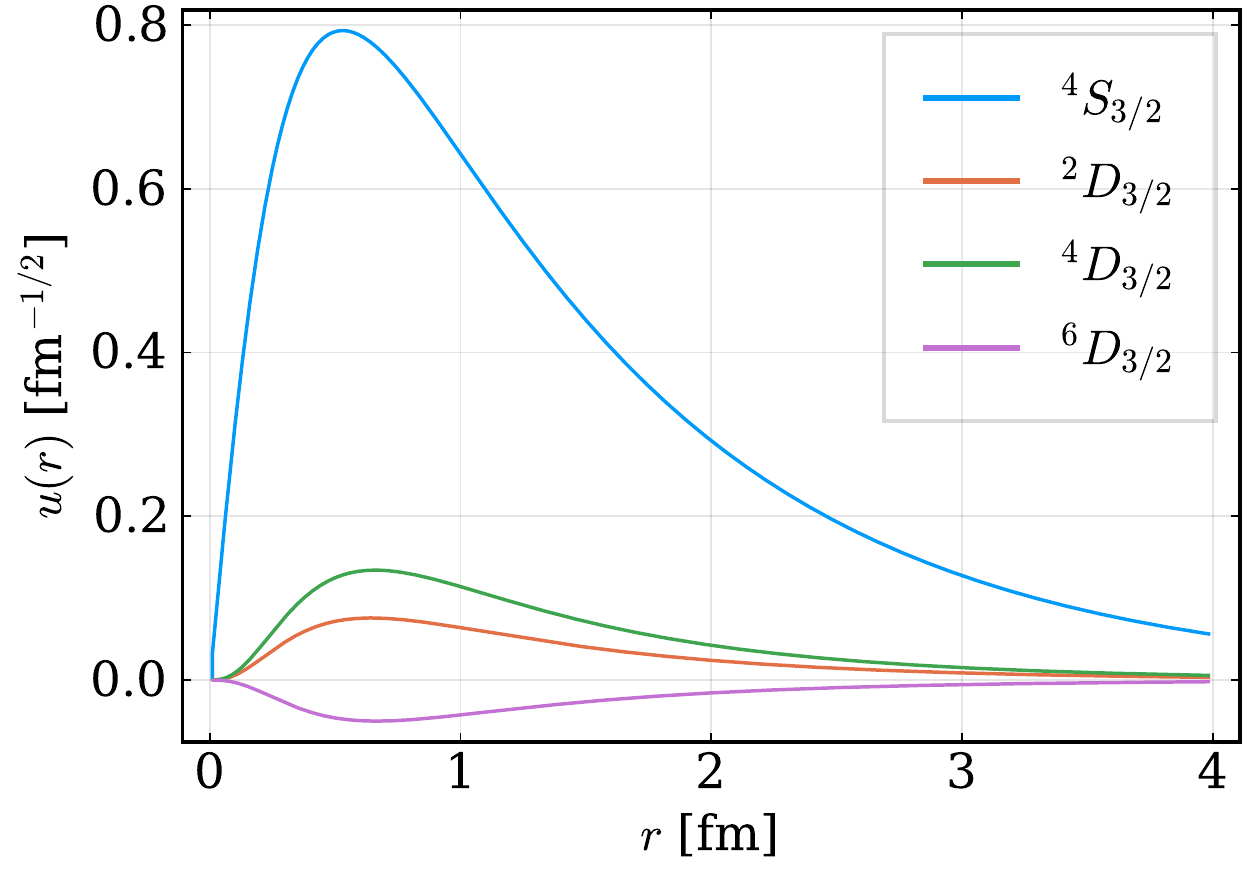}}
	\subfigure[$J^P=5/2^-(\Sigma_c^*\bar D^*)$]{\label{wf4498}\includegraphics[width=0.3\textwidth]{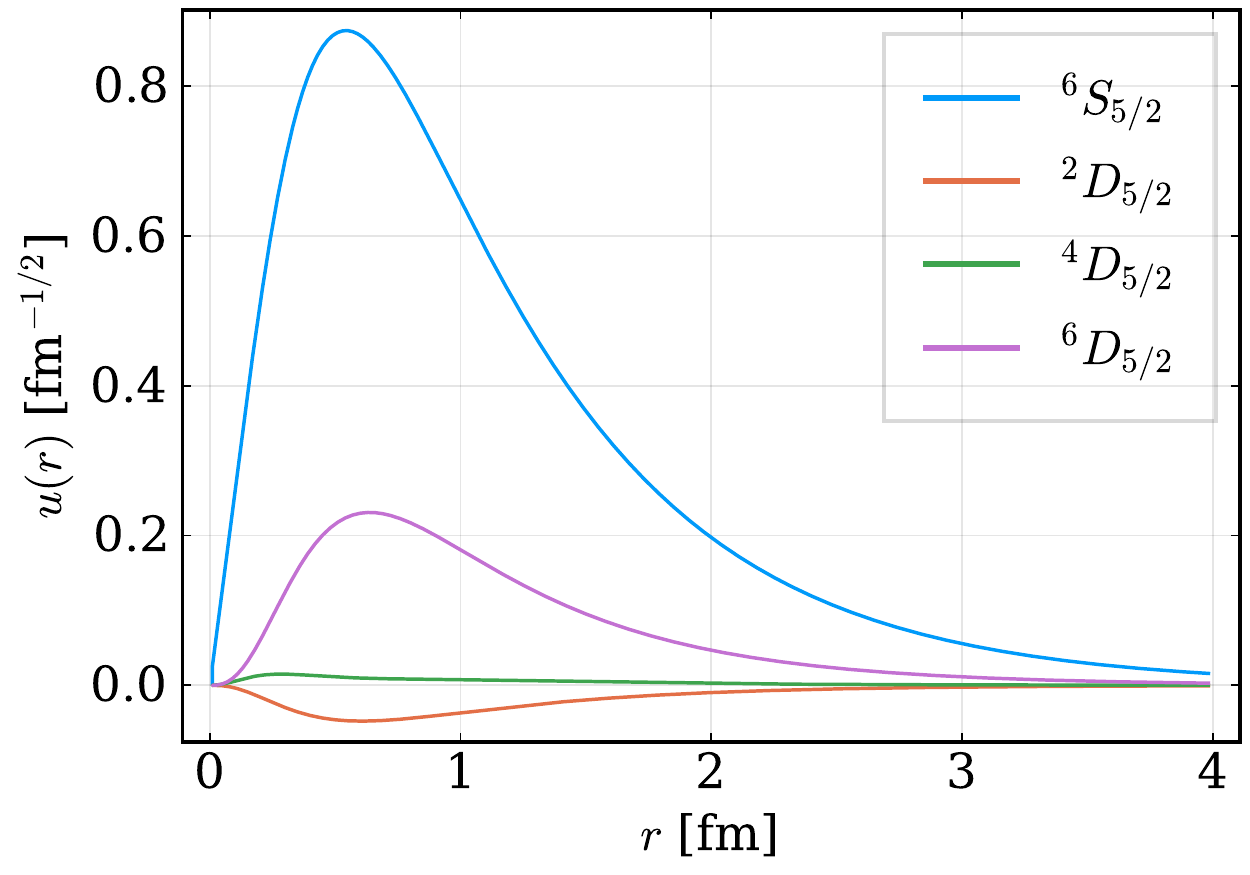}}
	\caption{Reduced wave functions $u(r)=rR(r)$ corresponding to the partial wave components considered in the single-channel   analysis with   $\Lambda=1.6$~GeV and $a=0.78$. The hadron pairs inside the parentheses are the corresponding channels.}\label{wavefig_for_single078}
\end{figure}

Then, let us move to the cases with the value of the reduction parameter $a$ taking a value somewhere between 0 and 1. The results with $a=0.78$ and $\Lambda=1.6$~GeV are obtained as
\begin{eqnarray}
	\Sigma_c\bar D  &:& M(1/2^-)=4317.38,\quad E(1/2^-)=-3.73,\nonumber\\
	\Sigma_c^*\bar D&:& M(3/2^-)=4381.34,\quad E(3/2^-)=-4.03,\nonumber\\
	\Sigma_c\bar D^*&:& M(3/2^-)=4441.01,\quad E(3/2^-)=-21.41,\nonumber\\
	\Sigma_c\bar D^*&:& M(1/2^-)=4458.44,\quad E(1/2^-)=-3.98,\nonumber\\
	\Sigma_c^*\bar D^*&:& M(1/2^-)=4518.17,\quad E(1/2^-)=-8.51,\nonumber\\
	\Sigma_c^*\bar D^*&:& M(3/2^-)=4514.67,\quad E(3/2^-)=-12.02,\nonumber\\
	\Sigma_c^*\bar D^*&:& M(5/2^-)=4498.19,\quad E(5/2^-)=-28.49,\nonumber
\end{eqnarray}       
where both the mass $M$ and the binding energy $E$ are in the units of MeV. The wave functions for the $P_c(4312)$, $P_c(4380)$, $P_c(4440)$, and $P_c(4457)$ as well as the other pentaquarks located below the $\Sigma_c^*\bar D^*$ threshold are shown in Fig.~\ref{wavefig_for_single078}. The $P_c(4312)$ and $P_c(4380)$ are pure $S$ wave molecules. 
For the $P_c(4440)$ and $P_c(4457)$, the $S$ wave components are dominant and mixed with a few percent of the $D$ wave components.         

If we take $a=0.58$, the $P_c(4440)$ and $P_c(4457)$ masses can be well reproduced in the $\Sigma_c\bar D^*$ channel with the same cutoff $\Lambda=1.4$~GeV. Their binding energies are solved as $E(1/2^-)=-23.81$~MeV and $E(3/2^-)=-4.51$~MeV, respectively. But there are no bound states for the lower $\Sigma_c\bar D$ and $\Sigma_c^*\bar D$ channels with the same parameters. For the $\Sigma_c^*\bar D^*$ system, three bound states with binding energies $E(1/2^-)=-46.21$~MeV, $E(3/2^-)=-12.98$~MeV, and $E(5/2^-)=-6.26$~MeV can be obtained with that set of parameters.

Finally, we consider the  $\Sigma_c\bar D$-$\Sigma_c^*\bar D$-$\Sigma_c\bar D^*$-$\Sigma_c^*\bar D^*$ coupled-channel  system with the $S$-$D$ wave mixing effects. 
We try to reproduce the $P_c(4312)$, $P_c(4380)$, $P_c(4440)$, and $P_c(4457)$ states by varying the cutoff $\Lambda$ and the reduction parameter $a$. 
As mentioned at the beginning of this section, the masses of $P_c(4380)$, $P_c(4440)$, and $P_c(4457)$ lie above the threshold of the $\Sigma_c\bar D$ channel. Then, these three $P_c$ states should be solved as resonances in the current coupled-channel system, that is, the eigenenergies of $P_c(4380)$, $P_c(4440)$, and $P_c(4457)$ will take complex values. 
Going to the appropriate Riemann sheets, one can find the complex poles of the $S$ matrix, which can be interpreted as resonances.
We interpret the real and imaginary parts of the pole position as the mass and half width of the corresponding resonance.\footnote{Since only the $\Sigma_c^{(*)}\bar D^{(*)}$ channels are included, and the finite widths of the $\Sigma_c^{(*)}$ and $\bar D^*$ are not considered, the width obtained here should be understood as a partial width into the channels considered. For an analysis with lower channels $\Lambda_c\bar D^{(*)}$, $\eta_cN$ and $J/\psi N$ included, see Ref.~\cite{Du:2021fmf}.}

\begin{figure}[tb]
	\centering
	\subfigure[ $a=0.55$]{\label{coupl055}\includegraphics[width=0.48\textwidth]{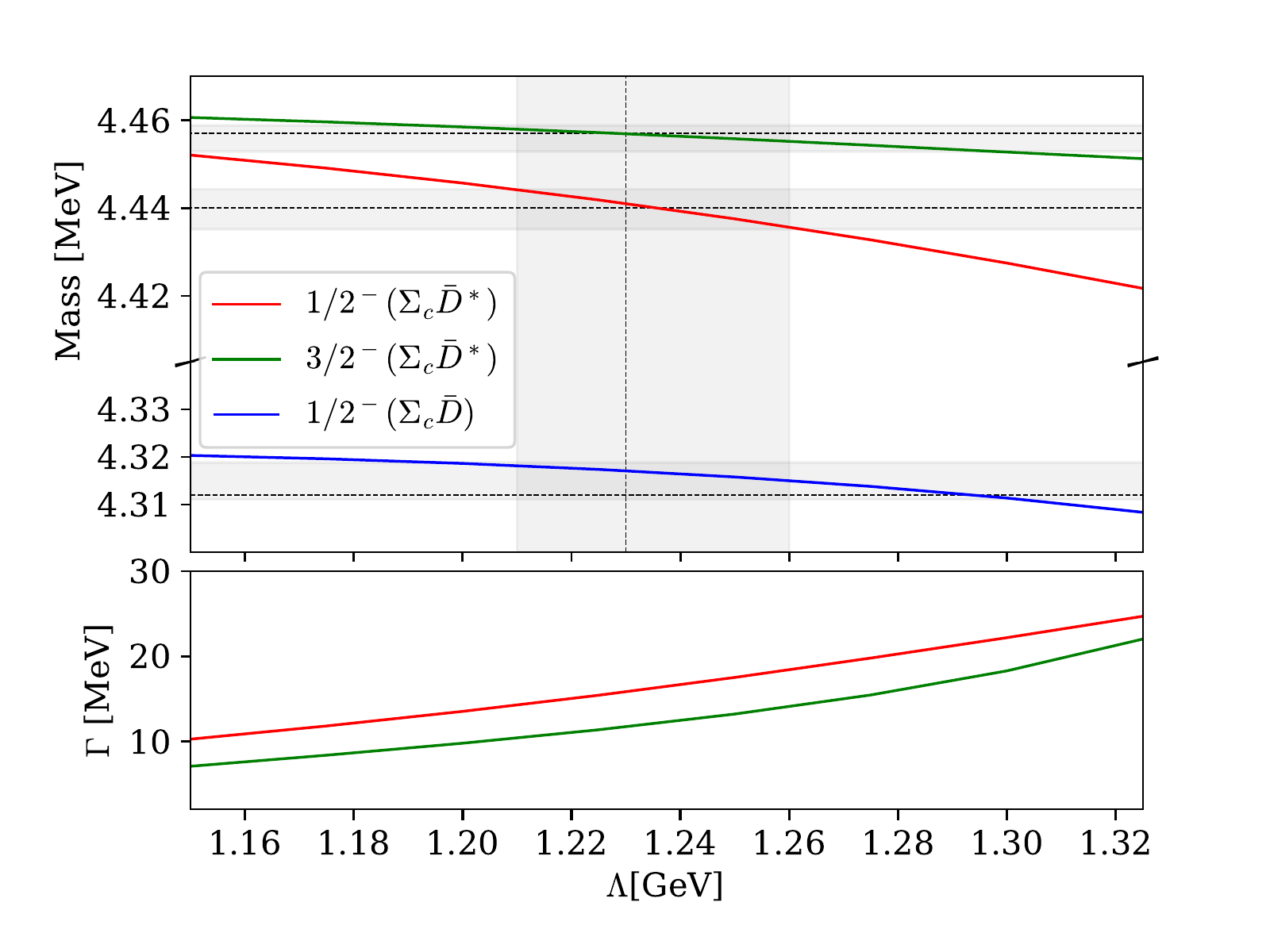}}
	\subfigure[ $a=0.79$]{\label{coupl079}\includegraphics[width=0.48\textwidth]{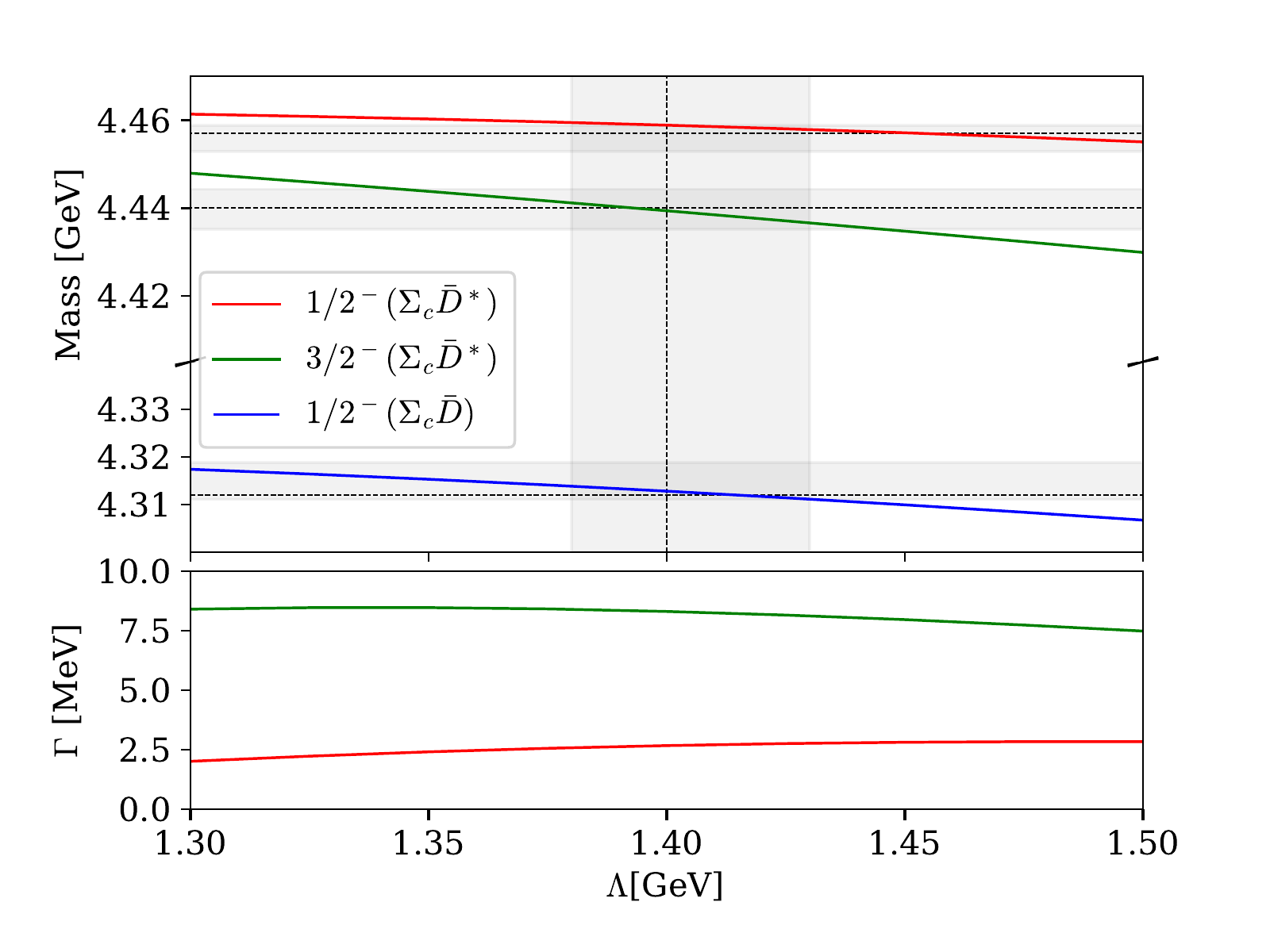}}
	\caption{Coupled-channel  results of isodoublet system $\Sigma_c^{(*)}\bar D^{(*)}$ by varying the cutoff $\Lambda$. Upper (lower) panel shows the mass (decay width through $\Sigma_c\bar D$ and $\Sigma_c^*\bar D$ channel) of corresponding states.  Horizontal gray bands are representing the experimental uncertainties of $P_c$ masses, and vertical gray bands stand for the cutoff range where masses of $P_c$ states can be simultaneously fitted. Lines for the $J^P=3/2^-(\Sigma_c^*\bar D)$ state are not shown, and its mass always lies in the experimental mass of $P_c(4380)$.}\label{fig_mass_coupl}
\end{figure}

Two sets of solutions are found that can reproduce the  $P_c(4312)$, $P_c(4380)$, $P_c(4440)$, and $P_c(4457)$ masses simultaneously. They are marked by  different reduction values, $a=0.55$ and $a=0.79$.
Figure~\ref{fig_mass_coupl} shows the masses (upper panel) and widths (lower panel) of the $1/2^-(\Sigma_c\bar D)$, $1/2^-(\Sigma_c\bar D^*)$ and $3/2^-(\Sigma_c\bar D^*)$ states as functions of $\Lambda$ for each value of $a$. 
The horizontal gray bands represent the experimental uncertainties of $P_c$ masses~\cite{LHCb:2019kea}, and the vertical gray bands stand for the cutoff range where masses of all $P_c$ states can be simultaneously reproduced. In this figure, we do not include the curves of the $3/2^-(\Sigma_c^*\bar D)$ molecule since its mass is always in line with the  $P_c(4380)$ within the whole cutoff range covered by the plot (and thus higher than the $P_c(4337)$ structure reported recently in Ref.~\cite{LHCb:2021chn}).
The vertical dashed lines are the best-fit solutions with $\Lambda=1.23\,\rm{GeV}$ for $a=0.55$ and $\Lambda=1.40\,\rm{GeV}$ for $a=0.79$, which are obtained by minimizing the $\chi^2$  that represents the deviation between our solved  $P_c$ masses and the LHCb measurements. 

The masses of $P_c$ states for the best-fit solutions are listed in Table~\ref{tab_coupl_res}. 
Note that the state with spin parity $J^P=1/2^-$ near the $\Sigma_c\bar D$ threshold does not have decay width since it emerges as a bound state in our calculation where the channel coupling to the lower channels is omitted.
As we can see from Table~\ref{tab_coupl_res}, the $P_c(4312)$ can be interpreted as the $1/2^-(\Sigma_c\bar D)$ molecule in both solutions, and it is consistent with the single-channel result. 
The spin-parity assignments for the $P_c(4440)$ and $P_c(4457)$ states are interchanged between these two solutions. In the solution with $a=0.55$, the masses of $P_c(4440)$ and $P_c(4457)$ can be reproduced well by the $1/2^-$ and $3/2^-$ $\Sigma_c\bar D^*$ molecules, respectively. However, the $P_c(4440)$ and $P_c(4457)$ are described as the $3/2^-$ and $1/2^-$ $\Sigma_c\bar D^*$ molecules in the solution with $a=0.79$ as in the single-channel case. 

\begin{figure}[tb]
	\includegraphics[width=0.9\textwidth]{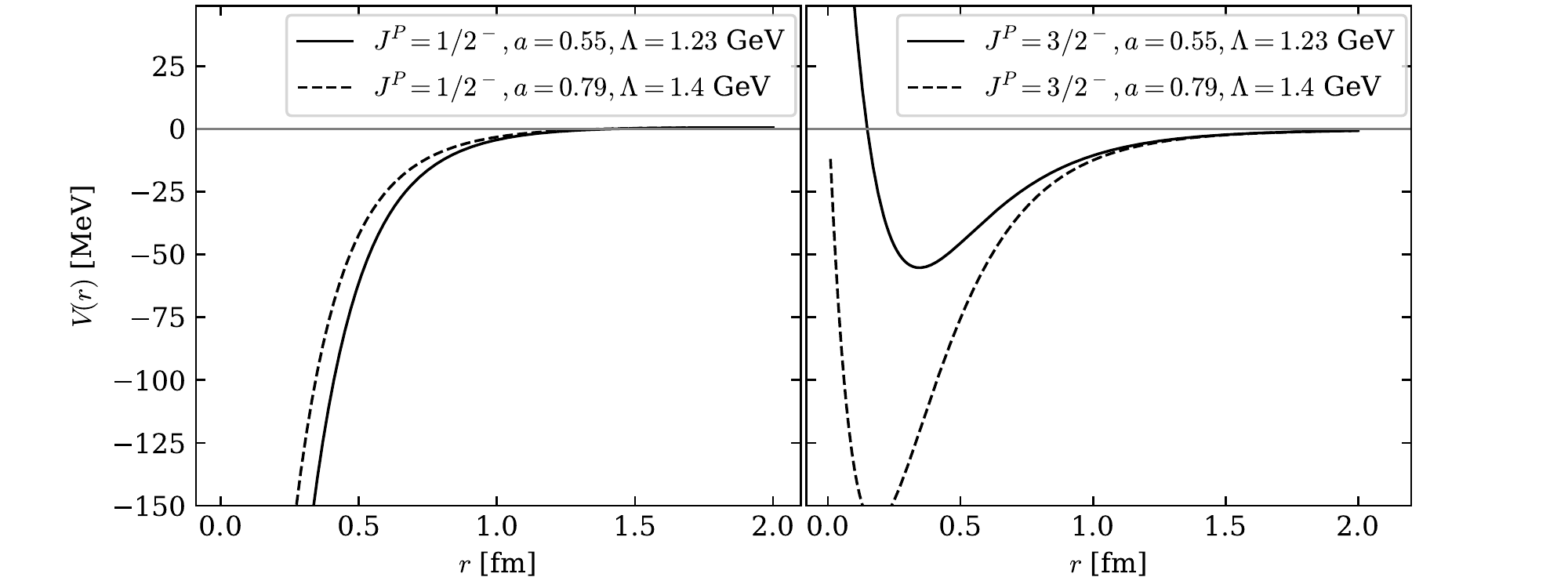}
	\caption{The $S$ wave potentials for the $\Sigma_c\bar D^*$ channel with spin parities of $J^P=1/2^-$ and $J^P=3/2^-$, where the total potentials (summing up all light-meson-exchange potentials) are shown.}\label{fig:pot}
\end{figure}      

Such spin-parity interchange can be understood as the dependence behavior of $1/2^-$ and $3/2^-$ $\Sigma_c\bar D^*$ elastic potential on the parameter $a$. 
We plot the impact of the parameter $a$ on the $\Sigma_c\bar D^*$ elastic potential for $J^P=1/2^-$ and $3/2^-$ in Fig.~\ref{fig:pot}. 
As one can see, the $1/2^-$ potential gets shallower as $a$ increases, leading to a smaller binding energy (absolute value of $E_B$) of the $1/2^-$ bound state, while the situation is reversed for the $3/2^-$ potential---the $3/2^-$ potential becomes deeper as $a$ increases, and the $3/2^-$ bound state will have a larger binding energy. It results in $3/2^-$-$P_c(4440)$ with larger $a$ and $1/2^-$-$P_c(4440)$ with smaller $a$. This behavior is originated from the sign difference of the $\delta(r)$ term in the $1/2^-$ and $3/2^-$ $\Sigma_c\bar D^*$ elastic potentials; see the value of the spin operator $\mathcal{O}_3^{44}\cdot\mathcal{O}_4^{44}$ in  Appendix~\ref{app_spin}.

In our model, we may distinguish the two possible solutions by the decays of the two states into the subdominant channels $\Sigma_c\bar D$ and $\Sigma_c^*\bar D$, which behave differently in these two spin-parity assignments. 
In the model calculation of Ref.~\cite{Lin:2019qiv}, the dominant decay channel for both $P_c(4440)$ and $P_c(4457)$ is suggested to be $\Lambda_c\bar D^{*}$.
For the solution with $a=0.55$, the partial decay width of the $3/2^-$ $\Sigma_c\bar D^*$ state corresponding to the $P_c(4457)$ is already larger than the central value of the experimentally measured total decay width of the $P_c(4457)$, $\Gamma_\text{exp}=6.4\pm2.0^{+5.7}_{-1.9}$~MeV,  and marginally consistent within $1\sigma$. 
It indicates that scenario I with a $3/2^-$ $P_c(4457)$, corresponding to the $a=0.55$ solution, is not favored. 
For the solution with $a=0.79$, corresponding to scenario II, the spin parities of $P_c(4440)$ and $P_c(4457)$ are $\{3/2^-,1/2^-\}$, and their partial decay widths through the subdominant $\Sigma_c^{(*)}\bar D$ channels are much smaller than the measured total widths and thus could be compatible with the experimental observations once lower channels such as $\Lambda_c\bar D^{(*)}$, $J/\psi N$, and $\eta_c N$ are considered.    

At last, let us mention that, since the widths of the $D^*$ and $\Sigma_c^{(*)}$ are not taken into account, the partial widths of the obtained states would be underestimated. It is expected that the widths for the states with $\Sigma_c \bar D^{(*)}$ as the main components are only marginally affected, while those for the $\Sigma_c^*\bar D^{(*)}$ can get a sizeable correction from the $\Sigma_c^*$ width (around 15~MeV).
In the favored scenario II, the $1/2^-$ and $3/2^-$ $\Sigma_c^*\bar D^*$ states have a small mass difference of only 7~MeV; considering further their decay widths, they could behave as a single structure around 4.52~GeV in the experimental data.
The $5/2^-$ state has a mass about 4.50~GeV.
These results are similar to those obtained from fitting to the LHCb data in Ref.~\cite{Du:2021fmf}.
It is worthwhile to notice that the LHCb data show a signal of nontrivial structures around 4.50 and 4.52~GeV in the $J/\psi p$ invariant mass distribution, in particular in the ``$m_{Kp}$ all" dataset~\cite{LHCb:2019kea}. Future data with higher statistics will be able to resolve the $\Sigma_c^*\bar D^*$ states.

\begin{table}[tb]\centering
	\caption{Poles of the $S$ matrix (corresponding to $M-i\Gamma/2$ ) closest to physical real axis  in the coupled-channel  analysis of the isodoublet $\Sigma_c^{(*)}\bar D^{(*)}$ systems with $(\Lambda,a)=(1.23\,\rm{    GeV},0.55)$ and $(\Lambda,a)=(1.4\,\rm{ GeV},0.79)$, corresponding to scenario I and scenario II, respectively. The channel closest to each pole is given in the parentheses in the first column. The results are in units of MeV.}\label{tab_coupl_res}
	\begin{ruledtabular}
	\begin{tabular}{l|c c}
		$J^P$ (dominant channel) &  
		$(\Lambda,a)=(1.23\,\rm{GeV},0.55)$ & $(\Lambda,a)=(1.4\,\rm{GeV},0.79)$\\\hline
$1/2^-(\Sigma_c\bar D)$        &$4317.1$       & $4312.8$\\
$3/2^-(\Sigma_c^*\bar D)$      &$4379.8-0.0i$      & $4375.6-0.1i$\\
$1/2^-(\Sigma_c\bar D^*)$      &$4441.0-8.0i$      & $4458.8-1.3i$\\
$3/2^-(\Sigma_c\bar D^*)$      &$4456.9-5.9i$       & $4439.4-4.2i$\\
$1/2^-(\Sigma_c^*\bar D^*)$    &$4498.6-6.6i$       & $4525.0-0.8i$\\
$3/2^-(\Sigma_c^*\bar D^*)$    &$4511.1-16.6i$      & $4518.0-4.2i$\\
$5/2^-(\Sigma_c^*\bar D^*)$    &$4521.9-5.1i$       & $4498.3-6.3i$
	\end{tabular}
	\end{ruledtabular}
\end{table}

\section{Conclusion}\label{sec:4}

We investigate the coupled-channel dynamics of the $\Lambda_{c1}\bar D$ and  $\Sigma_c^{(*)}\bar D^{(*)}$ channels within the OBE model to test the mechanism of a $J^P=1/2^+$ state being triggered by the $\Lambda_{c1}\bar D$ channel. It is found that the $J^P=1/2^+$ system cannot be bound with the OBE parameters constrained by other experimental sources with reasonable cutoff values  because the nondiagonal potentials in the $\Lambda_{c1}\bar D$ and $\Sigma_c\bar D^{(*)}$ coupled-channel system are not strong enough.
The situation does not change qualitatively when the $\Sigma_c^*\bar D^*$ channel is included in addition.  

We further investigate the role of the $\delta(\vec{r})$ term in the coupled-channel dynamics. Here, the $\delta(\vec{r})$ term comes from the constant term of the $t$-channel scattering amplitudes in the momentum space. 
Such a $\delta(\vec{r})$-term contribution is of a short-distance nature and needs to be regularized. Here the regularization is performed by introducing dipole form factors, the effects of which may be recognized as the short-range interactions derived by the exchange of mesons heavier than $\rho$ and $\omega$ mesons. 
In this work, as a phenomenological study, the $\delta(\vec{r})$ term is corrected by introducing a reduction factor $a$ that quantifies how much the $\delta(\vec{r})$ potential is reduced in the OBE potential. 
$a$ varies in the range of $[0,1]$ in our analysis. 
Two set of solutions for the parameters, the cutoff $\Lambda$ and the reduction parameter $a$, are found to be able to reproduce the masses of the observed $P_c$ states in the hadronic molecular picture. 
In the first solution, called scenario I, the best description of the $P_c$ masses is given by the parameters $\{\Lambda=1.23\,\rm{GeV},\,a=0.55\}$, where $P_c(4312)$, $P_c(4440)$, and $P_c(4457)$ are interpreted as the $1/2^-(\Sigma_c\bar D)$, $1/2^- (\Sigma_c\bar D^*$), and $3/2^-(\Sigma_c\bar D^*)$ molecules, respectively. 
The second solution, called scenario II, has $\{\Lambda=1.4\,\rm{GeV},\,a=0.79\}$, and the spin-parity quantum numbers of the $P_c(4440)$ and $P_c(4457)$ states are $3/2^-$ and $1/2^-$, respectively. 
Scenario II is favored since the partial decay width of the $P_c(4457)$ in scenario I is larger than the central value of the experimental value. This is consistent with previous analysis from an effective field theory point of view~\cite{Du:2021fmf}.
In this preferred scenario, the $P_c(4312)$, $P_c(4380)$, $P_c(4440)$, and $P_c(4457)$ states are the $1/2^-(\Sigma_c\bar D)$, $3/2^-(\Sigma_c^*\bar D)$, $3/2^-(\Sigma_c\bar D^*)$, and $1/2^-(\Sigma_c\bar D^*)$ molecules, respectively. 
Besides, another three $P_c$ states exist below the $\Sigma_c^*\bar D^*$ threshold, and their spin-parity quantum numbers and masses are  
$J^P(4525,4518,4498)=(1/2^-,3/2^-,5/2^-)$ in scenario II. 
These three states may show up as two structures at about 4.50 and 4.52~GeV. There are hints for their existence in the LHCb data, and their confirmation is expected with data of higher statistics. 
                      
\begin{acknowledgments}
 
 We thank Timothy Burns, Rui Chen,  Xiang-Kun Dong, Li-Sheng Geng, Hao-Jie Jing, Ming-Zhu Liu, Yakefu Reyimuaji, and Qiang Zhao for helpful discussions.  This work is supported by the National Natural Science Foundation of China (NSFC) under Grants No. 11835015, No. 12047503, and No.~11961141012; by the NSFC and the Deutsche Forschungsgemeinschaft (DFG) through the funds provided to the Sino-German Collaborative Research Center TRR110 ``Symmetries and the Emergence of Structure in QCD" (NSFC Grant No. 12070131001, DFG Project No. 196253076); by
 the Chinese Academy of Sciences (CAS) under Grants No.~XDPB15, No.~QYZDB-SSW-SYS013, and No.~XDB34030000; by the CAS President’s International Fellowship Initiative (PIFI) under Grant No.~2020PM0020; and by the China Postdoctoral Science Foundation under Grant No.~2020M680687.

\end{acknowledgments}

 \section*{Appendix}
 \begin{appendices}

 	\section{Spin Operators}\label{app_spin}

 	The spin wave functions for spin-$1/2$ and-$3/2$ particles are denoted with $\chi$ and $\vec\chi$, respectively, where $\chi$ is a two-component spinor. With the Clebsch-Gordan coefficients, the spin-$3/2$ spinor $\vec\chi$ for the $i$th particle can be decomposed as
 	\begin{eqnarray}
 		\vec\chi_i(h)=\mathbb{C}_{h_1,h_2}^{3/2,h}\vec\epsilon_i(h_2)\chi_i(h_1),
 	\end{eqnarray}
 where $\vec\epsilon$ is polarization vector and $\vec\epsilon(\pm1)=\mp 1/\sqrt{2}(1,\pm i,0)$, $\vec\epsilon(0)=(0,0,1)$.

 	Once we enumerate channels $\Sigma_c\bar{D}$, $\Sigma_c^*\bar{D}$, $\Lambda_{c1}\bar{D}$, $\Sigma_c\bar{D}^*$, and $\Sigma_c^*\bar{D}^*$ with upper indices $1,~2,~3,~4$, and $5$, respectively, all potentials for the coupled channels obtained by $t$-channel transitions $H_1H_2\to H_3H_4$ can be generated by the operators below. 
	$O_{1}^{ij}$ and $ O_{2}^{ij}$ are diagonal, and $O_1=O_2=(\chi^\dagger_3\chi_1,\vec\chi^\dagger_3\cdot\vec\chi_1,0,\chi^\dagger_3\chi_1,\vec\chi^\dagger_3\cdot\vec\chi_1)$. In the following, we only show the operators at the upper triangle of the coupled-channel potential matrix; the others can be obtained with the Hermitian condition of the potential matrix,    
 	\begin{eqnarray}
 		 	\vec O_3=\begin{pmatrix}
 			0&0&0&\chi^\dagger_3\vec\sigma\chi_1&i\vec\chi^\dagger_3\times\vec\sigma\chi_1\\
 			&0&0&i\vec\chi^\dagger_3\times\vec\sigma\chi_1&i\vec\chi^\dagger_3\times\vec\chi_1\\
 			& &0&0&0\\
 			& & &\chi^\dagger_3\vec\sigma\chi_1&i\vec\chi^\dagger_3\times\vec\sigma\chi_1\\
 			& & & &i\vec\chi^\dagger_3\times\vec\chi_1
 		\end{pmatrix},\quad
 		\vec O_4=\begin{pmatrix}
 			0&0&0&\vec\epsilon^*_4&\vec\epsilon^*_4\\
 			 &0&0&\vec\epsilon^*_4&\vec\epsilon^*_4\\
 			 & &0&0&0\\
 			 & & &i\vec\epsilon_2\times\vec\epsilon^*_4&i\vec\epsilon_2\times\vec\epsilon^*_4\\
 			 & & & &i\vec\epsilon_2\times\vec\epsilon^*_4
 		\end{pmatrix},\label{spin_matr}
 	\end{eqnarray}
 	and $\vec O_5^{34}=\vec\epsilon^*_4$, $\vec O_{6}^{34}=\chi^\dagger_3\vec\sigma\chi_1$, and $\vec O_{6}^{35}=\vec\chi^\dagger_3\times\vec\sigma\chi_1$. The other operators are zero. 

 	The partial wave projection of operators in the $J^P=1/2^-,3/2^-,1/2^+$ system is shown in the Table~\ref{tab_partial wave}, which is calculated by sandwiching the operators given above between the partial waves of the initial and final states~\cite{angular_momentom}. Every element of the spin operators is replaced by the corresponding partial wave projections in the actual calculation. 
	Here, we calculate the spin projection for the transition $ 1 \to 5$  ($\Sigma_c\bar D\to\Sigma_c^*\bar D^*$) as an example. For $J^P=1/2^-$, the partial waves for the initial and final states are
 	\begin{eqnarray}
 		\Sigma_c\bar D &:& ^2S_{1/2},\nonumber\\
 		\Sigma_c^*\bar D^* &:& ^2S_{1/2},^4D_{1/2},^6D_{1/2},\label{par_wae15}
 	\end{eqnarray}
    From Eq.~\eqref{spin_matr} together with Eqs.~\eqref{pi_r_pot} and \eqref{vec_r_pot}, we know that the spin operators, which are universal for the pseudoscalar and vector exchange potentials, are $\vec O_3^{15}\cdot \vec O_4^{15}=(i\vec\chi^\dagger_3\times\vec\sigma\chi_1)\cdot\vec\epsilon^*_4$ for the spin-spin coupling and $S(\vec O_3^{15},\vec O_4^{15},\hat r)=3(i\vec\chi^\dagger_3\times\vec\sigma\chi_1)\cdot \hat r \vec\epsilon^*_4\cdot\hat r-(i\vec\chi^\dagger_3\times\vec\sigma\chi_1)\cdot\vec\epsilon^*_4$ for the tensor coupling.
    Then it can be calculated as      
 	   \begin{eqnarray}
 	   	\langle^{2S'+1}L'_{J}|(i\vec\chi_3^\dagger\times\vec\sigma\chi_1)\cdot\vec\epsilon_4^*|^{2S+1}L_J\rangle&=&i\mathbb{C}_{ml',ms'}^{J,m}\mathbb{C}_{ml,ms}^{J,m}\mathbb{C}_{m3,m4}^{S',ms'}\mathbb{C}_{m1,m2}^{S,ms}\mathbb{C}_{h1,h2}^{s3,m3}\nonumber\\
 	   	&\times&\sum_{i,j,k=1}^3\varepsilon^{ijk}\epsilon^{*i}_3(h2)\chi_3^\dagger(h1)\sigma^j\chi_1(m1)\epsilon^{*k}_4(m4)\langle L',ml'|L,ml\rangle,\label{formula_par_15}\\
 	   	\langle^{2S'+1}L'_{J}|S(\vec O_3^{15},\vec O_4^{15},\hat r)|^{2S+1}L_J\rangle&=&i\mathbb{C}_{ml',ms'}^{J,m}\mathbb{C}_{ml,ms}^{J,m}\mathbb{C}_{m3,m4}^{S',ms'}\mathbb{C}_{m1,m2}^{S,ms}\mathbb{C}_{h1,h2}^{s3,m3}\nonumber\\
 	   	&\times&\{3\sum_{i,j,k,o=1}^3\varepsilon^{ijk}\epsilon^{*i}_3(h2)\chi^\dagger_3(h1)\sigma^j\chi_1(m1)\epsilon^{*o}_4(m4)\langle L',ml'| r^k  r^o|L,ml\rangle\nonumber\\ &-&\sum_{i,j,k=1}^3\varepsilon^{ijk}\epsilon^{*i}_3(h2)\chi_3^\dagger(h1)\sigma^j\chi_1(m1)\epsilon^{*k}_4(m4)\langle L',ml'|L,ml\rangle\},\label{formula_par_T_15}
 	   \end{eqnarray}
    where the lower indices of Clebsch-Gordan coefficients which represent the magnetic quantum numbers should be summed. 
    $\langle^{2S'+1}L'_{J}|$ and  $|^{2S+1}L_J\rangle$ stand for the partial waves for the final and initial states, respectively. The  spherical harmonics are integrated as   
    \begin{eqnarray}
    \langle L',ml'| r^k  r^o|L,ml\rangle=\int d\Omega Y^*_{L',ml'}(\theta,\phi) r^k r^oY_{L,ml}(\theta,\phi),     
    \end{eqnarray}
    where  $Y_{L,ml}(\theta,\phi)$ is the spherical harmonics and $r^k$ and $r^o$ are the components of the unit vector  $\hat r$ in the spherical coordinate.
    After having calculated Eqs.~\eqref{formula_par_15} and \eqref{formula_par_T_15} with the partial waves given Eq.~\eqref{par_wae15}, we get $(-\sqrt 2,0,0)$ and $(0,\sqrt{2/5},3\sqrt{2/5})$, respectively. One should be careful about the convention. If the ordering of the angular momenta in the Clebsch-Gordan coefficient $\mathbb{C}_{ml',ms'}^{J,m}$ changes to $\mathbb{C}_{ms',ml'}^{J,m}$, the result of Eq.~\eqref{formula_par_T_15} will change to $(0,-\sqrt{2/5},3\sqrt{2/5})$. All partial waves in our work are calculated with the convention of $\mathbb{C}_{ml',ms'}^{J,m}$, and the results are collected in Table~\ref{tab_partial wave}.      
\begin{center}
 	\begin{longtable}[H]{c|c|c|c}
 		\caption{Partial wave projection of the spin operators in the potentials.}\label{tab_partial wave} \\
 			\hline\hline
 			&$J^P=1/2^-$&$J^P=3/2^-$&$J^P=1/2^+$\\
 			\hline
 			\endfirsthead
 			\multicolumn{4}{c}%
 			{\tablename\ \thetable\ -- \text{Continued from previous page}} \\
 			\hline\hline
 			&$J^P=1/2^-$&$J^P=3/2^-$&$J^P=1/2^+$\\
 			\hline
			\endhead
			\hline \multicolumn{4}{c}{\text{Continued on next page}} \\
			\endfoot
			\hline
			\endlastfoot
 			$O_1^{11}$&$(1)$&$(1)$&$(1)$\\
 			$O_1^{22}$&$(1)$&diag$(1,1)$&$(1)$\\
 			$O_1^{44}$&diag$(1,1)$&diag$(1,1,1)$&diag$(1,1)$\\
 			$O_1^{55}$&diag$(1,1,1)$&diag$(1,1,1,1)$&diag$(1,1)$\\
 			$\vec O_3^{14}\cdot \vec O_4^{14}$&$(\sqrt 3 ,0)$&$(0,\sqrt 3 ,0)$&$(\sqrt 3 ,0)$\\
 			$S(\vec O_3^{14}\cdot,\vec O_4^{14},\hat r)$&$(0,\sqrt{6})$&$(-\sqrt 3 ,0,\sqrt 3 )$&$(0,\sqrt{6})$\\
 			$\vec O_3^{15}\cdot \vec O_4^{15}$&$(-\sqrt 2 ,0,0)$&$(0,-\sqrt 2,0,0)$&$(-\sqrt 2,0,0)$\\
			$S(\vec O_3^{15}\cdot,\vec O_4^{15},\hat r)$&$(0,\sqrt{2/5},3\sqrt{2/5})$&$(-\frac{1}{\sqrt 5},0,\frac{1}{\sqrt 5},-3\sqrt\frac{2}{35})$&$(0,\sqrt{2/5},3\sqrt{2/5})$\\
 			$\vec O_3^{24}\cdot \vec O_4^{24}$&$(0,1)$&$\begin{pmatrix}1&0&0\\0&0&1\end{pmatrix}$&$(0,1)$\\
			$S(\vec O_3^{24}\cdot,\vec O_4^{24},\hat r)$&$(\sqrt 2,1)$&$\begin{pmatrix}0&-1&-1\\-1&1&0\end{pmatrix}$&$(\sqrt 2,1)$\\
			$\vec O_3^{25}\cdot \vec O_4^{25}$&$(0,-\sqrt\frac{5}{3},0)$&$\begin{pmatrix}-\sqrt\frac{5}{3}&0&0&0\\0&0&-\sqrt\frac{5}{3}&0\end{pmatrix}$&$(0,-\sqrt\frac{5}{3},0)$\\
			$S(\vec O_3^{25}\cdot,\vec O_4^{25},\hat r)$&$(\sqrt\frac{1}{3},\frac{4}{\sqrt{15}},-\sqrt\frac{3}{5})$&$\begin{pmatrix}0&-\sqrt\frac{1}{6}&-\frac{4}{\sqrt{15}}&-\sqrt\frac{21}{10}\\-\frac{4}{\sqrt{15}}&\sqrt\frac{1}{6}&0&-\sqrt\frac{15}{14}\end{pmatrix}$&$(\sqrt\frac{1}{3},\frac{4}{\sqrt{15}},-\sqrt\frac{3}{5})$\\
$\vec O_3^{44}\cdot \vec O_4^{44}$&$\begin{pmatrix}-2&0\\0&1\end{pmatrix}$&$\begin{pmatrix}1&0&0\\0&-2&0\\0&0&1\end{pmatrix}$&$\begin{pmatrix}-2&0\\0&1\end{pmatrix}$\\
$S(\vec O_3^{44}\cdot,\vec O_4^{44},\hat r)$&$\begin{pmatrix}0&\sqrt 2\\\sqrt 2&-2\end{pmatrix}$&$\begin{pmatrix}0&-1&2\\-1&0&1\\2&1&0\end{pmatrix}$&$\begin{pmatrix}0&\sqrt 2\\\sqrt 2&-2\end{pmatrix}$\\ 			
$\vec O_3^{45}\cdot \vec O_4^{45}$&$\left(
\begin{array}{ccc}
	-\sqrt{\frac{2}{3}} & 0 & 0 \\
	0 & -\sqrt{\frac{5}{3}} & 0 \\
\end{array}
\right)$&$\left(
\begin{array}{cccc}
	-\sqrt{\frac{5}{3}} & 0 & 0 & 0  \\
	0 & -\sqrt{\frac{2}{3}} & 0 & 0  \\
	0 & 0 & -\sqrt{\frac{5}{3}} & 0  \\
\end{array}
\right)$&$\left(
\begin{array}{ccc}
	-\sqrt{\frac{2}{3}} & 0 & 0 \\
	0 & -\sqrt{\frac{5}{3}} & 0 \\
\end{array}
\right)$\\
$S(\vec O_3^{45}\cdot,\vec O_4^{45},\hat r)$&$\left(
\begin{array}{ccc}
	0 & 4 \sqrt{\frac{2}{15}} & -\sqrt{\frac{6}{5}} \\
	\frac{1}{\sqrt{3}} & \frac{1}{\sqrt{15}} & \sqrt{\frac{3}{5}} \\
\end{array}
\right)$&$\left(
\begin{array}{cccc}
	0 & -\frac{1}{\sqrt{6}} & -\frac{1}{\sqrt{15}} & \sqrt{\frac{21}{10}}  \\
	-\frac{4}{\sqrt{15}} & 0 & \frac{4}{\sqrt{15}} & \sqrt{\frac{6}{35}}  \\
	-\frac{1}{\sqrt{15}} & \frac{1}{\sqrt{6}} & 0 & \sqrt{\frac{15}{14}}  \\
\end{array}
\right)$&$\left(
\begin{array}{ccc}
	0 & 4 \sqrt{\frac{2}{15}} & -\sqrt{\frac{6}{5}} \\
	\frac{1}{\sqrt{3}} & \frac{1}{\sqrt{15}} & \sqrt{\frac{3}{5}} \\
\end{array}
\right)$\\ 			
$\vec O_3^{55}\cdot \vec O_4^{55}$&$\left(
\begin{array}{ccc}
	\frac{5}{3} & 0 & 0 \\
	0 & \frac{2}{3} & 0 \\
	0 & 0 & -1 \\
\end{array}
\right)$&$\left(
\begin{array}{cccc}
	\frac{2}{3} & 0 & 0 & 0  \\
	0 & \frac{5}{3} & 0 & 0  \\
	0 & 0 & \frac{2}{3} & 0 \\
	0 & 0 & 0 & -1  \\
\end{array}
\right)$&$\left(
\begin{array}{ccc}
	\frac{5}{3} & 0 & 0 \\
	0 & \frac{2}{3} & 0 \\
	0 & 0 & -1 \\
\end{array}
\right)$\\
$S(\vec O_3^{55}\cdot,\vec O_4^{55},\hat r)$&$\left(
\begin{array}{ccc}
	0 & \frac{7}{3 \sqrt{5}} & \frac{2}{\sqrt{5}} \\
	\frac{7}{3 \sqrt{5}} & \frac{16}{15} & \frac{1}{5} \\
	\frac{2}{\sqrt{5}} & \frac{1}{5} & \frac{8}{5} \\
\end{array}
\right)$&$\left(
\begin{array}{cccc}
	0 & -\frac{7}{3 \sqrt{10}} & -\frac{16}{15} & \frac{\sqrt{\frac{7}{2}}}{5}  \\
	-\frac{7}{3 \sqrt{10}} & 0 & \frac{7}{3 \sqrt{10}} & -\frac{2}{\sqrt{35}}  \\
	-\frac{16}{15} & \frac{7}{3 \sqrt{10}} & 0 & \frac{1}{\sqrt{14}}  \\
	\frac{\sqrt{\frac{7}{2}}}{5} & -\frac{2}{\sqrt{35}} & \frac{1}{\sqrt{14}} & \frac{4}{7}  
\end{array}
\right)$&$\left(
\begin{array}{ccc}
	0 & \frac{7}{3 \sqrt{5}} & \frac{2}{\sqrt{5}} \\
	\frac{7}{3 \sqrt{5}} & \frac{16}{15} & \frac{1}{5} \\
	\frac{2}{\sqrt{5}} & \frac{1}{5} & \frac{8}{5} \\
\end{array}
\right)$\\
$\vec O^{34}_5\cdot\hat r$&$(-1/\sqrt{3},-\sqrt{2/3})$&$(1/\sqrt{3},-1/\sqrt{3},-1/\sqrt{3})$&$(-1/\sqrt{3},-\sqrt{2/3})$\\
$\vec \vec O^{35}_6\cdot(\vec O^{35}_5\times\hat r)$&$(2i/\sqrt{3},-i\sqrt{2/3})$&$(i/\sqrt{3},2i/\sqrt{3},-i/\sqrt{3})$&$(2i/\sqrt{3},-i\sqrt{2/3})$\\
$\vec \vec O^{35}_6\cdot(\vec O^{35}_5\times\hat r)$&$(0,\sqrt{2}/3,\sqrt{3}/10)$&$(-\sqrt{5}/3,\sqrt{2}/3,\sqrt{5}/3,0)$&$(0,\sqrt{2}/3,\sqrt{3}/10)$\\
\hline		\hline	
 	\end{longtable}
 	\end{center}
 \end{appendices}
 
\balance
\bibliographystyle{apsrev4-1}
\bibliography{pc}

\begin{thebibliography}{109}%
\makeatletter
\providecommand \@ifxundefined [1]{%
 \@ifx{#1\undefined}
}%
\providecommand \@ifnum [1]{%
 \ifnum #1\expandafter \@firstoftwo
 \else \expandafter \@secondoftwo
 \fi
}%
\providecommand \@ifx [1]{%
 \ifx #1\expandafter \@firstoftwo
 \else \expandafter \@secondoftwo
 \fi
}%
\providecommand \natexlab [1]{#1}%
\providecommand \enquote  [1]{``#1''}%
\providecommand \bibnamefont  [1]{#1}%
\providecommand \bibfnamefont [1]{#1}%
\providecommand \citenamefont [1]{#1}%
\providecommand \href@noop [0]{\@secondoftwo}%
\providecommand \href [0]{\begingroup \@sanitize@url \@href}%
\providecommand \@href[1]{\@@startlink{#1}\@@href}%
\providecommand \@@href[1]{\endgroup#1\@@endlink}%
\providecommand \@sanitize@url [0]{\catcode `\\12\catcode `\$12\catcode
  `\&12\catcode `\#12\catcode `\^12\catcode `\_12\catcode `\%12\relax}%
\providecommand \@@startlink[1]{}%
\providecommand \@@endlink[0]{}%
\providecommand \url  [0]{\begingroup\@sanitize@url \@url }%
\providecommand \@url [1]{\endgroup\@href {#1}{\urlprefix }}%
\providecommand \urlprefix  [0]{URL }%
\providecommand \Eprint [0]{\href }%
\providecommand \doibase [0]{http://dx.doi.org/}%
\providecommand \selectlanguage [0]{\@gobble}%
\providecommand \bibinfo  [0]{\@secondoftwo}%
\providecommand \bibfield  [0]{\@secondoftwo}%
\providecommand \translation [1]{[#1]}%
\providecommand \BibitemOpen [0]{}%
\providecommand \bibitemStop [0]{}%
\providecommand \bibitemNoStop [0]{.\EOS\space}%
\providecommand \EOS [0]{\spacefactor3000\relax}%
\providecommand \BibitemShut  [1]{\csname bibitem#1\endcsname}%
\let\auto@bib@innerbib\@empty
\bibitem [{\citenamefont {Dong}\ \emph
  {et~al.}(2021{\natexlab{a}})\citenamefont {Dong}, \citenamefont {Guo},\ and\
  \citenamefont {Zou}}]{Dong:2020hxe}%
  \BibitemOpen
  \bibfield  {author} {\bibinfo {author} {\bibfnamefont {X.-K.}\ \bibnamefont
  {Dong}}, \bibinfo {author} {\bibfnamefont {F.-K.}\ \bibnamefont {Guo}}, \
  and\ \bibinfo {author} {\bibfnamefont {B.-S.}\ \bibnamefont {Zou}},\ }\href
  {\doibase 10.1103/PhysRevLett.126.152001} {\bibfield  {journal} {\bibinfo
  {journal} {Phys. Rev. Lett.}\ }\textbf {\bibinfo {volume} {126}},\ \bibinfo
  {pages} {152001} (\bibinfo {year} {2021}{\natexlab{a}})},\ \Eprint
  {http://arxiv.org/abs/2011.14517} {arXiv:2011.14517 [hep-ph]} \BibitemShut
  {NoStop}%
\bibitem [{\citenamefont {Chen}\ \emph {et~al.}(2016)\citenamefont {Chen},
  \citenamefont {Chen}, \citenamefont {Liu},\ and\ \citenamefont
  {Zhu}}]{Chen:2016qju}%
  \BibitemOpen
  \bibfield  {author} {\bibinfo {author} {\bibfnamefont {H.-X.}\ \bibnamefont
  {Chen}}, \bibinfo {author} {\bibfnamefont {W.}~\bibnamefont {Chen}}, \bibinfo
  {author} {\bibfnamefont {X.}~\bibnamefont {Liu}}, \ and\ \bibinfo {author}
  {\bibfnamefont {S.-L.}\ \bibnamefont {Zhu}},\ }\href {\doibase
  10.1016/j.physrep.2016.05.004} {\bibfield  {journal} {\bibinfo  {journal}
  {Phys. Rept.}\ }\textbf {\bibinfo {volume} {639}},\ \bibinfo {pages} {1}
  (\bibinfo {year} {2016})},\ \Eprint {http://arxiv.org/abs/1601.02092}
  {arXiv:1601.02092 [hep-ph]} \BibitemShut {NoStop}%
\bibitem [{\citenamefont {Guo}\ \emph {et~al.}(2018)\citenamefont {Guo},
  \citenamefont {Hanhart}, \citenamefont {Mei\ss{}ner}, \citenamefont {Wang},
  \citenamefont {Zhao},\ and\ \citenamefont {Zou}}]{Guo:2017jvc}%
  \BibitemOpen
  \bibfield  {author} {\bibinfo {author} {\bibfnamefont {F.-K.}\ \bibnamefont
  {Guo}}, \bibinfo {author} {\bibfnamefont {C.}~\bibnamefont {Hanhart}},
  \bibinfo {author} {\bibfnamefont {U.-G.}\ \bibnamefont {Mei\ss{}ner}},
  \bibinfo {author} {\bibfnamefont {Q.}~\bibnamefont {Wang}}, \bibinfo {author}
  {\bibfnamefont {Q.}~\bibnamefont {Zhao}}, \ and\ \bibinfo {author}
  {\bibfnamefont {B.-S.}\ \bibnamefont {Zou}},\ }\href {\doibase
  10.1103/RevModPhys.90.015004} {\bibfield  {journal} {\bibinfo  {journal}
  {Rev. Mod. Phys.}\ }\textbf {\bibinfo {volume} {90}},\ \bibinfo {pages}
  {015004} (\bibinfo {year} {2018})},\ \Eprint
  {http://arxiv.org/abs/1705.00141} {arXiv:1705.00141 [hep-ph]} \BibitemShut
  {NoStop}%
\bibitem [{\citenamefont {Brambilla}\ \emph {et~al.}(2020)\citenamefont
  {Brambilla}, \citenamefont {Eidelman}, \citenamefont {Hanhart}, \citenamefont
  {Nefediev}, \citenamefont {Shen}, \citenamefont {Thomas}, \citenamefont
  {Vairo},\ and\ \citenamefont {Yuan}}]{Brambilla:2019esw}%
  \BibitemOpen
  \bibfield  {author} {\bibinfo {author} {\bibfnamefont {N.}~\bibnamefont
  {Brambilla}}, \bibinfo {author} {\bibfnamefont {S.}~\bibnamefont {Eidelman}},
  \bibinfo {author} {\bibfnamefont {C.}~\bibnamefont {Hanhart}}, \bibinfo
  {author} {\bibfnamefont {A.}~\bibnamefont {Nefediev}}, \bibinfo {author}
  {\bibfnamefont {C.-P.}\ \bibnamefont {Shen}}, \bibinfo {author}
  {\bibfnamefont {C.~E.}\ \bibnamefont {Thomas}}, \bibinfo {author}
  {\bibfnamefont {A.}~\bibnamefont {Vairo}}, \ and\ \bibinfo {author}
  {\bibfnamefont {C.-Z.}\ \bibnamefont {Yuan}},\ }\href {\doibase
  10.1016/j.physrep.2020.05.001} {\bibfield  {journal} {\bibinfo  {journal}
  {Phys. Rept.}\ }\textbf {\bibinfo {volume} {873}},\ \bibinfo {pages} {1}
  (\bibinfo {year} {2020})},\ \Eprint {http://arxiv.org/abs/1907.07583}
  {arXiv:1907.07583 [hep-ex]} \BibitemShut {NoStop}%
\bibitem [{\citenamefont {Yamaguchi}\ \emph
  {et~al.}(2020{\natexlab{a}})\citenamefont {Yamaguchi}, \citenamefont
  {Hosaka}, \citenamefont {Takeuchi},\ and\ \citenamefont
  {Takizawa}}]{Yamaguchi:2019vea}%
  \BibitemOpen
  \bibfield  {author} {\bibinfo {author} {\bibfnamefont {Y.}~\bibnamefont
  {Yamaguchi}}, \bibinfo {author} {\bibfnamefont {A.}~\bibnamefont {Hosaka}},
  \bibinfo {author} {\bibfnamefont {S.}~\bibnamefont {Takeuchi}}, \ and\
  \bibinfo {author} {\bibfnamefont {M.}~\bibnamefont {Takizawa}},\ }\href
  {\doibase 10.1088/1361-6471/ab72b0} {\bibfield  {journal} {\bibinfo
  {journal} {J. Phys. G}\ }\textbf {\bibinfo {volume} {47}},\ \bibinfo {pages}
  {053001} (\bibinfo {year} {2020}{\natexlab{a}})},\ \Eprint
  {http://arxiv.org/abs/1908.08790} {arXiv:1908.08790 [hep-ph]} \BibitemShut
  {NoStop}%
\bibitem [{\citenamefont {Dong}\ \emph
  {et~al.}(2021{\natexlab{b}})\citenamefont {Dong}, \citenamefont {Guo},\ and\
  \citenamefont {Zou}}]{Dong:2021juy}%
  \BibitemOpen
  \bibfield  {author} {\bibinfo {author} {\bibfnamefont {X.-K.}\ \bibnamefont
  {Dong}}, \bibinfo {author} {\bibfnamefont {F.-K.}\ \bibnamefont {Guo}}, \
  and\ \bibinfo {author} {\bibfnamefont {B.-S.}\ \bibnamefont {Zou}},\ }\href
  {\doibase 10.13725/j.cnki.pip.2021.02.001} {\bibfield  {journal} {\bibinfo
  {journal} {Progr. Phys.}\ }\textbf {\bibinfo {volume} {41}},\ \bibinfo
  {pages} {65} (\bibinfo {year} {2021}{\natexlab{b}})},\ \Eprint
  {http://arxiv.org/abs/2101.01021} {arXiv:2101.01021 [hep-ph]} \BibitemShut
  {NoStop}%
\bibitem [{\citenamefont {Tornqvist}(1994)}]{Tornqvist:1993ng}%
  \BibitemOpen
  \bibfield  {author} {\bibinfo {author} {\bibfnamefont {N.~A.}\ \bibnamefont
  {Tornqvist}},\ }\href {\doibase 10.1007/BF01413192} {\bibfield  {journal}
  {\bibinfo  {journal} {Z. Phys. C}\ }\textbf {\bibinfo {volume} {61}},\
  \bibinfo {pages} {525} (\bibinfo {year} {1994})},\ \Eprint
  {http://arxiv.org/abs/hep-ph/9310247} {arXiv:hep-ph/9310247} \BibitemShut
  {NoStop}%
\bibitem [{\citenamefont {Wu}\ \emph {et~al.}(2010)\citenamefont {Wu},
  \citenamefont {Molina}, \citenamefont {Oset},\ and\ \citenamefont
  {Zou}}]{Wu:2010jy}%
  \BibitemOpen
  \bibfield  {author} {\bibinfo {author} {\bibfnamefont {J.-J.}\ \bibnamefont
  {Wu}}, \bibinfo {author} {\bibfnamefont {R.}~\bibnamefont {Molina}}, \bibinfo
  {author} {\bibfnamefont {E.}~\bibnamefont {Oset}}, \ and\ \bibinfo {author}
  {\bibfnamefont {B.~S.}\ \bibnamefont {Zou}},\ }\href {\doibase
  10.1103/PhysRevLett.105.232001} {\bibfield  {journal} {\bibinfo  {journal}
  {Phys. Rev. Lett.}\ }\textbf {\bibinfo {volume} {105}},\ \bibinfo {pages}
  {232001} (\bibinfo {year} {2010})},\ \Eprint {http://arxiv.org/abs/1007.0573}
  {arXiv:1007.0573 [nucl-th]} \BibitemShut {NoStop}%
\bibitem [{\citenamefont {Wu}\ \emph {et~al.}(2011)\citenamefont {Wu},
  \citenamefont {Molina}, \citenamefont {Oset},\ and\ \citenamefont
  {Zou}}]{Wu:2010vk}%
  \BibitemOpen
  \bibfield  {author} {\bibinfo {author} {\bibfnamefont {J.-J.}\ \bibnamefont
  {Wu}}, \bibinfo {author} {\bibfnamefont {R.}~\bibnamefont {Molina}}, \bibinfo
  {author} {\bibfnamefont {E.}~\bibnamefont {Oset}}, \ and\ \bibinfo {author}
  {\bibfnamefont {B.~S.}\ \bibnamefont {Zou}},\ }\href {\doibase
  10.1103/PhysRevC.84.015202} {\bibfield  {journal} {\bibinfo  {journal} {Phys.
  Rev. C}\ }\textbf {\bibinfo {volume} {84}},\ \bibinfo {pages} {015202}
  (\bibinfo {year} {2011})},\ \Eprint {http://arxiv.org/abs/1011.2399}
  {arXiv:1011.2399 [nucl-th]} \BibitemShut {NoStop}%
\bibitem [{\citenamefont {Wang}\ \emph {et~al.}(2011)\citenamefont {Wang},
  \citenamefont {Huang}, \citenamefont {Zhang},\ and\ \citenamefont
  {Zou}}]{Wang:2011rga}%
  \BibitemOpen
  \bibfield  {author} {\bibinfo {author} {\bibfnamefont {W.~L.}\ \bibnamefont
  {Wang}}, \bibinfo {author} {\bibfnamefont {F.}~\bibnamefont {Huang}},
  \bibinfo {author} {\bibfnamefont {Z.~Y.}\ \bibnamefont {Zhang}}, \ and\
  \bibinfo {author} {\bibfnamefont {B.~S.}\ \bibnamefont {Zou}},\ }\href
  {\doibase 10.1103/PhysRevC.84.015203} {\bibfield  {journal} {\bibinfo
  {journal} {Phys. Rev. C}\ }\textbf {\bibinfo {volume} {84}},\ \bibinfo
  {pages} {015203} (\bibinfo {year} {2011})},\ \Eprint
  {http://arxiv.org/abs/1101.0453} {arXiv:1101.0453 [nucl-th]} \BibitemShut
  {NoStop}%
\bibitem [{\citenamefont {Yang}\ \emph {et~al.}(2012)\citenamefont {Yang},
  \citenamefont {Sun}, \citenamefont {He}, \citenamefont {Liu},\ and\
  \citenamefont {Zhu}}]{Yang:2011wz}%
  \BibitemOpen
  \bibfield  {author} {\bibinfo {author} {\bibfnamefont {Z.-C.}\ \bibnamefont
  {Yang}}, \bibinfo {author} {\bibfnamefont {Z.-F.}\ \bibnamefont {Sun}},
  \bibinfo {author} {\bibfnamefont {J.}~\bibnamefont {He}}, \bibinfo {author}
  {\bibfnamefont {X.}~\bibnamefont {Liu}}, \ and\ \bibinfo {author}
  {\bibfnamefont {S.-L.}\ \bibnamefont {Zhu}},\ }\href {\doibase
  10.1088/1674-1137/36/1/002} {\bibfield  {journal} {\bibinfo  {journal} {Chin.
  Phys. C}\ }\textbf {\bibinfo {volume} {36}},\ \bibinfo {pages} {6} (\bibinfo
  {year} {2012})},\ \Eprint {http://arxiv.org/abs/1105.2901} {arXiv:1105.2901
  [hep-ph]} \BibitemShut {NoStop}%
\bibitem [{\citenamefont {Wu}\ \emph {et~al.}(2012)\citenamefont {Wu},
  \citenamefont {Lee},\ and\ \citenamefont {Zou}}]{Wu:2012md}%
  \BibitemOpen
  \bibfield  {author} {\bibinfo {author} {\bibfnamefont {J.-J.}\ \bibnamefont
  {Wu}}, \bibinfo {author} {\bibfnamefont {T.~S.~H.}\ \bibnamefont {Lee}}, \
  and\ \bibinfo {author} {\bibfnamefont {B.~S.}\ \bibnamefont {Zou}},\ }\href
  {\doibase 10.1103/PhysRevC.85.044002} {\bibfield  {journal} {\bibinfo
  {journal} {Phys. Rev. C}\ }\textbf {\bibinfo {volume} {85}},\ \bibinfo
  {pages} {044002} (\bibinfo {year} {2012})},\ \Eprint
  {http://arxiv.org/abs/1202.1036} {arXiv:1202.1036 [nucl-th]} \BibitemShut
  {NoStop}%
\bibitem [{\citenamefont {Xiao}\ \emph {et~al.}(2013)\citenamefont {Xiao},
  \citenamefont {Nieves},\ and\ \citenamefont {Oset}}]{Xiao:2013yca}%
  \BibitemOpen
  \bibfield  {author} {\bibinfo {author} {\bibfnamefont {C.~W.}\ \bibnamefont
  {Xiao}}, \bibinfo {author} {\bibfnamefont {J.}~\bibnamefont {Nieves}}, \ and\
  \bibinfo {author} {\bibfnamefont {E.}~\bibnamefont {Oset}},\ }\href {\doibase
  10.1103/PhysRevD.88.056012} {\bibfield  {journal} {\bibinfo  {journal} {Phys.
  Rev. D}\ }\textbf {\bibinfo {volume} {88}},\ \bibinfo {pages} {056012}
  (\bibinfo {year} {2013})},\ \Eprint {http://arxiv.org/abs/1304.5368}
  {arXiv:1304.5368 [hep-ph]} \BibitemShut {NoStop}%
\bibitem [{\citenamefont {Uchino}\ \emph {et~al.}(2016)\citenamefont {Uchino},
  \citenamefont {Liang},\ and\ \citenamefont {Oset}}]{Uchino:2015uha}%
  \BibitemOpen
  \bibfield  {author} {\bibinfo {author} {\bibfnamefont {T.}~\bibnamefont
  {Uchino}}, \bibinfo {author} {\bibfnamefont {W.-H.}\ \bibnamefont {Liang}}, \
  and\ \bibinfo {author} {\bibfnamefont {E.}~\bibnamefont {Oset}},\ }\href
  {\doibase 10.1140/epja/i2016-16043-0} {\bibfield  {journal} {\bibinfo
  {journal} {Eur. Phys. J. A}\ }\textbf {\bibinfo {volume} {52}},\ \bibinfo
  {pages} {43} (\bibinfo {year} {2016})},\ \Eprint
  {http://arxiv.org/abs/1504.05726} {arXiv:1504.05726 [hep-ph]} \BibitemShut
  {NoStop}%
\bibitem [{\citenamefont {Karliner}\ and\ \citenamefont
  {Rosner}(2015)}]{Karliner:2015ina}%
  \BibitemOpen
  \bibfield  {author} {\bibinfo {author} {\bibfnamefont {M.}~\bibnamefont
  {Karliner}}\ and\ \bibinfo {author} {\bibfnamefont {J.~L.}\ \bibnamefont
  {Rosner}},\ }\href {\doibase 10.1103/PhysRevLett.115.122001} {\bibfield
  {journal} {\bibinfo  {journal} {Phys. Rev. Lett.}\ }\textbf {\bibinfo
  {volume} {115}},\ \bibinfo {pages} {122001} (\bibinfo {year} {2015})},\
  \Eprint {http://arxiv.org/abs/1506.06386} {arXiv:1506.06386 [hep-ph]}
  \BibitemShut {NoStop}%
\bibitem [{\citenamefont {Aaij}\ \emph {et~al.}(2015)\citenamefont {Aaij} \emph
  {et~al.}}]{Aaij:2015tga}%
  \BibitemOpen
  \bibfield  {author} {\bibinfo {author} {\bibfnamefont {R.}~\bibnamefont
  {Aaij}} \emph {et~al.} (\bibinfo {collaboration} {LHCb}),\ }\href {\doibase
  10.1103/PhysRevLett.115.072001} {\bibfield  {journal} {\bibinfo  {journal}
  {Phys. Rev. Lett.}\ }\textbf {\bibinfo {volume} {115}},\ \bibinfo {pages}
  {072001} (\bibinfo {year} {2015})},\ \Eprint
  {http://arxiv.org/abs/1507.03414} {arXiv:1507.03414 [hep-ex]} \BibitemShut
  {NoStop}%
\bibitem [{\citenamefont {Aaij}\ \emph {et~al.}(2016)\citenamefont {Aaij} \emph
  {et~al.}}]{Aaij:2016phn}%
  \BibitemOpen
  \bibfield  {author} {\bibinfo {author} {\bibfnamefont {R.}~\bibnamefont
  {Aaij}} \emph {et~al.} (\bibinfo {collaboration} {LHCb}),\ }\href {\doibase
  10.1103/PhysRevLett.117.082002} {\bibfield  {journal} {\bibinfo  {journal}
  {Phys. Rev. Lett.}\ }\textbf {\bibinfo {volume} {117}},\ \bibinfo {pages}
  {082002} (\bibinfo {year} {2016})},\ \Eprint
  {http://arxiv.org/abs/1604.05708} {arXiv:1604.05708 [hep-ex]} \BibitemShut
  {NoStop}%
\bibitem [{\citenamefont {Yuan}\ \emph {et~al.}(2012)\citenamefont {Yuan},
  \citenamefont {Wei}, \citenamefont {He}, \citenamefont {Xu},\ and\
  \citenamefont {Zou}}]{Yuan:2012wz}%
  \BibitemOpen
  \bibfield  {author} {\bibinfo {author} {\bibfnamefont {S.~G.}\ \bibnamefont
  {Yuan}}, \bibinfo {author} {\bibfnamefont {K.~W.}\ \bibnamefont {Wei}},
  \bibinfo {author} {\bibfnamefont {J.}~\bibnamefont {He}}, \bibinfo {author}
  {\bibfnamefont {H.~S.}\ \bibnamefont {Xu}}, \ and\ \bibinfo {author}
  {\bibfnamefont {B.~S.}\ \bibnamefont {Zou}},\ }\href {\doibase
  10.1140/epja/i2012-12061-2} {\bibfield  {journal} {\bibinfo  {journal} {Eur.
  Phys. J. A}\ }\textbf {\bibinfo {volume} {48}},\ \bibinfo {pages} {61}
  (\bibinfo {year} {2012})},\ \Eprint {http://arxiv.org/abs/1201.0807}
  {arXiv:1201.0807 [nucl-th]} \BibitemShut {NoStop}%
\bibitem [{\citenamefont {Chen}\ \emph
  {et~al.}(2015{\natexlab{a}})\citenamefont {Chen}, \citenamefont {Chen},
  \citenamefont {Liu}, \citenamefont {Steele},\ and\ \citenamefont
  {Zhu}}]{Chen:2015moa}%
  \BibitemOpen
  \bibfield  {author} {\bibinfo {author} {\bibfnamefont {H.-X.}\ \bibnamefont
  {Chen}}, \bibinfo {author} {\bibfnamefont {W.}~\bibnamefont {Chen}}, \bibinfo
  {author} {\bibfnamefont {X.}~\bibnamefont {Liu}}, \bibinfo {author}
  {\bibfnamefont {T.~G.}\ \bibnamefont {Steele}}, \ and\ \bibinfo {author}
  {\bibfnamefont {S.-L.}\ \bibnamefont {Zhu}},\ }\href {\doibase
  10.1103/PhysRevLett.115.172001} {\bibfield  {journal} {\bibinfo  {journal}
  {Phys. Rev. Lett.}\ }\textbf {\bibinfo {volume} {115}},\ \bibinfo {pages}
  {172001} (\bibinfo {year} {2015}{\natexlab{a}})},\ \Eprint
  {http://arxiv.org/abs/1507.03717} {arXiv:1507.03717 [hep-ph]} \BibitemShut
  {NoStop}%
\bibitem [{\citenamefont {He}(2016)}]{He:2015cea}%
  \BibitemOpen
  \bibfield  {author} {\bibinfo {author} {\bibfnamefont {J.}~\bibnamefont
  {He}},\ }\href {\doibase 10.1016/j.physletb.2015.12.071} {\bibfield
  {journal} {\bibinfo  {journal} {Phys. Lett. B}\ }\textbf {\bibinfo {volume}
  {753}},\ \bibinfo {pages} {547} (\bibinfo {year} {2016})},\ \Eprint
  {http://arxiv.org/abs/1507.05200} {arXiv:1507.05200 [hep-ph]} \BibitemShut
  {NoStop}%
\bibitem [{\citenamefont {Roca}\ \emph {et~al.}(2015)\citenamefont {Roca},
  \citenamefont {Nieves},\ and\ \citenamefont {Oset}}]{Roca:2015dva}%
  \BibitemOpen
  \bibfield  {author} {\bibinfo {author} {\bibfnamefont {L.}~\bibnamefont
  {Roca}}, \bibinfo {author} {\bibfnamefont {J.}~\bibnamefont {Nieves}}, \ and\
  \bibinfo {author} {\bibfnamefont {E.}~\bibnamefont {Oset}},\ }\href {\doibase
  10.1103/PhysRevD.92.094003} {\bibfield  {journal} {\bibinfo  {journal} {Phys.
  Rev. D}\ }\textbf {\bibinfo {volume} {92}},\ \bibinfo {pages} {094003}
  (\bibinfo {year} {2015})},\ \Eprint {http://arxiv.org/abs/1507.04249}
  {arXiv:1507.04249 [hep-ph]} \BibitemShut {NoStop}%
\bibitem [{\citenamefont {Chen}\ \emph
  {et~al.}(2015{\natexlab{b}})\citenamefont {Chen}, \citenamefont {Liu},
  \citenamefont {Li},\ and\ \citenamefont {Zhu}}]{Chen:2015loa}%
  \BibitemOpen
  \bibfield  {author} {\bibinfo {author} {\bibfnamefont {R.}~\bibnamefont
  {Chen}}, \bibinfo {author} {\bibfnamefont {X.}~\bibnamefont {Liu}}, \bibinfo
  {author} {\bibfnamefont {X.-Q.}\ \bibnamefont {Li}}, \ and\ \bibinfo {author}
  {\bibfnamefont {S.-L.}\ \bibnamefont {Zhu}},\ }\href {\doibase
  10.1103/PhysRevLett.115.132002} {\bibfield  {journal} {\bibinfo  {journal}
  {Phys. Rev. Lett.}\ }\textbf {\bibinfo {volume} {115}},\ \bibinfo {pages}
  {132002} (\bibinfo {year} {2015}{\natexlab{b}})},\ \Eprint
  {http://arxiv.org/abs/1507.03704} {arXiv:1507.03704 [hep-ph]} \BibitemShut
  {NoStop}%
\bibitem [{\citenamefont {Xiao}\ and\ \citenamefont
  {Mei\ss{}ner}(2015)}]{Xiao:2015fia}%
  \BibitemOpen
  \bibfield  {author} {\bibinfo {author} {\bibfnamefont {C.~W.}\ \bibnamefont
  {Xiao}}\ and\ \bibinfo {author} {\bibfnamefont {U.~G.}\ \bibnamefont
  {Mei\ss{}ner}},\ }\href {\doibase 10.1103/PhysRevD.92.114002} {\bibfield
  {journal} {\bibinfo  {journal} {Phys. Rev. D}\ }\textbf {\bibinfo {volume}
  {92}},\ \bibinfo {pages} {114002} (\bibinfo {year} {2015})},\ \Eprint
  {http://arxiv.org/abs/1508.00924} {arXiv:1508.00924 [hep-ph]} \BibitemShut
  {NoStop}%
\bibitem [{\citenamefont {Burns}(2015)}]{Burns:2015dwa}%
  \BibitemOpen
  \bibfield  {author} {\bibinfo {author} {\bibfnamefont {T.~J.}\ \bibnamefont
  {Burns}},\ }\href {\doibase 10.1140/epja/i2015-15152-6} {\bibfield  {journal}
  {\bibinfo  {journal} {Eur. Phys. J. A}\ }\textbf {\bibinfo {volume} {51}},\
  \bibinfo {pages} {152} (\bibinfo {year} {2015})},\ \Eprint
  {http://arxiv.org/abs/1509.02460} {arXiv:1509.02460 [hep-ph]} \BibitemShut
  {NoStop}%
\bibitem [{\citenamefont {Mironov}\ and\ \citenamefont
  {Morozov}(2015)}]{Mironov:2015ica}%
  \BibitemOpen
  \bibfield  {author} {\bibinfo {author} {\bibfnamefont {A.}~\bibnamefont
  {Mironov}}\ and\ \bibinfo {author} {\bibfnamefont {A.}~\bibnamefont
  {Morozov}},\ }\href {\doibase 10.7868/S0370274X15170038} {\bibfield
  {journal} {\bibinfo  {journal} {JETP Lett.}\ }\textbf {\bibinfo {volume}
  {102}},\ \bibinfo {pages} {271} (\bibinfo {year} {2015})},\ \Eprint
  {http://arxiv.org/abs/1507.04694} {arXiv:1507.04694 [hep-ph]} \BibitemShut
  {NoStop}%
\bibitem [{\citenamefont {Mei\ss{}ner}\ and\ \citenamefont
  {Oller}(2015)}]{Meissner:2015mza}%
  \BibitemOpen
  \bibfield  {author} {\bibinfo {author} {\bibfnamefont {U.-G.}\ \bibnamefont
  {Mei\ss{}ner}}\ and\ \bibinfo {author} {\bibfnamefont {J.~A.}\ \bibnamefont
  {Oller}},\ }\href {\doibase 10.1016/j.physletb.2015.10.015} {\bibfield
  {journal} {\bibinfo  {journal} {Phys. Lett. B}\ }\textbf {\bibinfo {volume}
  {751}},\ \bibinfo {pages} {59} (\bibinfo {year} {2015})},\ \Eprint
  {http://arxiv.org/abs/1507.07478} {arXiv:1507.07478 [hep-ph]} \BibitemShut
  {NoStop}%
\bibitem [{\citenamefont {L\"u}\ and\ \citenamefont {Dong}(2016)}]{Lu:2016nnt}%
  \BibitemOpen
  \bibfield  {author} {\bibinfo {author} {\bibfnamefont {Q.-F.}\ \bibnamefont
  {L\"u}}\ and\ \bibinfo {author} {\bibfnamefont {Y.-B.}\ \bibnamefont
  {Dong}},\ }\href {\doibase 10.1103/PhysRevD.93.074020} {\bibfield  {journal}
  {\bibinfo  {journal} {Phys. Rev. D}\ }\textbf {\bibinfo {volume} {93}},\
  \bibinfo {pages} {074020} (\bibinfo {year} {2016})},\ \Eprint
  {http://arxiv.org/abs/1603.00559} {arXiv:1603.00559 [hep-ph]} \BibitemShut
  {NoStop}%
\bibitem [{\citenamefont {Shen}\ \emph {et~al.}(2016)\citenamefont {Shen},
  \citenamefont {Guo}, \citenamefont {Xie},\ and\ \citenamefont
  {Zou}}]{Shen:2016tzq}%
  \BibitemOpen
  \bibfield  {author} {\bibinfo {author} {\bibfnamefont {C.-W.}\ \bibnamefont
  {Shen}}, \bibinfo {author} {\bibfnamefont {F.-K.}\ \bibnamefont {Guo}},
  \bibinfo {author} {\bibfnamefont {J.-J.}\ \bibnamefont {Xie}}, \ and\
  \bibinfo {author} {\bibfnamefont {B.-S.}\ \bibnamefont {Zou}},\ }\href
  {\doibase 10.1016/j.nuclphysa.2016.04.034} {\bibfield  {journal} {\bibinfo
  {journal} {Nucl. Phys. A}\ }\textbf {\bibinfo {volume} {954}},\ \bibinfo
  {pages} {393} (\bibinfo {year} {2016})},\ \Eprint
  {http://arxiv.org/abs/1603.04672} {arXiv:1603.04672 [hep-ph]} \BibitemShut
  {NoStop}%
\bibitem [{\citenamefont {Kang}\ \emph {et~al.}(2016)\citenamefont {Kang},
  \citenamefont {Guo},\ and\ \citenamefont {Oller}}]{Kang:2016ezb}%
  \BibitemOpen
  \bibfield  {author} {\bibinfo {author} {\bibfnamefont {X.-W.}\ \bibnamefont
  {Kang}}, \bibinfo {author} {\bibfnamefont {Z.-H.}\ \bibnamefont {Guo}}, \
  and\ \bibinfo {author} {\bibfnamefont {J.~A.}\ \bibnamefont {Oller}},\ }\href
  {\doibase 10.1103/PhysRevD.94.014012} {\bibfield  {journal} {\bibinfo
  {journal} {Phys. Rev. D}\ }\textbf {\bibinfo {volume} {94}},\ \bibinfo
  {pages} {014012} (\bibinfo {year} {2016})},\ \Eprint
  {http://arxiv.org/abs/1603.05546} {arXiv:1603.05546 [hep-ph]} \BibitemShut
  {NoStop}%
\bibitem [{\citenamefont {Shimizu}\ \emph {et~al.}(2016)\citenamefont
  {Shimizu}, \citenamefont {Suenaga},\ and\ \citenamefont
  {Harada}}]{Shimizu:2016rrd}%
  \BibitemOpen
  \bibfield  {author} {\bibinfo {author} {\bibfnamefont {Y.}~\bibnamefont
  {Shimizu}}, \bibinfo {author} {\bibfnamefont {D.}~\bibnamefont {Suenaga}}, \
  and\ \bibinfo {author} {\bibfnamefont {M.}~\bibnamefont {Harada}},\ }\href
  {\doibase 10.1103/PhysRevD.93.114003} {\bibfield  {journal} {\bibinfo
  {journal} {Phys. Rev. D}\ }\textbf {\bibinfo {volume} {93}},\ \bibinfo
  {pages} {114003} (\bibinfo {year} {2016})},\ \Eprint
  {http://arxiv.org/abs/1603.02376} {arXiv:1603.02376 [hep-ph]} \BibitemShut
  {NoStop}%
\bibitem [{\citenamefont {Yamaguchi}\ and\ \citenamefont
  {Santopinto}(2017)}]{Yamaguchi:2016ote}%
  \BibitemOpen
  \bibfield  {author} {\bibinfo {author} {\bibfnamefont {Y.}~\bibnamefont
  {Yamaguchi}}\ and\ \bibinfo {author} {\bibfnamefont {E.}~\bibnamefont
  {Santopinto}},\ }\href {\doibase 10.1103/PhysRevD.96.014018} {\bibfield
  {journal} {\bibinfo  {journal} {Phys. Rev. D}\ }\textbf {\bibinfo {volume}
  {96}},\ \bibinfo {pages} {014018} (\bibinfo {year} {2017})},\ \Eprint
  {http://arxiv.org/abs/1606.08330} {arXiv:1606.08330 [hep-ph]} \BibitemShut
  {NoStop}%
\bibitem [{\citenamefont {Lin}\ \emph {et~al.}(2017)\citenamefont {Lin},
  \citenamefont {Shen}, \citenamefont {Guo},\ and\ \citenamefont
  {Zou}}]{Lin:2017mtz}%
  \BibitemOpen
  \bibfield  {author} {\bibinfo {author} {\bibfnamefont {Y.-H.}\ \bibnamefont
  {Lin}}, \bibinfo {author} {\bibfnamefont {C.-W.}\ \bibnamefont {Shen}},
  \bibinfo {author} {\bibfnamefont {F.-K.}\ \bibnamefont {Guo}}, \ and\
  \bibinfo {author} {\bibfnamefont {B.-S.}\ \bibnamefont {Zou}},\ }\href
  {\doibase 10.1103/PhysRevD.95.114017} {\bibfield  {journal} {\bibinfo
  {journal} {Phys. Rev. D}\ }\textbf {\bibinfo {volume} {95}},\ \bibinfo
  {pages} {114017} (\bibinfo {year} {2017})},\ \Eprint
  {http://arxiv.org/abs/1703.01045} {arXiv:1703.01045 [hep-ph]} \BibitemShut
  {NoStop}%
\bibitem [{\citenamefont {Shimizu}\ and\ \citenamefont
  {Harada}(2017)}]{Shimizu:2017xrg}%
  \BibitemOpen
  \bibfield  {author} {\bibinfo {author} {\bibfnamefont {Y.}~\bibnamefont
  {Shimizu}}\ and\ \bibinfo {author} {\bibfnamefont {M.}~\bibnamefont
  {Harada}},\ }\href {\doibase 10.1103/PhysRevD.96.094012} {\bibfield
  {journal} {\bibinfo  {journal} {Phys. Rev. D}\ }\textbf {\bibinfo {volume}
  {96}},\ \bibinfo {pages} {094012} (\bibinfo {year} {2017})},\ \Eprint
  {http://arxiv.org/abs/1708.04743} {arXiv:1708.04743 [hep-ph]} \BibitemShut
  {NoStop}%
\bibitem [{\citenamefont {Voloshin}(2019)}]{Voloshin:2019aut}%
  \BibitemOpen
  \bibfield  {author} {\bibinfo {author} {\bibfnamefont {M.~B.}\ \bibnamefont
  {Voloshin}},\ }\href {\doibase 10.1103/PhysRevD.100.034020} {\bibfield
  {journal} {\bibinfo  {journal} {Phys. Rev. D}\ }\textbf {\bibinfo {volume}
  {100}},\ \bibinfo {pages} {034020} (\bibinfo {year} {2019})},\ \Eprint
  {http://arxiv.org/abs/1907.01476} {arXiv:1907.01476 [hep-ph]} \BibitemShut
  {NoStop}%
\bibitem [{\citenamefont {Gutsche}\ and\ \citenamefont
  {Lyubovitskij}(2019)}]{Gutsche:2019mkg}%
  \BibitemOpen
  \bibfield  {author} {\bibinfo {author} {\bibfnamefont {T.}~\bibnamefont
  {Gutsche}}\ and\ \bibinfo {author} {\bibfnamefont {V.~E.}\ \bibnamefont
  {Lyubovitskij}},\ }\href {\doibase 10.1103/PhysRevD.100.094031} {\bibfield
  {journal} {\bibinfo  {journal} {Phys. Rev. D}\ }\textbf {\bibinfo {volume}
  {100}},\ \bibinfo {pages} {094031} (\bibinfo {year} {2019})},\ \Eprint
  {http://arxiv.org/abs/1910.03984} {arXiv:1910.03984 [hep-ph]} \BibitemShut
  {NoStop}%
\bibitem [{\citenamefont {Ali}\ \emph {et~al.}(2016)\citenamefont {Ali},
  \citenamefont {Ahmed}, \citenamefont {Aslam},\ and\ \citenamefont
  {Rehman}}]{Ali:2016dkf}%
  \BibitemOpen
  \bibfield  {author} {\bibinfo {author} {\bibfnamefont {A.}~\bibnamefont
  {Ali}}, \bibinfo {author} {\bibfnamefont {I.}~\bibnamefont {Ahmed}}, \bibinfo
  {author} {\bibfnamefont {M.~J.}\ \bibnamefont {Aslam}}, \ and\ \bibinfo
  {author} {\bibfnamefont {A.}~\bibnamefont {Rehman}},\ }\href {\doibase
  10.1103/PhysRevD.94.054001} {\bibfield  {journal} {\bibinfo  {journal} {Phys.
  Rev. D}\ }\textbf {\bibinfo {volume} {94}},\ \bibinfo {pages} {054001}
  (\bibinfo {year} {2016})},\ \Eprint {http://arxiv.org/abs/1607.00987}
  {arXiv:1607.00987 [hep-ph]} \BibitemShut {NoStop}%
\bibitem [{\citenamefont {Ali}\ and\ \citenamefont
  {Parkhomenko}(2019)}]{Ali:2019npk}%
  \BibitemOpen
  \bibfield  {author} {\bibinfo {author} {\bibfnamefont {A.}~\bibnamefont
  {Ali}}\ and\ \bibinfo {author} {\bibfnamefont {A.~Y.}\ \bibnamefont
  {Parkhomenko}},\ }\href {\doibase 10.1016/j.physletb.2019.05.002} {\bibfield
  {journal} {\bibinfo  {journal} {Phys. Lett. B}\ }\textbf {\bibinfo {volume}
  {793}},\ \bibinfo {pages} {365} (\bibinfo {year} {2019})},\ \Eprint
  {http://arxiv.org/abs/1904.00446} {arXiv:1904.00446 [hep-ph]} \BibitemShut
  {NoStop}%
\bibitem [{\citenamefont {Maiani}\ \emph {et~al.}(2015)\citenamefont {Maiani},
  \citenamefont {Polosa},\ and\ \citenamefont {Riquer}}]{Maiani:2015vwa}%
  \BibitemOpen
  \bibfield  {author} {\bibinfo {author} {\bibfnamefont {L.}~\bibnamefont
  {Maiani}}, \bibinfo {author} {\bibfnamefont {A.~D.}\ \bibnamefont {Polosa}},
  \ and\ \bibinfo {author} {\bibfnamefont {V.}~\bibnamefont {Riquer}},\ }\href
  {\doibase 10.1016/j.physletb.2015.08.008} {\bibfield  {journal} {\bibinfo
  {journal} {Phys. Lett. B}\ }\textbf {\bibinfo {volume} {749}},\ \bibinfo
  {pages} {289} (\bibinfo {year} {2015})},\ \Eprint
  {http://arxiv.org/abs/1507.04980} {arXiv:1507.04980 [hep-ph]} \BibitemShut
  {NoStop}%
\bibitem [{\citenamefont {Li}\ \emph {et~al.}(2015)\citenamefont {Li},
  \citenamefont {He},\ and\ \citenamefont {He}}]{Li:2015gta}%
  \BibitemOpen
  \bibfield  {author} {\bibinfo {author} {\bibfnamefont {G.-N.}\ \bibnamefont
  {Li}}, \bibinfo {author} {\bibfnamefont {X.-G.}\ \bibnamefont {He}}, \ and\
  \bibinfo {author} {\bibfnamefont {M.}~\bibnamefont {He}},\ }\href {\doibase
  10.1007/JHEP12(2015)128} {\bibfield  {journal} {\bibinfo  {journal} {JHEP}\
  }\textbf {\bibinfo {volume} {12}},\ \bibinfo {pages} {128} (\bibinfo {year}
  {2015})},\ \Eprint {http://arxiv.org/abs/1507.08252} {arXiv:1507.08252
  [hep-ph]} \BibitemShut {NoStop}%
\bibitem [{\citenamefont {Anisovich}\ \emph {et~al.}(2015)\citenamefont
  {Anisovich}, \citenamefont {Matveev}, \citenamefont {Nyiri}, \citenamefont
  {Sarantsev},\ and\ \citenamefont {Semenova}}]{Anisovich:2015cia}%
  \BibitemOpen
  \bibfield  {author} {\bibinfo {author} {\bibfnamefont {V.~V.}\ \bibnamefont
  {Anisovich}}, \bibinfo {author} {\bibfnamefont {M.~A.}\ \bibnamefont
  {Matveev}}, \bibinfo {author} {\bibfnamefont {J.}~\bibnamefont {Nyiri}},
  \bibinfo {author} {\bibfnamefont {A.~V.}\ \bibnamefont {Sarantsev}}, \ and\
  \bibinfo {author} {\bibfnamefont {A.~N.}\ \bibnamefont {Semenova}},\
  }\href@noop {} {\  (\bibinfo {year} {2015})},\ \Eprint
  {http://arxiv.org/abs/1507.07652} {arXiv:1507.07652 [hep-ph]} \BibitemShut
  {NoStop}%
\bibitem [{\citenamefont {Ghosh}\ \emph {et~al.}(2017)\citenamefont {Ghosh},
  \citenamefont {Bhattacharya},\ and\ \citenamefont
  {Chakrabarti}}]{Ghosh:2017fwg}%
  \BibitemOpen
  \bibfield  {author} {\bibinfo {author} {\bibfnamefont {R.}~\bibnamefont
  {Ghosh}}, \bibinfo {author} {\bibfnamefont {A.}~\bibnamefont {Bhattacharya}},
  \ and\ \bibinfo {author} {\bibfnamefont {B.}~\bibnamefont {Chakrabarti}},\
  }\href {\doibase 10.1134/S1547477117040100} {\bibfield  {journal} {\bibinfo
  {journal} {Phys. Part. Nucl. Lett.}\ }\textbf {\bibinfo {volume} {14}},\
  \bibinfo {pages} {550} (\bibinfo {year} {2017})},\ \Eprint
  {http://arxiv.org/abs/1508.00356} {arXiv:1508.00356 [hep-ph]} \BibitemShut
  {NoStop}%
\bibitem [{\citenamefont {Wang}(2016)}]{Wang:2015epa}%
  \BibitemOpen
  \bibfield  {author} {\bibinfo {author} {\bibfnamefont {Z.-G.}\ \bibnamefont
  {Wang}},\ }\href {\doibase 10.1140/epjc/s10052-016-3920-4} {\bibfield
  {journal} {\bibinfo  {journal} {Eur. Phys. J. C}\ }\textbf {\bibinfo {volume}
  {76}},\ \bibinfo {pages} {70} (\bibinfo {year} {2016})},\ \Eprint
  {http://arxiv.org/abs/1508.01468} {arXiv:1508.01468 [hep-ph]} \BibitemShut
  {NoStop}%
\bibitem [{\citenamefont {Hiyama}\ \emph {et~al.}(2018)\citenamefont {Hiyama},
  \citenamefont {Hosaka}, \citenamefont {Oka},\ and\ \citenamefont
  {Richard}}]{Hiyama:2018ukv}%
  \BibitemOpen
  \bibfield  {author} {\bibinfo {author} {\bibfnamefont {E.}~\bibnamefont
  {Hiyama}}, \bibinfo {author} {\bibfnamefont {A.}~\bibnamefont {Hosaka}},
  \bibinfo {author} {\bibfnamefont {M.}~\bibnamefont {Oka}}, \ and\ \bibinfo
  {author} {\bibfnamefont {J.-M.}\ \bibnamefont {Richard}},\ }\href {\doibase
  10.1103/PhysRevC.98.045208} {\bibfield  {journal} {\bibinfo  {journal} {Phys.
  Rev. C}\ }\textbf {\bibinfo {volume} {98}},\ \bibinfo {pages} {045208}
  (\bibinfo {year} {2018})},\ \Eprint {http://arxiv.org/abs/1803.11369}
  {arXiv:1803.11369 [nucl-th]} \BibitemShut {NoStop}%
\bibitem [{\citenamefont {Dubynskiy}\ and\ \citenamefont
  {Voloshin}(2008)}]{Dubynskiy:2008mq}%
  \BibitemOpen
  \bibfield  {author} {\bibinfo {author} {\bibfnamefont {S.}~\bibnamefont
  {Dubynskiy}}\ and\ \bibinfo {author} {\bibfnamefont {M.~B.}\ \bibnamefont
  {Voloshin}},\ }\href {\doibase 10.1016/j.physletb.2008.07.086} {\bibfield
  {journal} {\bibinfo  {journal} {Phys. Lett. B}\ }\textbf {\bibinfo {volume}
  {666}},\ \bibinfo {pages} {344} (\bibinfo {year} {2008})},\ \Eprint
  {http://arxiv.org/abs/0803.2224} {arXiv:0803.2224 [hep-ph]} \BibitemShut
  {NoStop}%
\bibitem [{\citenamefont {Kubarovsky}\ and\ \citenamefont
  {Voloshin}(2015)}]{Kubarovsky:2015aaa}%
  \BibitemOpen
  \bibfield  {author} {\bibinfo {author} {\bibfnamefont {V.}~\bibnamefont
  {Kubarovsky}}\ and\ \bibinfo {author} {\bibfnamefont {M.~B.}\ \bibnamefont
  {Voloshin}},\ }\href {\doibase 10.1103/PhysRevD.92.031502} {\bibfield
  {journal} {\bibinfo  {journal} {Phys. Rev. D}\ }\textbf {\bibinfo {volume}
  {92}},\ \bibinfo {pages} {031502} (\bibinfo {year} {2015})},\ \Eprint
  {http://arxiv.org/abs/1508.00888} {arXiv:1508.00888 [hep-ph]} \BibitemShut
  {NoStop}%
\bibitem [{\citenamefont {Perevalova}\ \emph {et~al.}(2016)\citenamefont
  {Perevalova}, \citenamefont {Polyakov},\ and\ \citenamefont
  {Schweitzer}}]{Perevalova:2016dln}%
  \BibitemOpen
  \bibfield  {author} {\bibinfo {author} {\bibfnamefont {I.~A.}\ \bibnamefont
  {Perevalova}}, \bibinfo {author} {\bibfnamefont {M.~V.}\ \bibnamefont
  {Polyakov}}, \ and\ \bibinfo {author} {\bibfnamefont {P.}~\bibnamefont
  {Schweitzer}},\ }\href {\doibase 10.1103/PhysRevD.94.054024} {\bibfield
  {journal} {\bibinfo  {journal} {Phys. Rev. D}\ }\textbf {\bibinfo {volume}
  {94}},\ \bibinfo {pages} {054024} (\bibinfo {year} {2016})},\ \Eprint
  {http://arxiv.org/abs/1607.07008} {arXiv:1607.07008 [hep-ph]} \BibitemShut
  {NoStop}%
\bibitem [{\citenamefont {Eides}\ \emph {et~al.}(2018)\citenamefont {Eides},
  \citenamefont {Petrov},\ and\ \citenamefont {Polyakov}}]{Eides:2017xnt}%
  \BibitemOpen
  \bibfield  {author} {\bibinfo {author} {\bibfnamefont {M.~I.}\ \bibnamefont
  {Eides}}, \bibinfo {author} {\bibfnamefont {V.~Y.}\ \bibnamefont {Petrov}}, \
  and\ \bibinfo {author} {\bibfnamefont {M.~V.}\ \bibnamefont {Polyakov}},\
  }\href {\doibase 10.1140/epjc/s10052-018-5530-9} {\bibfield  {journal}
  {\bibinfo  {journal} {Eur. Phys. J. C}\ }\textbf {\bibinfo {volume} {78}},\
  \bibinfo {pages} {36} (\bibinfo {year} {2018})},\ \Eprint
  {http://arxiv.org/abs/1709.09523} {arXiv:1709.09523 [hep-ph]} \BibitemShut
  {NoStop}%
\bibitem [{\citenamefont {Eides}\ and\ \citenamefont
  {Petrov}(2018)}]{Eides:2018lqg}%
  \BibitemOpen
  \bibfield  {author} {\bibinfo {author} {\bibfnamefont {M.~I.}\ \bibnamefont
  {Eides}}\ and\ \bibinfo {author} {\bibfnamefont {V.~Y.}\ \bibnamefont
  {Petrov}},\ }\href {\doibase 10.1103/PhysRevD.98.114037} {\bibfield
  {journal} {\bibinfo  {journal} {Phys. Rev. D}\ }\textbf {\bibinfo {volume}
  {98}},\ \bibinfo {pages} {114037} (\bibinfo {year} {2018})},\ \Eprint
  {http://arxiv.org/abs/1811.01691} {arXiv:1811.01691 [hep-ph]} \BibitemShut
  {NoStop}%
\bibitem [{\citenamefont {Eides}\ \emph {et~al.}(2020)\citenamefont {Eides},
  \citenamefont {Petrov},\ and\ \citenamefont {Polyakov}}]{Eides:2019tgv}%
  \BibitemOpen
  \bibfield  {author} {\bibinfo {author} {\bibfnamefont {M.~I.}\ \bibnamefont
  {Eides}}, \bibinfo {author} {\bibfnamefont {V.~Y.}\ \bibnamefont {Petrov}}, \
  and\ \bibinfo {author} {\bibfnamefont {M.~V.}\ \bibnamefont {Polyakov}},\
  }\href {\doibase 10.1142/S0217732320501515} {\bibfield  {journal} {\bibinfo
  {journal} {Mod. Phys. Lett. A}\ }\textbf {\bibinfo {volume} {35}},\ \bibinfo
  {pages} {2050151} (\bibinfo {year} {2020})},\ \Eprint
  {http://arxiv.org/abs/1904.11616} {arXiv:1904.11616 [hep-ph]} \BibitemShut
  {NoStop}%
\bibitem [{\citenamefont {Guo}\ \emph {et~al.}(2015)\citenamefont {Guo},
  \citenamefont {Mei\ss{}ner}, \citenamefont {Wang},\ and\ \citenamefont
  {Yang}}]{Guo:2015umn}%
  \BibitemOpen
  \bibfield  {author} {\bibinfo {author} {\bibfnamefont {F.-K.}\ \bibnamefont
  {Guo}}, \bibinfo {author} {\bibfnamefont {U.-G.}\ \bibnamefont
  {Mei\ss{}ner}}, \bibinfo {author} {\bibfnamefont {W.}~\bibnamefont {Wang}}, \
  and\ \bibinfo {author} {\bibfnamefont {Z.}~\bibnamefont {Yang}},\ }\href
  {\doibase 10.1103/PhysRevD.92.071502} {\bibfield  {journal} {\bibinfo
  {journal} {Phys. Rev. D}\ }\textbf {\bibinfo {volume} {92}},\ \bibinfo
  {pages} {071502} (\bibinfo {year} {2015})},\ \Eprint
  {http://arxiv.org/abs/1507.04950} {arXiv:1507.04950 [hep-ph]} \BibitemShut
  {NoStop}%
\bibitem [{\citenamefont {Liu}\ \emph {et~al.}(2016)\citenamefont {Liu},
  \citenamefont {Wang},\ and\ \citenamefont {Zhao}}]{Liu:2015fea}%
  \BibitemOpen
  \bibfield  {author} {\bibinfo {author} {\bibfnamefont {X.-H.}\ \bibnamefont
  {Liu}}, \bibinfo {author} {\bibfnamefont {Q.}~\bibnamefont {Wang}}, \ and\
  \bibinfo {author} {\bibfnamefont {Q.}~\bibnamefont {Zhao}},\ }\href {\doibase
  10.1016/j.physletb.2016.03.089} {\bibfield  {journal} {\bibinfo  {journal}
  {Phys. Lett. B}\ }\textbf {\bibinfo {volume} {757}},\ \bibinfo {pages} {231}
  (\bibinfo {year} {2016})},\ \Eprint {http://arxiv.org/abs/1507.05359}
  {arXiv:1507.05359 [hep-ph]} \BibitemShut {NoStop}%
\bibitem [{\citenamefont {Bayar}\ \emph {et~al.}(2016)\citenamefont {Bayar},
  \citenamefont {Aceti}, \citenamefont {Guo},\ and\ \citenamefont
  {Oset}}]{Bayar:2016ftu}%
  \BibitemOpen
  \bibfield  {author} {\bibinfo {author} {\bibfnamefont {M.}~\bibnamefont
  {Bayar}}, \bibinfo {author} {\bibfnamefont {F.}~\bibnamefont {Aceti}},
  \bibinfo {author} {\bibfnamefont {F.-K.}\ \bibnamefont {Guo}}, \ and\
  \bibinfo {author} {\bibfnamefont {E.}~\bibnamefont {Oset}},\ }\href {\doibase
  10.1103/PhysRevD.94.074039} {\bibfield  {journal} {\bibinfo  {journal} {Phys.
  Rev. D}\ }\textbf {\bibinfo {volume} {94}},\ \bibinfo {pages} {074039}
  (\bibinfo {year} {2016})},\ \Eprint {http://arxiv.org/abs/1609.04133}
  {arXiv:1609.04133 [hep-ph]} \BibitemShut {NoStop}%
\bibitem [{\citenamefont {Aaij}\ \emph {et~al.}(2019)\citenamefont {Aaij} \emph
  {et~al.}}]{LHCb:2019kea}%
  \BibitemOpen
  \bibfield  {author} {\bibinfo {author} {\bibfnamefont {R.}~\bibnamefont
  {Aaij}} \emph {et~al.} (\bibinfo {collaboration} {LHCb}),\ }\href {\doibase
  10.1103/PhysRevLett.122.222001} {\bibfield  {journal} {\bibinfo  {journal}
  {Phys. Rev. Lett.}\ }\textbf {\bibinfo {volume} {122}},\ \bibinfo {pages}
  {222001} (\bibinfo {year} {2019})},\ \Eprint
  {http://arxiv.org/abs/1904.03947} {arXiv:1904.03947 [hep-ex]} \BibitemShut
  {NoStop}%
\bibitem [{\citenamefont {Guo}\ and\ \citenamefont
  {Oller}(2019)}]{Guo:2019kdc}%
  \BibitemOpen
  \bibfield  {author} {\bibinfo {author} {\bibfnamefont {Z.-H.}\ \bibnamefont
  {Guo}}\ and\ \bibinfo {author} {\bibfnamefont {J.~A.}\ \bibnamefont
  {Oller}},\ }\href {\doibase 10.1016/j.physletb.2019.04.053} {\bibfield
  {journal} {\bibinfo  {journal} {Phys. Lett. B}\ }\textbf {\bibinfo {volume}
  {793}},\ \bibinfo {pages} {144} (\bibinfo {year} {2019})},\ \Eprint
  {http://arxiv.org/abs/1904.00851} {arXiv:1904.00851 [hep-ph]} \BibitemShut
  {NoStop}%
\bibitem [{\citenamefont {Xiao}\ \emph
  {et~al.}(2019{\natexlab{a}})\citenamefont {Xiao}, \citenamefont {Huang},
  \citenamefont {Dong}, \citenamefont {Geng},\ and\ \citenamefont
  {Chen}}]{Xiao:2019mvs}%
  \BibitemOpen
  \bibfield  {author} {\bibinfo {author} {\bibfnamefont {C.-J.}\ \bibnamefont
  {Xiao}}, \bibinfo {author} {\bibfnamefont {Y.}~\bibnamefont {Huang}},
  \bibinfo {author} {\bibfnamefont {Y.-B.}\ \bibnamefont {Dong}}, \bibinfo
  {author} {\bibfnamefont {L.-S.}\ \bibnamefont {Geng}}, \ and\ \bibinfo
  {author} {\bibfnamefont {D.-Y.}\ \bibnamefont {Chen}},\ }\href {\doibase
  10.1103/PhysRevD.100.014022} {\bibfield  {journal} {\bibinfo  {journal}
  {Phys. Rev. D}\ }\textbf {\bibinfo {volume} {100}},\ \bibinfo {pages}
  {014022} (\bibinfo {year} {2019}{\natexlab{a}})},\ \Eprint
  {http://arxiv.org/abs/1904.00872} {arXiv:1904.00872 [hep-ph]} \BibitemShut
  {NoStop}%
\bibitem [{\citenamefont {Xiao}\ \emph {et~al.}(2020)\citenamefont {Xiao},
  \citenamefont {Lu}, \citenamefont {Wu},\ and\ \citenamefont
  {Geng}}]{Xiao:2020frg}%
  \BibitemOpen
  \bibfield  {author} {\bibinfo {author} {\bibfnamefont {C.~W.}\ \bibnamefont
  {Xiao}}, \bibinfo {author} {\bibfnamefont {J.~X.}\ \bibnamefont {Lu}},
  \bibinfo {author} {\bibfnamefont {J.~J.}\ \bibnamefont {Wu}}, \ and\ \bibinfo
  {author} {\bibfnamefont {L.~S.}\ \bibnamefont {Geng}},\ }\href {\doibase
  10.1103/PhysRevD.102.056018} {\bibfield  {journal} {\bibinfo  {journal}
  {Phys. Rev. D}\ }\textbf {\bibinfo {volume} {102}},\ \bibinfo {pages}
  {056018} (\bibinfo {year} {2020})},\ \Eprint
  {http://arxiv.org/abs/2007.12106} {arXiv:2007.12106 [hep-ph]} \BibitemShut
  {NoStop}%
\bibitem [{\citenamefont {Xiao}\ \emph
  {et~al.}(2019{\natexlab{b}})\citenamefont {Xiao}, \citenamefont {Nieves},\
  and\ \citenamefont {Oset}}]{Xiao:2019aya}%
  \BibitemOpen
  \bibfield  {author} {\bibinfo {author} {\bibfnamefont {C.~W.}\ \bibnamefont
  {Xiao}}, \bibinfo {author} {\bibfnamefont {J.}~\bibnamefont {Nieves}}, \ and\
  \bibinfo {author} {\bibfnamefont {E.}~\bibnamefont {Oset}},\ }\href {\doibase
  10.1103/PhysRevD.100.014021} {\bibfield  {journal} {\bibinfo  {journal}
  {Phys. Rev. D}\ }\textbf {\bibinfo {volume} {100}},\ \bibinfo {pages}
  {014021} (\bibinfo {year} {2019}{\natexlab{b}})},\ \Eprint
  {http://arxiv.org/abs/1904.01296} {arXiv:1904.01296 [hep-ph]} \BibitemShut
  {NoStop}%
\bibitem [{\citenamefont {Guo}\ \emph {et~al.}(2019)\citenamefont {Guo},
  \citenamefont {Jing}, \citenamefont {Mei\ss{}ner},\ and\ \citenamefont
  {Sakai}}]{Guo:2019fdo}%
  \BibitemOpen
  \bibfield  {author} {\bibinfo {author} {\bibfnamefont {F.-K.}\ \bibnamefont
  {Guo}}, \bibinfo {author} {\bibfnamefont {H.-J.}\ \bibnamefont {Jing}},
  \bibinfo {author} {\bibfnamefont {U.-G.}\ \bibnamefont {Mei\ss{}ner}}, \ and\
  \bibinfo {author} {\bibfnamefont {S.}~\bibnamefont {Sakai}},\ }\href
  {\doibase 10.1103/PhysRevD.99.091501} {\bibfield  {journal} {\bibinfo
  {journal} {Phys. Rev. D}\ }\textbf {\bibinfo {volume} {99}},\ \bibinfo
  {pages} {091501} (\bibinfo {year} {2019})},\ \Eprint
  {http://arxiv.org/abs/1903.11503} {arXiv:1903.11503 [hep-ph]} \BibitemShut
  {NoStop}%
\bibitem [{\citenamefont {Liu}\ \emph {et~al.}(2019{\natexlab{a}})\citenamefont
  {Liu}, \citenamefont {Pan}, \citenamefont {Peng}, \citenamefont
  {S\'anchez~S\'anchez}, \citenamefont {Geng}, \citenamefont {Hosaka},\ and\
  \citenamefont {Pavon~Valderrama}}]{Liu:2019tjn}%
  \BibitemOpen
  \bibfield  {author} {\bibinfo {author} {\bibfnamefont {M.-Z.}\ \bibnamefont
  {Liu}}, \bibinfo {author} {\bibfnamefont {Y.-W.}\ \bibnamefont {Pan}},
  \bibinfo {author} {\bibfnamefont {F.-Z.}\ \bibnamefont {Peng}}, \bibinfo
  {author} {\bibfnamefont {M.}~\bibnamefont {S\'anchez~S\'anchez}}, \bibinfo
  {author} {\bibfnamefont {L.-S.}\ \bibnamefont {Geng}}, \bibinfo {author}
  {\bibfnamefont {A.}~\bibnamefont {Hosaka}}, \ and\ \bibinfo {author}
  {\bibfnamefont {M.}~\bibnamefont {Pavon~Valderrama}},\ }\href {\doibase
  10.1103/PhysRevLett.122.242001} {\bibfield  {journal} {\bibinfo  {journal}
  {Phys. Rev. Lett.}\ }\textbf {\bibinfo {volume} {122}},\ \bibinfo {pages}
  {242001} (\bibinfo {year} {2019}{\natexlab{a}})},\ \Eprint
  {http://arxiv.org/abs/1903.11560} {arXiv:1903.11560 [hep-ph]} \BibitemShut
  {NoStop}%
\bibitem [{\citenamefont {He}(2019)}]{He:2019ify}%
  \BibitemOpen
  \bibfield  {author} {\bibinfo {author} {\bibfnamefont {J.}~\bibnamefont
  {He}},\ }\href {\doibase 10.1140/epjc/s10052-019-6906-1} {\bibfield
  {journal} {\bibinfo  {journal} {Eur. Phys. J. C}\ }\textbf {\bibinfo {volume}
  {79}},\ \bibinfo {pages} {393} (\bibinfo {year} {2019})},\ \Eprint
  {http://arxiv.org/abs/1903.11872} {arXiv:1903.11872 [hep-ph]} \BibitemShut
  {NoStop}%
\bibitem [{\citenamefont {Liu}\ \emph {et~al.}(2019{\natexlab{b}})\citenamefont
  {Liu}, \citenamefont {Chen}, \citenamefont {Chen}, \citenamefont {Liu},\ and\
  \citenamefont {Zhu}}]{Liu:2019zoy}%
  \BibitemOpen
  \bibfield  {author} {\bibinfo {author} {\bibfnamefont {Y.-R.}\ \bibnamefont
  {Liu}}, \bibinfo {author} {\bibfnamefont {H.-X.}\ \bibnamefont {Chen}},
  \bibinfo {author} {\bibfnamefont {W.}~\bibnamefont {Chen}}, \bibinfo {author}
  {\bibfnamefont {X.}~\bibnamefont {Liu}}, \ and\ \bibinfo {author}
  {\bibfnamefont {S.-L.}\ \bibnamefont {Zhu}},\ }\href {\doibase
  10.1016/j.ppnp.2019.04.003} {\bibfield  {journal} {\bibinfo  {journal} {Prog.
  Part. Nucl. Phys.}\ }\textbf {\bibinfo {volume} {107}},\ \bibinfo {pages}
  {237} (\bibinfo {year} {2019}{\natexlab{b}})},\ \Eprint
  {http://arxiv.org/abs/1903.11976} {arXiv:1903.11976 [hep-ph]} \BibitemShut
  {NoStop}%
\bibitem [{\citenamefont {Shimizu}\ \emph {et~al.}(2019)\citenamefont
  {Shimizu}, \citenamefont {Yamaguchi},\ and\ \citenamefont
  {Harada}}]{Shimizu:2019jfy}%
  \BibitemOpen
  \bibfield  {author} {\bibinfo {author} {\bibfnamefont {Y.}~\bibnamefont
  {Shimizu}}, \bibinfo {author} {\bibfnamefont {Y.}~\bibnamefont {Yamaguchi}},
  \ and\ \bibinfo {author} {\bibfnamefont {M.}~\bibnamefont {Harada}},\ }\href
  {\doibase 10.1093/ptep/ptz146} {\bibfield  {journal} {\bibinfo  {journal}
  {PTEP}\ }\textbf {\bibinfo {volume} {2019}},\ \bibinfo {pages} {123D01}
  (\bibinfo {year} {2019})},\ \Eprint {http://arxiv.org/abs/1901.09215}
  {arXiv:1901.09215 [hep-ph]} \BibitemShut {NoStop}%
\bibitem [{\citenamefont {Weng}\ \emph {et~al.}(2019)\citenamefont {Weng},
  \citenamefont {Chen}, \citenamefont {Deng},\ and\ \citenamefont
  {Zhu}}]{Weng:2019ynv}%
  \BibitemOpen
  \bibfield  {author} {\bibinfo {author} {\bibfnamefont {X.-Z.}\ \bibnamefont
  {Weng}}, \bibinfo {author} {\bibfnamefont {X.-L.}\ \bibnamefont {Chen}},
  \bibinfo {author} {\bibfnamefont {W.-Z.}\ \bibnamefont {Deng}}, \ and\
  \bibinfo {author} {\bibfnamefont {S.-L.}\ \bibnamefont {Zhu}},\ }\href
  {\doibase 10.1103/PhysRevD.100.016014} {\bibfield  {journal} {\bibinfo
  {journal} {Phys. Rev. D}\ }\textbf {\bibinfo {volume} {100}},\ \bibinfo
  {pages} {016014} (\bibinfo {year} {2019})},\ \Eprint
  {http://arxiv.org/abs/1904.09891} {arXiv:1904.09891 [hep-ph]} \BibitemShut
  {NoStop}%
\bibitem [{\citenamefont {Wang}\ \emph {et~al.}(2019)\citenamefont {Wang},
  \citenamefont {He}, \citenamefont {Chen}, \citenamefont {Wang},\ and\
  \citenamefont {Zhu}}]{Wang:2019dsi}%
  \BibitemOpen
  \bibfield  {author} {\bibinfo {author} {\bibfnamefont {X.-Y.}\ \bibnamefont
  {Wang}}, \bibinfo {author} {\bibfnamefont {J.}~\bibnamefont {He}}, \bibinfo
  {author} {\bibfnamefont {X.-R.}\ \bibnamefont {Chen}}, \bibinfo {author}
  {\bibfnamefont {Q.}~\bibnamefont {Wang}}, \ and\ \bibinfo {author}
  {\bibfnamefont {X.}~\bibnamefont {Zhu}},\ }\href {\doibase
  10.1016/j.physletb.2019.134862} {\bibfield  {journal} {\bibinfo  {journal}
  {Phys. Lett. B}\ }\textbf {\bibinfo {volume} {797}},\ \bibinfo {pages}
  {134862} (\bibinfo {year} {2019})},\ \Eprint
  {http://arxiv.org/abs/1906.04044} {arXiv:1906.04044 [hep-ph]} \BibitemShut
  {NoStop}%
\bibitem [{\citenamefont {Cheng}\ and\ \citenamefont
  {Liu}(2019)}]{Cheng:2019obk}%
  \BibitemOpen
  \bibfield  {author} {\bibinfo {author} {\bibfnamefont {J.-B.}\ \bibnamefont
  {Cheng}}\ and\ \bibinfo {author} {\bibfnamefont {Y.-R.}\ \bibnamefont
  {Liu}},\ }\href {\doibase 10.1103/PhysRevD.100.054002} {\bibfield  {journal}
  {\bibinfo  {journal} {Phys. Rev. D}\ }\textbf {\bibinfo {volume} {100}},\
  \bibinfo {pages} {054002} (\bibinfo {year} {2019})},\ \Eprint
  {http://arxiv.org/abs/1905.08605} {arXiv:1905.08605 [hep-ph]} \BibitemShut
  {NoStop}%
\bibitem [{\citenamefont {Du}\ \emph {et~al.}(2020)\citenamefont {Du},
  \citenamefont {Baru}, \citenamefont {Guo}, \citenamefont {Hanhart},
  \citenamefont {Mei\ss{}ner}, \citenamefont {Oller},\ and\ \citenamefont
  {Wang}}]{Du:2019pij}%
  \BibitemOpen
  \bibfield  {author} {\bibinfo {author} {\bibfnamefont {M.-L.}\ \bibnamefont
  {Du}}, \bibinfo {author} {\bibfnamefont {V.}~\bibnamefont {Baru}}, \bibinfo
  {author} {\bibfnamefont {F.-K.}\ \bibnamefont {Guo}}, \bibinfo {author}
  {\bibfnamefont {C.}~\bibnamefont {Hanhart}}, \bibinfo {author} {\bibfnamefont
  {U.-G.}\ \bibnamefont {Mei\ss{}ner}}, \bibinfo {author} {\bibfnamefont
  {J.~A.}\ \bibnamefont {Oller}}, \ and\ \bibinfo {author} {\bibfnamefont
  {Q.}~\bibnamefont {Wang}},\ }\href {\doibase 10.1103/PhysRevLett.124.072001}
  {\bibfield  {journal} {\bibinfo  {journal} {Phys. Rev. Lett.}\ }\textbf
  {\bibinfo {volume} {124}},\ \bibinfo {pages} {072001} (\bibinfo {year}
  {2020})},\ \Eprint {http://arxiv.org/abs/1910.11846} {arXiv:1910.11846
  [hep-ph]} \BibitemShut {NoStop}%
\bibitem [{\citenamefont {Pavon~Valderrama}(2019)}]{PavonValderrama:2019nbk}%
  \BibitemOpen
  \bibfield  {author} {\bibinfo {author} {\bibfnamefont {M.}~\bibnamefont
  {Pavon~Valderrama}},\ }\href {\doibase 10.1103/PhysRevD.100.094028}
  {\bibfield  {journal} {\bibinfo  {journal} {Phys. Rev. D}\ }\textbf {\bibinfo
  {volume} {100}},\ \bibinfo {pages} {094028} (\bibinfo {year} {2019})},\
  \Eprint {http://arxiv.org/abs/1907.05294} {arXiv:1907.05294 [hep-ph]}
  \BibitemShut {NoStop}%
\bibitem [{\citenamefont {Liu}\ \emph {et~al.}(2021{\natexlab{a}})\citenamefont
  {Liu}, \citenamefont {Wu}, \citenamefont {S\'anchez~S\'anchez}, \citenamefont
  {Valderrama}, \citenamefont {Geng},\ and\ \citenamefont {Xie}}]{Liu:2019zvb}%
  \BibitemOpen
  \bibfield  {author} {\bibinfo {author} {\bibfnamefont {M.-Z.}\ \bibnamefont
  {Liu}}, \bibinfo {author} {\bibfnamefont {T.-W.}\ \bibnamefont {Wu}},
  \bibinfo {author} {\bibfnamefont {M.}~\bibnamefont {S\'anchez~S\'anchez}},
  \bibinfo {author} {\bibfnamefont {M.~P.}\ \bibnamefont {Valderrama}},
  \bibinfo {author} {\bibfnamefont {L.-S.}\ \bibnamefont {Geng}}, \ and\
  \bibinfo {author} {\bibfnamefont {J.-J.}\ \bibnamefont {Xie}},\ }\href
  {\doibase 10.1103/PhysRevD.103.054004} {\bibfield  {journal} {\bibinfo
  {journal} {Phys. Rev. D}\ }\textbf {\bibinfo {volume} {103}},\ \bibinfo
  {pages} {054004} (\bibinfo {year} {2021}{\natexlab{a}})},\ \Eprint
  {http://arxiv.org/abs/1907.06093} {arXiv:1907.06093 [hep-ph]} \BibitemShut
  {NoStop}%
\bibitem [{\citenamefont {Yang}\ \emph
  {et~al.}(2020{\natexlab{a}})\citenamefont {Yang}, \citenamefont {Meng},\ and\
  \citenamefont {Zhu}}]{Yang:2019rgw}%
  \BibitemOpen
  \bibfield  {author} {\bibinfo {author} {\bibfnamefont {B.}~\bibnamefont
  {Yang}}, \bibinfo {author} {\bibfnamefont {L.}~\bibnamefont {Meng}}, \ and\
  \bibinfo {author} {\bibfnamefont {S.-L.}\ \bibnamefont {Zhu}},\ }\href
  {\doibase 10.1140/epja/s10050-020-00028-9} {\bibfield  {journal} {\bibinfo
  {journal} {Eur. Phys. J. A}\ }\textbf {\bibinfo {volume} {56}},\ \bibinfo
  {pages} {67} (\bibinfo {year} {2020}{\natexlab{a}})},\ \Eprint
  {http://arxiv.org/abs/1906.04956} {arXiv:1906.04956 [hep-ph]} \BibitemShut
  {NoStop}%
\bibitem [{\citenamefont {Xu}\ \emph {et~al.}(2020{\natexlab{a}})\citenamefont
  {Xu}, \citenamefont {Cui}, \citenamefont {Liu},\ and\ \citenamefont
  {Huang}}]{Xu:2019zme}%
  \BibitemOpen
  \bibfield  {author} {\bibinfo {author} {\bibfnamefont {Y.-J.}\ \bibnamefont
  {Xu}}, \bibinfo {author} {\bibfnamefont {C.-Y.}\ \bibnamefont {Cui}},
  \bibinfo {author} {\bibfnamefont {Y.-L.}\ \bibnamefont {Liu}}, \ and\
  \bibinfo {author} {\bibfnamefont {M.-Q.}\ \bibnamefont {Huang}},\ }\href
  {\doibase 10.1103/PhysRevD.102.034028} {\bibfield  {journal} {\bibinfo
  {journal} {Phys. Rev. D}\ }\textbf {\bibinfo {volume} {102}},\ \bibinfo
  {pages} {034028} (\bibinfo {year} {2020}{\natexlab{a}})},\ \Eprint
  {http://arxiv.org/abs/1907.05097} {arXiv:1907.05097 [hep-ph]} \BibitemShut
  {NoStop}%
\bibitem [{\citenamefont {Xu}\ \emph {et~al.}(2020{\natexlab{b}})\citenamefont
  {Xu}, \citenamefont {Li}, \citenamefont {Chang},\ and\ \citenamefont
  {Wang}}]{Xu:2020gjl}%
  \BibitemOpen
  \bibfield  {author} {\bibinfo {author} {\bibfnamefont {H.}~\bibnamefont
  {Xu}}, \bibinfo {author} {\bibfnamefont {Q.}~\bibnamefont {Li}}, \bibinfo
  {author} {\bibfnamefont {C.-H.}\ \bibnamefont {Chang}}, \ and\ \bibinfo
  {author} {\bibfnamefont {G.-L.}\ \bibnamefont {Wang}},\ }\href {\doibase
  10.1103/PhysRevD.101.054037} {\bibfield  {journal} {\bibinfo  {journal}
  {Phys. Rev. D}\ }\textbf {\bibinfo {volume} {101}},\ \bibinfo {pages}
  {054037} (\bibinfo {year} {2020}{\natexlab{b}})},\ \Eprint
  {http://arxiv.org/abs/2001.02980} {arXiv:2001.02980 [hep-ph]} \BibitemShut
  {NoStop}%
\bibitem [{\citenamefont {Yamaguchi}\ \emph
  {et~al.}(2020{\natexlab{b}})\citenamefont {Yamaguchi}, \citenamefont
  {Garc\'\i{}a-Tecocoatzi}, \citenamefont {Giachino}, \citenamefont {Hosaka},
  \citenamefont {Santopinto}, \citenamefont {Takeuchi},\ and\ \citenamefont
  {Takizawa}}]{Yamaguchi:2019seo}%
  \BibitemOpen
  \bibfield  {author} {\bibinfo {author} {\bibfnamefont {Y.}~\bibnamefont
  {Yamaguchi}}, \bibinfo {author} {\bibfnamefont {H.}~\bibnamefont
  {Garc\'\i{}a-Tecocoatzi}}, \bibinfo {author} {\bibfnamefont {A.}~\bibnamefont
  {Giachino}}, \bibinfo {author} {\bibfnamefont {A.}~\bibnamefont {Hosaka}},
  \bibinfo {author} {\bibfnamefont {E.}~\bibnamefont {Santopinto}}, \bibinfo
  {author} {\bibfnamefont {S.}~\bibnamefont {Takeuchi}}, \ and\ \bibinfo
  {author} {\bibfnamefont {M.}~\bibnamefont {Takizawa}},\ }\href {\doibase
  10.1103/PhysRevD.101.091502} {\bibfield  {journal} {\bibinfo  {journal}
  {Phys. Rev. D}\ }\textbf {\bibinfo {volume} {101}},\ \bibinfo {pages}
  {091502} (\bibinfo {year} {2020}{\natexlab{b}})},\ \Eprint
  {http://arxiv.org/abs/1907.04684} {arXiv:1907.04684 [hep-ph]} \BibitemShut
  {NoStop}%
\bibitem [{\citenamefont {Ke}\ \emph {et~al.}(2020)\citenamefont {Ke},
  \citenamefont {Li}, \citenamefont {Liu},\ and\ \citenamefont
  {Li}}]{Ke:2019bkf}%
  \BibitemOpen
  \bibfield  {author} {\bibinfo {author} {\bibfnamefont {H.-W.}\ \bibnamefont
  {Ke}}, \bibinfo {author} {\bibfnamefont {M.}~\bibnamefont {Li}}, \bibinfo
  {author} {\bibfnamefont {X.-H.}\ \bibnamefont {Liu}}, \ and\ \bibinfo
  {author} {\bibfnamefont {X.-Q.}\ \bibnamefont {Li}},\ }\href {\doibase
  10.1103/PhysRevD.101.014024} {\bibfield  {journal} {\bibinfo  {journal}
  {Phys. Rev. D}\ }\textbf {\bibinfo {volume} {101}},\ \bibinfo {pages}
  {014024} (\bibinfo {year} {2020})},\ \Eprint
  {http://arxiv.org/abs/1909.12509} {arXiv:1909.12509 [hep-ph]} \BibitemShut
  {NoStop}%
\bibitem [{\citenamefont {Giachino}\ \emph {et~al.}(2020)\citenamefont
  {Giachino}, \citenamefont {Hosaka}, \citenamefont {Santopinto}, \citenamefont
  {Takeuchi}, \citenamefont {Takizawa},\ and\ \citenamefont
  {Yamaguchi}}]{Giachino:2020rkj}%
  \BibitemOpen
  \bibfield  {author} {\bibinfo {author} {\bibfnamefont {A.}~\bibnamefont
  {Giachino}}, \bibinfo {author} {\bibfnamefont {A.}~\bibnamefont {Hosaka}},
  \bibinfo {author} {\bibfnamefont {E.}~\bibnamefont {Santopinto}}, \bibinfo
  {author} {\bibfnamefont {S.}~\bibnamefont {Takeuchi}}, \bibinfo {author}
  {\bibfnamefont {M.}~\bibnamefont {Takizawa}}, \ and\ \bibinfo {author}
  {\bibfnamefont {Y.}~\bibnamefont {Yamaguchi}},\ }\href {\doibase
  10.1007/978-3-030-32357-8_98} {\bibfield  {journal} {\bibinfo  {journal}
  {Springer Proc. Phys.}\ }\textbf {\bibinfo {volume} {238}},\ \bibinfo {pages}
  {621} (\bibinfo {year} {2020})}\BibitemShut {NoStop}%
\bibitem [{\citenamefont {Yang}\ \emph
  {et~al.}(2020{\natexlab{b}})\citenamefont {Yang}, \citenamefont {Ping},\ and\
  \citenamefont {Segovia}}]{Yang:2020twg}%
  \BibitemOpen
  \bibfield  {author} {\bibinfo {author} {\bibfnamefont {G.}~\bibnamefont
  {Yang}}, \bibinfo {author} {\bibfnamefont {J.}~\bibnamefont {Ping}}, \ and\
  \bibinfo {author} {\bibfnamefont {J.}~\bibnamefont {Segovia}},\ }\href
  {\doibase 10.1103/PhysRevD.101.074030} {\bibfield  {journal} {\bibinfo
  {journal} {Phys. Rev. D}\ }\textbf {\bibinfo {volume} {101}},\ \bibinfo
  {pages} {074030} (\bibinfo {year} {2020}{\natexlab{b}})},\ \Eprint
  {http://arxiv.org/abs/2003.05253} {arXiv:2003.05253 [hep-ph]} \BibitemShut
  {NoStop}%
\bibitem [{\citenamefont {Azizi}\ \emph {et~al.}(2021)\citenamefont {Azizi},
  \citenamefont {Sarac},\ and\ \citenamefont {Sundu}}]{Azizi:2020ogm}%
  \BibitemOpen
  \bibfield  {author} {\bibinfo {author} {\bibfnamefont {K.}~\bibnamefont
  {Azizi}}, \bibinfo {author} {\bibfnamefont {Y.}~\bibnamefont {Sarac}}, \ and\
  \bibinfo {author} {\bibfnamefont {H.}~\bibnamefont {Sundu}},\ }\href
  {\doibase 10.1088/1674-1137/abe8ce} {\bibfield  {journal} {\bibinfo
  {journal} {Chin. Phys. C}\ }\textbf {\bibinfo {volume} {45}},\ \bibinfo
  {pages} {053103} (\bibinfo {year} {2021})},\ \Eprint
  {http://arxiv.org/abs/2011.05828} {arXiv:2011.05828 [hep-ph]} \BibitemShut
  {NoStop}%
\bibitem [{\citenamefont {Liu}\ \emph {et~al.}(2021{\natexlab{b}})\citenamefont
  {Liu}, \citenamefont {Pan},\ and\ \citenamefont {Geng}}]{Liu:2020hcv}%
  \BibitemOpen
  \bibfield  {author} {\bibinfo {author} {\bibfnamefont {M.-Z.}\ \bibnamefont
  {Liu}}, \bibinfo {author} {\bibfnamefont {Y.-W.}\ \bibnamefont {Pan}}, \ and\
  \bibinfo {author} {\bibfnamefont {L.-S.}\ \bibnamefont {Geng}},\ }\href
  {\doibase 10.1103/PhysRevD.103.034003} {\bibfield  {journal} {\bibinfo
  {journal} {Phys. Rev. D}\ }\textbf {\bibinfo {volume} {103}},\ \bibinfo
  {pages} {034003} (\bibinfo {year} {2021}{\natexlab{b}})},\ \Eprint
  {http://arxiv.org/abs/2011.07935} {arXiv:2011.07935 [hep-ph]} \BibitemShut
  {NoStop}%
\bibitem [{\citenamefont {Peng}\ \emph {et~al.}(2021)\citenamefont {Peng},
  \citenamefont {Lu}, \citenamefont {S\'anchez~S\'anchez}, \citenamefont
  {Yan},\ and\ \citenamefont {Pavon~Valderrama}}]{Peng:2020gwk}%
  \BibitemOpen
  \bibfield  {author} {\bibinfo {author} {\bibfnamefont {F.-Z.}\ \bibnamefont
  {Peng}}, \bibinfo {author} {\bibfnamefont {J.-X.}\ \bibnamefont {Lu}},
  \bibinfo {author} {\bibfnamefont {M.}~\bibnamefont {S\'anchez~S\'anchez}},
  \bibinfo {author} {\bibfnamefont {M.-J.}\ \bibnamefont {Yan}}, \ and\
  \bibinfo {author} {\bibfnamefont {M.}~\bibnamefont {Pavon~Valderrama}},\
  }\href {\doibase 10.1103/PhysRevD.103.014023} {\bibfield  {journal} {\bibinfo
   {journal} {Phys. Rev. D}\ }\textbf {\bibinfo {volume} {103}},\ \bibinfo
  {pages} {014023} (\bibinfo {year} {2021})},\ \Eprint
  {http://arxiv.org/abs/2007.01198} {arXiv:2007.01198 [hep-ph]} \BibitemShut
  {NoStop}%
\bibitem [{\citenamefont {Chen}\ \emph {et~al.}(2021)\citenamefont {Chen},
  \citenamefont {Wang},\ and\ \citenamefont {Zhu}}]{Chen:2021htr}%
  \BibitemOpen
  \bibfield  {author} {\bibinfo {author} {\bibfnamefont {K.}~\bibnamefont
  {Chen}}, \bibinfo {author} {\bibfnamefont {B.}~\bibnamefont {Wang}}, \ and\
  \bibinfo {author} {\bibfnamefont {S.-L.}\ \bibnamefont {Zhu}},\ }\href
  {\doibase 10.1103/PhysRevD.103.116017} {\bibfield  {journal} {\bibinfo
  {journal} {Phys. Rev. D}\ }\textbf {\bibinfo {volume} {103}},\ \bibinfo
  {pages} {116017} (\bibinfo {year} {2021})},\ \Eprint
  {http://arxiv.org/abs/2102.05868} {arXiv:2102.05868 [hep-ph]} \BibitemShut
  {NoStop}%
\bibitem [{\citenamefont {Phumphan}\ \emph {et~al.}(2021)\citenamefont
  {Phumphan}, \citenamefont {Ruangyoo}, \citenamefont {Chen}, \citenamefont
  {Limphirat},\ and\ \citenamefont {Yan}}]{Phumphan:2021tta}%
  \BibitemOpen
  \bibfield  {author} {\bibinfo {author} {\bibfnamefont {K.}~\bibnamefont
  {Phumphan}}, \bibinfo {author} {\bibfnamefont {W.}~\bibnamefont {Ruangyoo}},
  \bibinfo {author} {\bibfnamefont {C.-C.}\ \bibnamefont {Chen}}, \bibinfo
  {author} {\bibfnamefont {A.}~\bibnamefont {Limphirat}}, \ and\ \bibinfo
  {author} {\bibfnamefont {Y.}~\bibnamefont {Yan}},\ }\href@noop {} {\
  (\bibinfo {year} {2021})},\ \Eprint {http://arxiv.org/abs/2105.03150}
  {arXiv:2105.03150 [hep-ph]} \BibitemShut {NoStop}%
\bibitem [{\citenamefont {Du}\ \emph {et~al.}(2021)\citenamefont {Du},
  \citenamefont {Baru}, \citenamefont {Guo}, \citenamefont {Hanhart},
  \citenamefont {Mei\ss{}ner}, \citenamefont {Oller},\ and\ \citenamefont
  {Wang}}]{Du:2021fmf}%
  \BibitemOpen
  \bibfield  {author} {\bibinfo {author} {\bibfnamefont {M.-L.}\ \bibnamefont
  {Du}}, \bibinfo {author} {\bibfnamefont {V.}~\bibnamefont {Baru}}, \bibinfo
  {author} {\bibfnamefont {F.-K.}\ \bibnamefont {Guo}}, \bibinfo {author}
  {\bibfnamefont {C.}~\bibnamefont {Hanhart}}, \bibinfo {author} {\bibfnamefont
  {U.-G.}\ \bibnamefont {Mei\ss{}ner}}, \bibinfo {author} {\bibfnamefont
  {J.~A.}\ \bibnamefont {Oller}}, \ and\ \bibinfo {author} {\bibfnamefont
  {Q.}~\bibnamefont {Wang}},\ }\href {\doibase 10.1007/JHEP08(2021)157}
  {\bibfield  {journal} {\bibinfo  {journal} {JHEP}\ }\textbf {\bibinfo
  {volume} {08}},\ \bibinfo {pages} {157} (\bibinfo {year} {2021})},\ \Eprint
  {http://arxiv.org/abs/2102.07159} {arXiv:2102.07159 [hep-ph]} \BibitemShut
  {NoStop}%
\bibitem [{\citenamefont {Chen}\ \emph {et~al.}(2019)\citenamefont {Chen},
  \citenamefont {Sun}, \citenamefont {Liu},\ and\ \citenamefont
  {Zhu}}]{Chen:2019asm}%
  \BibitemOpen
  \bibfield  {author} {\bibinfo {author} {\bibfnamefont {R.}~\bibnamefont
  {Chen}}, \bibinfo {author} {\bibfnamefont {Z.-F.}\ \bibnamefont {Sun}},
  \bibinfo {author} {\bibfnamefont {X.}~\bibnamefont {Liu}}, \ and\ \bibinfo
  {author} {\bibfnamefont {S.-L.}\ \bibnamefont {Zhu}},\ }\href {\doibase
  10.1103/PhysRevD.100.011502} {\bibfield  {journal} {\bibinfo  {journal}
  {Phys. Rev. D}\ }\textbf {\bibinfo {volume} {100}},\ \bibinfo {pages}
  {011502} (\bibinfo {year} {2019})},\ \Eprint
  {http://arxiv.org/abs/1903.11013} {arXiv:1903.11013 [hep-ph]} \BibitemShut
  {NoStop}%
\bibitem [{\citenamefont {Sakai}\ \emph {et~al.}(2019)\citenamefont {Sakai},
  \citenamefont {Jing},\ and\ \citenamefont {Guo}}]{Sakai:2019qph}%
  \BibitemOpen
  \bibfield  {author} {\bibinfo {author} {\bibfnamefont {S.}~\bibnamefont
  {Sakai}}, \bibinfo {author} {\bibfnamefont {H.-J.}\ \bibnamefont {Jing}}, \
  and\ \bibinfo {author} {\bibfnamefont {F.-K.}\ \bibnamefont {Guo}},\ }\href
  {\doibase 10.1103/PhysRevD.100.074007} {\bibfield  {journal} {\bibinfo
  {journal} {Phys. Rev. D}\ }\textbf {\bibinfo {volume} {100}},\ \bibinfo
  {pages} {074007} (\bibinfo {year} {2019})},\ \Eprint
  {http://arxiv.org/abs/1907.03414} {arXiv:1907.03414 [hep-ph]} \BibitemShut
  {NoStop}%
\bibitem [{\citenamefont {Burns}\ and\ \citenamefont
  {Swanson}(2019)}]{Burns:2019iih}%
  \BibitemOpen
  \bibfield  {author} {\bibinfo {author} {\bibfnamefont {T.~J.}\ \bibnamefont
  {Burns}}\ and\ \bibinfo {author} {\bibfnamefont {E.~S.}\ \bibnamefont
  {Swanson}},\ }\href {\doibase 10.1103/PhysRevD.100.114033} {\bibfield
  {journal} {\bibinfo  {journal} {Phys. Rev. D}\ }\textbf {\bibinfo {volume}
  {100}},\ \bibinfo {pages} {114033} (\bibinfo {year} {2019})},\ \Eprint
  {http://arxiv.org/abs/1908.03528} {arXiv:1908.03528 [hep-ph]} \BibitemShut
  {NoStop}%
\bibitem [{\citenamefont {Geng}\ \emph {et~al.}(2018)\citenamefont {Geng},
  \citenamefont {Lu},\ and\ \citenamefont {Valderrama}}]{Geng:2017hxc}%
  \BibitemOpen
  \bibfield  {author} {\bibinfo {author} {\bibfnamefont {L.}~\bibnamefont
  {Geng}}, \bibinfo {author} {\bibfnamefont {J.}~\bibnamefont {Lu}}, \ and\
  \bibinfo {author} {\bibfnamefont {M.~P.}\ \bibnamefont {Valderrama}},\ }\href
  {\doibase 10.1103/PhysRevD.97.094036} {\bibfield  {journal} {\bibinfo
  {journal} {Phys. Rev. D}\ }\textbf {\bibinfo {volume} {97}},\ \bibinfo
  {pages} {094036} (\bibinfo {year} {2018})},\ \Eprint
  {http://arxiv.org/abs/1704.06123} {arXiv:1704.06123 [hep-ph]} \BibitemShut
  {NoStop}%
\bibitem [{\citenamefont {Zyla}\ \emph {et~al.}(2020)\citenamefont {Zyla} \emph
  {et~al.}}]{Zyla:2020zbs}%
  \BibitemOpen
  \bibfield  {author} {\bibinfo {author} {\bibfnamefont {P.~A.}\ \bibnamefont
  {Zyla}} \emph {et~al.} (\bibinfo {collaboration} {Particle Data Group}),\
  }\href {\doibase 10.1093/ptep/ptaa104} {\bibfield  {journal} {\bibinfo
  {journal} {PTEP}\ }\textbf {\bibinfo {volume} {2020}},\ \bibinfo {pages}
  {083C01} (\bibinfo {year} {2020})}\BibitemShut {NoStop}%
\bibitem [{\citenamefont {He}\ and\ \citenamefont {Chen}(2019)}]{He:2019rva}%
  \BibitemOpen
  \bibfield  {author} {\bibinfo {author} {\bibfnamefont {J.}~\bibnamefont
  {He}}\ and\ \bibinfo {author} {\bibfnamefont {D.-Y.}\ \bibnamefont {Chen}},\
  }\href {\doibase 10.1140/epjc/s10052-019-7419-7} {\bibfield  {journal}
  {\bibinfo  {journal} {Eur. Phys. J. C}\ }\textbf {\bibinfo {volume} {79}},\
  \bibinfo {pages} {887} (\bibinfo {year} {2019})},\ \Eprint
  {http://arxiv.org/abs/1909.05681} {arXiv:1909.05681 [hep-ph]} \BibitemShut
  {NoStop}%
\bibitem [{\citenamefont {Cheng}\ \emph {et~al.}(1993)\citenamefont {Cheng},
  \citenamefont {Cheung}, \citenamefont {Lin}, \citenamefont {Lin},
  \citenamefont {Yan},\ and\ \citenamefont {Yu}}]{Cheng:1992xi}%
  \BibitemOpen
  \bibfield  {author} {\bibinfo {author} {\bibfnamefont {H.-Y.}\ \bibnamefont
  {Cheng}}, \bibinfo {author} {\bibfnamefont {C.-Y.}\ \bibnamefont {Cheung}},
  \bibinfo {author} {\bibfnamefont {G.-L.}\ \bibnamefont {Lin}}, \bibinfo
  {author} {\bibfnamefont {Y.~C.}\ \bibnamefont {Lin}}, \bibinfo {author}
  {\bibfnamefont {T.-M.}\ \bibnamefont {Yan}}, \ and\ \bibinfo {author}
  {\bibfnamefont {H.-L.}\ \bibnamefont {Yu}},\ }\href {\doibase
  10.1103/PhysRevD.47.1030} {\bibfield  {journal} {\bibinfo  {journal} {Phys.
  Rev. D}\ }\textbf {\bibinfo {volume} {47}},\ \bibinfo {pages} {1030}
  (\bibinfo {year} {1993})},\ \Eprint {http://arxiv.org/abs/hep-ph/9209262}
  {arXiv:hep-ph/9209262} \BibitemShut {NoStop}%
\bibitem [{\citenamefont {Yan}\ \emph {et~al.}(1992)\citenamefont {Yan},
  \citenamefont {Cheng}, \citenamefont {Cheung}, \citenamefont {Lin},
  \citenamefont {Lin},\ and\ \citenamefont {Yu}}]{Yan:1992gz}%
  \BibitemOpen
  \bibfield  {author} {\bibinfo {author} {\bibfnamefont {T.-M.}\ \bibnamefont
  {Yan}}, \bibinfo {author} {\bibfnamefont {H.-Y.}\ \bibnamefont {Cheng}},
  \bibinfo {author} {\bibfnamefont {C.-Y.}\ \bibnamefont {Cheung}}, \bibinfo
  {author} {\bibfnamefont {G.-L.}\ \bibnamefont {Lin}}, \bibinfo {author}
  {\bibfnamefont {Y.~C.}\ \bibnamefont {Lin}}, \ and\ \bibinfo {author}
  {\bibfnamefont {H.-L.}\ \bibnamefont {Yu}},\ }\href {\doibase
  10.1103/PhysRevD.46.1148} {\bibfield  {journal} {\bibinfo  {journal} {Phys.
  Rev. D}\ }\textbf {\bibinfo {volume} {46}},\ \bibinfo {pages} {1148}
  (\bibinfo {year} {1992})},\ \bibinfo {note} {[Erratum: Phys.Rev.D 55, 5851
  (1997)]}\BibitemShut {NoStop}%
\bibitem [{\citenamefont {Wise}(1992)}]{Wise:1992hn}%
  \BibitemOpen
  \bibfield  {author} {\bibinfo {author} {\bibfnamefont {M.~B.}\ \bibnamefont
  {Wise}},\ }\href {\doibase 10.1103/PhysRevD.45.R2188} {\bibfield  {journal}
  {\bibinfo  {journal} {Phys. Rev. D}\ }\textbf {\bibinfo {volume} {45}},\
  \bibinfo {pages} {R2188} (\bibinfo {year} {1992})}\BibitemShut {NoStop}%
\bibitem [{\citenamefont {Liu}\ and\ \citenamefont {Oka}(2012)}]{Liu:2011xc}%
  \BibitemOpen
  \bibfield  {author} {\bibinfo {author} {\bibfnamefont {Y.-R.}\ \bibnamefont
  {Liu}}\ and\ \bibinfo {author} {\bibfnamefont {M.}~\bibnamefont {Oka}},\
  }\href {\doibase 10.1103/PhysRevD.85.014015} {\bibfield  {journal} {\bibinfo
  {journal} {Phys. Rev. D}\ }\textbf {\bibinfo {volume} {85}},\ \bibinfo
  {pages} {014015} (\bibinfo {year} {2012})},\ \Eprint
  {http://arxiv.org/abs/1103.4624} {arXiv:1103.4624 [hep-ph]} \BibitemShut
  {NoStop}%
\bibitem [{\citenamefont {Cho}(1994)}]{Cho:1994vg}%
  \BibitemOpen
  \bibfield  {author} {\bibinfo {author} {\bibfnamefont {P.~L.}\ \bibnamefont
  {Cho}},\ }\href {\doibase 10.1103/PhysRevD.50.3295} {\bibfield  {journal}
  {\bibinfo  {journal} {Phys. Rev. D}\ }\textbf {\bibinfo {volume} {50}},\
  \bibinfo {pages} {3295} (\bibinfo {year} {1994})},\ \Eprint
  {http://arxiv.org/abs/hep-ph/9401276} {arXiv:hep-ph/9401276} \BibitemShut
  {NoStop}%
\bibitem [{\citenamefont {Casalbuoni}\ \emph {et~al.}(1997)\citenamefont
  {Casalbuoni}, \citenamefont {Deandrea}, \citenamefont {Di~Bartolomeo},
  \citenamefont {Gatto}, \citenamefont {Feruglio},\ and\ \citenamefont
  {Nardulli}}]{Casalbuoni:1996pg}%
  \BibitemOpen
  \bibfield  {author} {\bibinfo {author} {\bibfnamefont {R.}~\bibnamefont
  {Casalbuoni}}, \bibinfo {author} {\bibfnamefont {A.}~\bibnamefont
  {Deandrea}}, \bibinfo {author} {\bibfnamefont {N.}~\bibnamefont
  {Di~Bartolomeo}}, \bibinfo {author} {\bibfnamefont {R.}~\bibnamefont
  {Gatto}}, \bibinfo {author} {\bibfnamefont {F.}~\bibnamefont {Feruglio}}, \
  and\ \bibinfo {author} {\bibfnamefont {G.}~\bibnamefont {Nardulli}},\ }\href
  {\doibase 10.1016/S0370-1573(96)00027-0} {\bibfield  {journal} {\bibinfo
  {journal} {Phys. Rept.}\ }\textbf {\bibinfo {volume} {281}},\ \bibinfo
  {pages} {145} (\bibinfo {year} {1997})},\ \Eprint
  {http://arxiv.org/abs/hep-ph/9605342} {arXiv:hep-ph/9605342} \BibitemShut
  {NoStop}%
\bibitem [{\citenamefont {Pirjol}\ and\ \citenamefont
  {Yan}(1997)}]{Pirjol:1997nh}%
  \BibitemOpen
  \bibfield  {author} {\bibinfo {author} {\bibfnamefont {D.}~\bibnamefont
  {Pirjol}}\ and\ \bibinfo {author} {\bibfnamefont {T.-M.}\ \bibnamefont
  {Yan}},\ }\href {\doibase 10.1103/PhysRevD.56.5483} {\bibfield  {journal}
  {\bibinfo  {journal} {Phys. Rev. D}\ }\textbf {\bibinfo {volume} {56}},\
  \bibinfo {pages} {5483} (\bibinfo {year} {1997})},\ \Eprint
  {http://arxiv.org/abs/hep-ph/9701291} {arXiv:hep-ph/9701291} \BibitemShut
  {NoStop}%
\bibitem [{\citenamefont {Ding}(2009)}]{Ding:2008gr}%
  \BibitemOpen
  \bibfield  {author} {\bibinfo {author} {\bibfnamefont {G.-J.}\ \bibnamefont
  {Ding}},\ }\href {\doibase 10.1103/PhysRevD.79.014001} {\bibfield  {journal}
  {\bibinfo  {journal} {Phys. Rev. D}\ }\textbf {\bibinfo {volume} {79}},\
  \bibinfo {pages} {014001} (\bibinfo {year} {2009})},\ \Eprint
  {http://arxiv.org/abs/0809.4818} {arXiv:0809.4818 [hep-ph]} \BibitemShut
  {NoStop}%
\bibitem [{\citenamefont {Meng}\ \emph {et~al.}(2019)\citenamefont {Meng},
  \citenamefont {Wang}, \citenamefont {Wang},\ and\ \citenamefont
  {Zhu}}]{Meng:2019ilv}%
  \BibitemOpen
  \bibfield  {author} {\bibinfo {author} {\bibfnamefont {L.}~\bibnamefont
  {Meng}}, \bibinfo {author} {\bibfnamefont {B.}~\bibnamefont {Wang}}, \bibinfo
  {author} {\bibfnamefont {G.-J.}\ \bibnamefont {Wang}}, \ and\ \bibinfo
  {author} {\bibfnamefont {S.-L.}\ \bibnamefont {Zhu}},\ }\href {\doibase
  10.1103/PhysRevD.100.014031} {\bibfield  {journal} {\bibinfo  {journal}
  {Phys. Rev. D}\ }\textbf {\bibinfo {volume} {100}},\ \bibinfo {pages}
  {014031} (\bibinfo {year} {2019})},\ \Eprint
  {http://arxiv.org/abs/1905.04113} {arXiv:1905.04113 [hep-ph]} \BibitemShut
  {NoStop}%
\bibitem [{\citenamefont {Isola}\ \emph {et~al.}(2003)\citenamefont {Isola},
  \citenamefont {Ladisa}, \citenamefont {Nardulli},\ and\ \citenamefont
  {Santorelli}}]{Isola:2003fh}%
  \BibitemOpen
  \bibfield  {author} {\bibinfo {author} {\bibfnamefont {C.}~\bibnamefont
  {Isola}}, \bibinfo {author} {\bibfnamefont {M.}~\bibnamefont {Ladisa}},
  \bibinfo {author} {\bibfnamefont {G.}~\bibnamefont {Nardulli}}, \ and\
  \bibinfo {author} {\bibfnamefont {P.}~\bibnamefont {Santorelli}},\ }\href
  {\doibase 10.1103/PhysRevD.68.114001} {\bibfield  {journal} {\bibinfo
  {journal} {Phys. Rev. D}\ }\textbf {\bibinfo {volume} {68}},\ \bibinfo
  {pages} {114001} (\bibinfo {year} {2003})},\ \Eprint
  {http://arxiv.org/abs/hep-ph/0307367} {arXiv:hep-ph/0307367} \BibitemShut
  {NoStop}%
\bibitem [{\citenamefont {Cheng}\ and\ \citenamefont
  {Chua}(2015)}]{Cheng:2015naa}%
  \BibitemOpen
  \bibfield  {author} {\bibinfo {author} {\bibfnamefont {H.-Y.}\ \bibnamefont
  {Cheng}}\ and\ \bibinfo {author} {\bibfnamefont {C.-K.}\ \bibnamefont
  {Chua}},\ }\href {\doibase 10.1103/PhysRevD.92.074014} {\bibfield  {journal}
  {\bibinfo  {journal} {Phys. Rev. D}\ }\textbf {\bibinfo {volume} {92}},\
  \bibinfo {pages} {074014} (\bibinfo {year} {2015})},\ \Eprint
  {http://arxiv.org/abs/1508.05653} {arXiv:1508.05653 [hep-ph]} \BibitemShut
  {NoStop}%
\bibitem [{\citenamefont {Gell-Mann}\ and\ \citenamefont
  {Levy}(1960)}]{GellMann:1960np}%
  \BibitemOpen
  \bibfield  {author} {\bibinfo {author} {\bibfnamefont {M.}~\bibnamefont
  {Gell-Mann}}\ and\ \bibinfo {author} {\bibfnamefont {M.}~\bibnamefont
  {Levy}},\ }\href {\doibase 10.1007/BF02859738} {\bibfield  {journal}
  {\bibinfo  {journal} {Nuovo Cim.}\ }\textbf {\bibinfo {volume} {16}},\
  \bibinfo {pages} {705} (\bibinfo {year} {1960})}\BibitemShut {NoStop}%
\bibitem [{\citenamefont {Bardeen}\ \emph {et~al.}(2003)\citenamefont
  {Bardeen}, \citenamefont {Eichten},\ and\ \citenamefont
  {Hill}}]{Bardeen:2003kt}%
  \BibitemOpen
  \bibfield  {author} {\bibinfo {author} {\bibfnamefont {W.~A.}\ \bibnamefont
  {Bardeen}}, \bibinfo {author} {\bibfnamefont {E.~J.}\ \bibnamefont
  {Eichten}}, \ and\ \bibinfo {author} {\bibfnamefont {C.~T.}\ \bibnamefont
  {Hill}},\ }\href {\doibase 10.1103/PhysRevD.68.054024} {\bibfield  {journal}
  {\bibinfo  {journal} {Phys. Rev. D}\ }\textbf {\bibinfo {volume} {68}},\
  \bibinfo {pages} {054024} (\bibinfo {year} {2003})},\ \Eprint
  {http://arxiv.org/abs/hep-ph/0305049} {arXiv:hep-ph/0305049} \BibitemShut
  {NoStop}%
\bibitem [{\citenamefont {Bando}\ \emph {et~al.}(1988)\citenamefont {Bando},
  \citenamefont {Kugo},\ and\ \citenamefont {Yamawaki}}]{Bando:1987br}%
  \BibitemOpen
  \bibfield  {author} {\bibinfo {author} {\bibfnamefont {M.}~\bibnamefont
  {Bando}}, \bibinfo {author} {\bibfnamefont {T.}~\bibnamefont {Kugo}}, \ and\
  \bibinfo {author} {\bibfnamefont {K.}~\bibnamefont {Yamawaki}},\ }\href
  {\doibase 10.1016/0370-1573(88)90019-1} {\bibfield  {journal} {\bibinfo
  {journal} {Phys. Rept.}\ }\textbf {\bibinfo {volume} {164}},\ \bibinfo
  {pages} {217} (\bibinfo {year} {1988})}\BibitemShut {NoStop}%
\bibitem [{\citenamefont {Nagahiro}\ \emph {et~al.}(2009)\citenamefont
  {Nagahiro}, \citenamefont {Roca}, \citenamefont {Hosaka},\ and\ \citenamefont
  {Oset}}]{Nagahiro:2008cv}%
  \BibitemOpen
  \bibfield  {author} {\bibinfo {author} {\bibfnamefont {H.}~\bibnamefont
  {Nagahiro}}, \bibinfo {author} {\bibfnamefont {L.}~\bibnamefont {Roca}},
  \bibinfo {author} {\bibfnamefont {A.}~\bibnamefont {Hosaka}}, \ and\ \bibinfo
  {author} {\bibfnamefont {E.}~\bibnamefont {Oset}},\ }\href {\doibase
  10.1103/PhysRevD.79.014015} {\bibfield  {journal} {\bibinfo  {journal} {Phys.
  Rev. D}\ }\textbf {\bibinfo {volume} {79}},\ \bibinfo {pages} {014015}
  (\bibinfo {year} {2009})},\ \Eprint {http://arxiv.org/abs/0809.0943}
  {arXiv:0809.0943 [hep-ph]} \BibitemShut {NoStop}%
\bibitem [{\citenamefont {Tawfiq}\ \emph {et~al.}(2001)\citenamefont {Tawfiq},
  \citenamefont {Korner},\ and\ \citenamefont {O'Donnell}}]{Tawfiq:1999cf}%
  \BibitemOpen
  \bibfield  {author} {\bibinfo {author} {\bibfnamefont {S.}~\bibnamefont
  {Tawfiq}}, \bibinfo {author} {\bibfnamefont {J.~G.}\ \bibnamefont {Korner}},
  \ and\ \bibinfo {author} {\bibfnamefont {P.~J.}\ \bibnamefont {O'Donnell}},\
  }\href {\doibase 10.1103/PhysRevD.63.034005} {\bibfield  {journal} {\bibinfo
  {journal} {Phys. Rev. D}\ }\textbf {\bibinfo {volume} {63}},\ \bibinfo
  {pages} {034005} (\bibinfo {year} {2001})},\ \Eprint
  {http://arxiv.org/abs/hep-ph/9909444} {arXiv:hep-ph/9909444} \BibitemShut
  {NoStop}%
\bibitem [{\citenamefont {Hiyama}\ \emph {et~al.}(2003)\citenamefont {Hiyama},
  \citenamefont {Kino},\ and\ \citenamefont {Kamimura}}]{Hiyama:2003cu}%
  \BibitemOpen
  \bibfield  {author} {\bibinfo {author} {\bibfnamefont {E.}~\bibnamefont
  {Hiyama}}, \bibinfo {author} {\bibfnamefont {Y.}~\bibnamefont {Kino}}, \ and\
  \bibinfo {author} {\bibfnamefont {M.}~\bibnamefont {Kamimura}},\ }\href
  {\doibase 10.1016/S0146-6410(03)90015-9} {\bibfield  {journal} {\bibinfo
  {journal} {Prog. Part. Nucl. Phys.}\ }\textbf {\bibinfo {volume} {51}},\
  \bibinfo {pages} {223} (\bibinfo {year} {2003})}\BibitemShut {NoStop}%
\bibitem [{\citenamefont {Wang}\ \emph {et~al.}(2020)\citenamefont {Wang},
  \citenamefont {Xiao}, \citenamefont {Chen}, \citenamefont {Liu},
  \citenamefont {Liu},\ and\ \citenamefont {Zhu}}]{Wang:2019spc}%
  \BibitemOpen
  \bibfield  {author} {\bibinfo {author} {\bibfnamefont {G.-J.}\ \bibnamefont
  {Wang}}, \bibinfo {author} {\bibfnamefont {L.-Y.}\ \bibnamefont {Xiao}},
  \bibinfo {author} {\bibfnamefont {R.}~\bibnamefont {Chen}}, \bibinfo {author}
  {\bibfnamefont {X.-H.}\ \bibnamefont {Liu}}, \bibinfo {author} {\bibfnamefont
  {X.}~\bibnamefont {Liu}}, \ and\ \bibinfo {author} {\bibfnamefont {S.-L.}\
  \bibnamefont {Zhu}},\ }\href {\doibase 10.1103/PhysRevD.102.036012}
  {\bibfield  {journal} {\bibinfo  {journal} {Phys. Rev. D}\ }\textbf {\bibinfo
  {volume} {102}},\ \bibinfo {pages} {036012} (\bibinfo {year} {2020})},\
  \Eprint {http://arxiv.org/abs/1911.09613} {arXiv:1911.09613 [hep-ph]}
  \BibitemShut {NoStop}%
\bibitem [{\citenamefont {Taylor}(1972)}]{osti_4661960}%
  \BibitemOpen
  \bibfield  {author} {\bibinfo {author} {\bibfnamefont {J.~R.}\ \bibnamefont
  {Taylor}},\ }\href {https://www.osti.gov/biblio/4661960} {\emph {\bibinfo
  {title} {Scattering Theory: The Quantum Theory on Nonrelativistic
  Collisions}}}\ (\bibinfo  {publisher} {New York, John Wiley \& Sons, Inc},\
  \bibinfo {year} {1972})\ pp.\ \bibinfo {pages} {382--410}\BibitemShut
  {NoStop}%
\bibitem [{\citenamefont {Aaij}\ \emph {et~al.}(2021)\citenamefont {Aaij} \emph
  {et~al.}}]{LHCb:2021chn}%
  \BibitemOpen
  \bibfield  {author} {\bibinfo {author} {\bibfnamefont {R.}~\bibnamefont
  {Aaij}} \emph {et~al.} (\bibinfo {collaboration} {LHCb}),\ }\href@noop {} {\
  (\bibinfo {year} {2021})},\ \Eprint {http://arxiv.org/abs/2108.04720}
  {arXiv:2108.04720 [hep-ex]} \BibitemShut {NoStop}%
\bibitem [{\citenamefont {Lin}\ and\ \citenamefont {Zou}(2019)}]{Lin:2019qiv}%
  \BibitemOpen
  \bibfield  {author} {\bibinfo {author} {\bibfnamefont {Y.-H.}\ \bibnamefont
  {Lin}}\ and\ \bibinfo {author} {\bibfnamefont {B.-S.}\ \bibnamefont {Zou}},\
  }\href {\doibase 10.1103/PhysRevD.100.056005} {\bibfield  {journal} {\bibinfo
   {journal} {Phys. Rev. D}\ }\textbf {\bibinfo {volume} {100}},\ \bibinfo
  {pages} {056005} (\bibinfo {year} {2019})},\ \Eprint
  {http://arxiv.org/abs/1908.05309} {arXiv:1908.05309 [hep-ph]} \BibitemShut
  {NoStop}%
\bibitem [{\citenamefont {Devanathan}(2002)}]{angular_momentom}%
  \BibitemOpen
  \bibfield  {author} {\bibinfo {author} {\bibfnamefont {V.}~\bibnamefont
  {Devanathan}},\ }\href {\doibase 10.1007/0-306-47123-x} {\emph {\bibinfo
  {title} {Angular Momentum Techniques in Quantum Mechanics}}}\ (\bibinfo
  {publisher} {Springer, Dordrecht},\ \bibinfo {year} {2002})\ pp.\ \bibinfo
  {pages} {24--99}\BibitemShut {NoStop}%
\end{thebibliography}%
\end{document}